\documentclass[12pt]{article}
\usepackage{amsfonts}
\usepackage{amssymb}
\usepackage{amsmath}
\usepackage{amsfonts}
\usepackage{amssymb}
\def\llsymbol#1{\@llsymbol{\@nameuse{c@#1}}}
\def\@llsymbol#1{\ifcase#1\or {}\or {'}\or {''}\or {'''}\or
{''''}\or {'''''}\or \else\@ctrerr\fi\relaz}
\newcounter{contador}
\newcommand{\letra}{
\stepcounter{equation}
\setcounter{contador}{\value{equation}}
\setcounter{equation}{0}
\renewcommand{\theequation}{\thecontador\alph{equation}}}
\newcommand{\antiletra}{
\renewcommand{\theequation}{\arabic{equation}}
\setcounter{equation}{\value{contador}}}

\usepackage{enumerate}
\usepackage{color}
\usepackage{indentfirst}
\usepackage{eucal}
\usepackage{amsmath}
\usepackage{amsfonts}
\usepackage{amssymb}
\usepackage{mathrsfs}
\usepackage{latexsym}
\usepackage{latexsym}
\begin{document}
%
%
\begin{center}{\bf Confluent and Double-Confluent Heun 
Equations: Convergence
of Solutions in 
Series of Coulomb Wavefunctions}\\
\vskip 0.4cm {L\'ea Jaccoud El-Jaick\footnote{Electronic adress: leajj@cbpf.br, tel: 55-21-21417178, fax: 55-21-21417160} and Bartolomeu D. B. Figueiredo\footnote{Electronic adress: barto@cbpf.br}} \\
{Centro Brasileiro de Pesquisas F\'{\i}sicas (CBPF)\\
 Rua Dr. Xavier Sigaud, 150 - 22290-180 - Rio de Janeiro, RJ, Brasil}
\end{center}

%
\begin{abstract}
\noindent
{The Leaver 
solutions in series of Coulomb wave functions for the
confluent Heun equation (CHE) are given by two-sided 
infinite series,
that is, by series where the summation
index $n$ runs from minus to plus infinity 
[E. W. Leaver, {J. Math. Phys.}
\textbf{27}, 1238 (1986)].  First we show that,  
in contrast to the D'Alembert test,  under certain conditions 
the Raabe test assures that the domains of convergence
of these solutions include an additional singular point. }
Further, by using a limit proposed by Leaver, 
we obtain solutions for the double-confluent Heun equation (DCHE).
In addition, we get solutions 
for the so-called Whittaker-Ince limit of the CHE and DCHE.  
For these four equations, new solutions are
generated by transformations of variables. 
In the second place, for each of the above equations 
we obtain one-sided series solutions
($n\geq 0$) by truncating on the left 
the two-sided series. {Finally we discuss the time 
dependence of the Klein-Gordon equation in 
two cosmological models and the spatial dependence
of the Schr\"{o}dinger
equation to a family of quasi-exactly solvable potentials.
For a subfamily of these potentials
we obtain infinite-series solutions which converge
and are bounded for all values of the independent variable, 
in opposition to a common belief}.

\tableofcontents

\end{abstract}
%
%
%
%

\section{Introduction}

\noindent

In 1986 Leaver \cite{leaver} found two types of solutions
in series of confluent hypergeometric functions
for the confluent Heun equation (CHE) and presented 
a limit procedure to generate solutions for the double-confluent
Heun equation (DCHE) out of solutions for the CHE. 
Later on \cite{eu,lea-1,lea-2} we have found that there 
are two other physically
relevant equations whose solutions can also be derived 
{from the Leaver solutions for the
CHE and DCHE by means of a procedure called Whittaker-Ince limit.}  
Further, from solutions of the CHE and/or DCHE,
we can find solutions for the Mathieu, Whittaker-Hill and spheroidal
equations \cite{leaver,decarreau1}.

In  view of the above {connections}, from
the inception we establish {
the convergence properties of Leaver's solutions}. 
We consider only the expansions in series of Coulomb 
wave functions which are given by a set of three solutions, one 
in series of regular confluent hypergeometric functions
and two in series of irregular functions. {By  
redefining the Coulomb functions, we avoid difficulties arising 
from the Leaver definitions and } find that the convergence of the solutions for the 
CHE and its Whittaker-Ince limit follows from the Raabe test. 
{We investigate as well the transformations 
of the Leaver solutions 
and their limiting and special cases}.

First we write the aforementioned equations, 
present the connections among them 
and call attention for the fact that there are 
three types of series expansions whose convergence
require different treatments. 
After that we introduce the D'Alembert and Raabe
tests for convergence and outline the structure of the article.

The Leaver form for the CHE is \cite{leaver}
\begin{eqnarray}
\label{gswe}
z(z-z_{0})\frac{d^{2}U}{dz^{2}}+(B_{1}+B_{2}z)
\frac{dU}{dz}+
\left[B_{3}-2\eta
\omega(z-z_{0})+\omega^{2}z(z-z_{0})\right]U=0, 
\end{eqnarray}
where $B_{i}$, $\eta$ and $\omega$ are constants. The points
$z=0$ and $z=z_{0}$ (if $z_{0}\neq 0$) are regular singular
points, whereas $z=\infty$ is an irregular point.
Since $z_0$ is free, by taking $z_{0}=0$ Leaver
obtained the DCHE
\begin{eqnarray}\label{dche}
z^{2}\frac{d^{2}U}{dz^{2}}+
\left(B_{1}+B_{2}z\right)\frac{dU}{dz}+
\left(B_{3}-2\eta \omega z+\omega^{2}z^{2}\right)U=0,
\qquad \left[B_{1}\neq 0, \ \omega\neq 0\right]
\end{eqnarray}
in which case $z=0$ and $z=\infty$ are both irregular
singularities.

In addition, the CHE and the DCHE admit a
(Whittaker-Ince) limit which changes
the nature of the singularity at $z=\infty$,
keeping unaltered the other singular points.
This limit is given by \cite{eu,lea-1}
\begin{eqnarray}\label{ince}
\omega\rightarrow 0, \ \
\eta\rightarrow
\infty, \ \mbox{such that }\  \ 2\eta \omega =-q,\ \
(\mbox{Whittaker-Ince limit})
\end{eqnarray}
where $q$ is a nonvanishing constant.
The Whittaker-Ince limit of the CHE is
\begin{eqnarray}\label{incegswe}
z(z-z_{0})\frac{d^{2}U}{dz^{2}}+(B_{1}+B_{2}z)
\frac{dU}{dz}+\left[B_{3}+q(z-z_{0})\right]U=0,\qquad (q\neq 0)
\end{eqnarray}
(if $q=0$ this equation can be transformed into a hypergeometric
equation), while the limit of the DCHE is
\begin{eqnarray}
\label{incedche}
z^2\frac{d^{2}U}{dz^{2}}+(B_{1}+B_{2}z)
\frac{dU}{dz}+
\left(B_{3}+qz\right)U=0,\qquad (q\neq 0, \ B_{1}\neq 0)
\end{eqnarray}
(if $q=0$ and/or $B_1=0$
the equation degenerates into a confluent hypergeometric
equation or simpler equations). This last
equation also follows from Eq. (\ref{incegswe})
when $z_0=0$ (Leaver's limit).

The Mathieu, Whittaker-Hill and spheroidal equations 
have been studied by themselves, but they are
particular cases of the above equations. The Mathieu equation
reads \cite{McLachlan}
\begin{eqnarray}\label{mathieu}
\frac{d^2w}{du^2}+\sigma^2[\mathrm{a}-2k^2\cos(2\sigma u)]w=0,
\quad (\mbox{Mathieu equation})
\end{eqnarray}
where $\mathrm{a}$ and $k$ are constants,
while $\sigma=1$ or $\sigma=i$ for the Mathieu or
modified Mathieu equation, respectively. This
equation is transformed into particular instances of
Eqs. (\ref{gswe}), (\ref{dche})
and (\ref{incegswe}) by the substitutions of variables
given in Eqs.  (\ref{mathieu-as-che}), (\ref{mathieu3})
and (\ref{mathieu-as-wil}), respectively. 
The Whittaker-Hill equation (WHE) can be
written in the form \cite{arscott,arscott3}
 \begin{eqnarray}\label{whe}
 \frac{d^2W}{du^2}+\varsigma^2\left[\vartheta-\frac{1}{8}\xi^{2}
 -(p+1)\xi\cos(2\varsigma u)+
 \frac{1}{8}\xi^{2}\cos(4\varsigma u)\right]W=0, \ \ (\mbox{WHE}).
 \end{eqnarray}
where $\vartheta$, $\xi$ and $p$ are parameters;
if $u$ is a real variable, this represents the WHE
when $\varsigma=1$ and the modified WHE when $\varsigma=i$.
The WHE reduces to the CHE (\ref{gswe}) and DCHE (\ref{dche})
by the substitutions (\ref{wheasgswe}) and (\ref{whe2}), respectively. 
Finally, the spheroidal equation reads \cite{meixner}
\begin{eqnarray}\label{esferoidal}
\frac{d}{dy}\left[\left( 1-y^2\right) \frac{dS(y)}{dy} \right]+
\left[\lambda+ \gamma^2(1-y^2)-
\frac{\mu^2}{1-\
y^2}\right] S(y)=0,
\end{eqnarray}
where $\gamma$, $\lambda$ and $\mu$ are constants. The
substitutions (\ref {esferoidal-1}) transform this into
a special case of the CHE (\ref{gswe}). 

On the other side, in general there are solutions given by
three different types of series, called 
two-sided infinite series, one-sided infinite series 
and finite series. These
take, respectively, the forms 
\begin{eqnarray}\label{tipos-de-solucoes}
\displaystyle
\sum_n a_{n}^{\nu}\ f_n^{\nu}(z):
=
\sum_{n=-\infty}^{\infty} a_{n}^{\nu}\ f_n^{\nu}(z),\qquad
 \sum_{n=0}^{\infty} b_n\ g_n(z),\qquad
\displaystyle \sum_{n=0}^{N} b_n\ g_n(z),
\end{eqnarray}
where $f_{n}^{\nu}(z)$
and $g_{n}(z)$ are functions of the independent
variable $z$,  $N$ is a non-negative
integer and $\nu$ is a 
parameter which does not appear in the differential equations. 
In the present case, the series
coefficients $a_{n}$ and $b_{n}$   
satisfy three-term recurrence relations. 
No finite-series solutions are known for the Mathieu 
equation (\ref{mathieu}) nor for 
the Whittaker-Ince limit (\ref{incedche})
of the DCHE. 

Two-sided infinite series, the only  
considered by Leaver, are necessary to assure the 
convergence of solutions of equations in which 
there is no free parameter,
as in scattering problems \cite{eu} or in some wave equations 
in curved spacetimes \cite{eu2}. In such cases
the parameter $\nu$ must be determined as solutions
of a transcendental (characteristic) equation. 
However, when truncated on the left ($n\geq 0$), 
the two-sided infinite series give one-sided infinite series
which are useful for equations having a free parameter; 
in turn, these lead to solutions given by finite series
if the parameters of the equation satisfy certain constraints.

Finite-series solutions are suitable for  
quasi-exactly solvable (QES) problems, that is, 
for quantum-mechanical problems where one part of 
energy spectrum and the respective eigenfunctions
can be computed explicitly \cite{turbiner,ushveridze1}.
For QES problems obeying equations of the
Heun family \cite{lea-1}, that part of
the spectrum may be derived from finite-series 
solutions if these are known. Indeed, a problem 
is also said to be QES if it 
admits solutions given by finite series whose coefficients
necessarily satisfy three-term  or higher order
recurrence relations \cite{kalnins}, and 
is said to be exactly solvable if its 
solutions can be expressed by hypergeometric functions.

The convergence of two-sided infinity series is 
obtained from { ratios between successive terms 
of the series, namely, from
\begin{eqnarray}
L_1(z)=\left|\frac{a_{n+1}^{\nu}f_{n+1}^{\nu}(z)}
{a_{n}^{\nu}f_{n}^{\nu}(z)}\right| \text{ when } n\to\infty,
\qquad
L_2(z)=\left|\frac{a_{n-1}^{\nu}f_{n-1}^{\nu}(z)}
{a_{n}^{\nu}f_{n}^{\nu}(z)}\right|\text{ when } n\to-\infty.
\end{eqnarray}
By the D'Alembert test (used by Leaver)  
the series converges in the intersection of the regions of $z$ where 
$L_1<1$ and $L_2<1$, and diverges otherwise (except if 
$L_1=L_2=1$, in which case the test is inconclusive). However, Leaver's definitions for the 
Coulomb wave functions give ratios 
presenting square roots (except if $\eta=0$), a fact that makes     
difficult to deal with the convergence. We avoid this by using 
alternative definitions that, in addition, enable us to apply the Raabe test \cite{watson,knopp} 
to decide about the convergence when $L_1=L_2=1$. In effect, if for some value of $z$, 
\begin{eqnarray}\label{Raabe}
L_1(z)=1+\frac{A}{n}+O\left(\frac{1}{n^2}\right),
\qquad
L_2(z)=1+\frac{B}{|n|}+O\left(\frac{1}{n^2}\right),\quad [n\to\infty]
\end{eqnarray}
where $A$ and $B$ are constants, then the Raabe test states that 
the series converges if   
$A<-1$ and $B<-1$,
and diverges otherwise (for  $A=B=-1$ the test is
inconclusive).} For one-sided series 
the convergence may be enhanced since we use only
the limit $L_1$, while for finite series the convergence must
be decided from the behavior of each term of the series.

Furthermore, by using transformations of variables 
we find four sets of two-sided solutions 
instead of one set as in Leaver. 
By the Raabe test, under certain conditions 
these solutions converge 
absolutely for $|z|\geq|z_0|$ or $|z-z_0|\geq|z_0|$ 
rather than for $|z|>|z_0|$ or $|z-z_0|>|z_0|$;  
the one-sided solutions given by series of regular
confluent hypergeometric functions converge 
for $|z|\geq 0$. 
Nevertheless, the behavior of each 
solution for $z\to\infty$ 
must be analyzed carefully because, in
computing  $L_1(z)$ and $L_2(z)$, we assume that
$z$ is bounded. We have also to examine the behavior of the solutions
at the finite singular points because the series appear multiplied
by factors which may become unbounded at
such points.

Taking into account the above considerations, 
we proceed as follows. In Section 2 we deal  with 
the two-sided
infinite expansions for CHE (\ref{gswe}), its limiting and 
particular cases. For the spheroidal equation,
we get the Meixner solutions in series of
Bessel functions \cite{meixner} instead of the Chu and Stratton 
solutions \cite{chu} mentioned by Leaver.
%

In Section 3 we 
apply the analysis of section 2 to one-sided expansions. 
These are generated by truncating  on the left-hand side the two-sided series:
by requiring that $n\geq 0$ we
determine the parameter $\nu$ of 
the two-sided solutions as function of the
parameters of the differential equations. 
For $\eta\neq0$ we find three types of recurrence relations 
for the series coefficients; for $\eta=0$, only two types.
Conditions for finite-series solutions are obtained.

In section 4, we consider examples which
illustrate the consequences of Raabe test. In particular, we show that 
the Schr\"{o}dinger equation for some quasi-exactly solvable 
potentials admits infinite-series solutions which are convergent
and bounded for all values of the independent variable. Thus, 
in addition to the energy levels resulting from finite series, 
in principle it is possible to get additional energy levels as 
solutions of characteristic equations
corresponding to the infinite series.

In section 5 there are some final considerations and remarks
on open problems concerning the Leaver solutions. Appendix A gives 
the normalization used for the Coulomb functions
and takes the case $\eta=0$ as a criterion to decide 
in favor of one of two possibilities for 
the ratio between successive Coulomb functions. 
The derivation of the recurrence relations
for one-sided and two-sided series  
is given in Appendix B. Appendix C
lists one-sided solutions for the
CHE, omitted in section III.

\section{Two-sided Series Expansions}
In this section we examine separately the 
two-sided expansions 
for each of the four equations: CHE, DCHE and their 
Whittaker-Ince limits. In the Whittaker-Ince limit, 
the expansions in series of Coulomb functions give
solutions in series of Bessel functions,
excepting the solutions (\ref{incedche-3})
for the Whittaker-Ince limit of the DCHE. 
The solutions of the CHE with $\eta=0$ are also expressible by 
series of Bessel functions. 
In all cases, given an initial set of solutions,
new sets are generated by transformations of
variables which preserve the form of
the differential equations. 
Notice that the linear independence of the functions used as
basis for the series expansions will impose
restrictions on the characteristic parameter
$\nu$ and/or on some parameters of the differential equations.

In Eqs. (\ref{Meixner-1}-\ref{Meixner-3}) we recover
the Meixner solutions for the spheroidal equation
as particular cases of the solutions for the CHE, while 
in Eqs. (\ref{Heine-1}-\ref{Heine-3}) we recover 
the usual solutions in series of Bessel functions for the Mathieu
equation as particular cases of the solutions for 
the Whittaker-Ince limit of the CHE. In both cases
the Raabe test modifies the domains of convergence
of some solutions.

\subsection{ Solutions for the CHE}

The initial set of solutions, $\mathbb{U}_1(z)$, is
reconstructed in Appendix B. It reads 
%
\begin{eqnarray}\label{forma-de-leaver}
\mathbb{U}_{1}(z) =z^{-\frac{B_2}{2}}\displaystyle
\sum_{n}  b_{n}^{1}
\mathscr{U}_{n+\nu}\left(\eta,\omega{z}\right),
\quad
\mathscr{U}_{n+\nu}\left(\eta,\omega{z}\right)=
\left(\phi_{n+\nu},
\psi_{n+\nu}^{+},
\psi_{n+\nu}^{-}\right)\left(\eta,\omega{z}\right),
\end{eqnarray}
where $\sum_n$ denotes two-sided series,  
$\phi_{n+\nu}$ and $\psi_{n+\nu}^{\pm}$
represent the Coulomb wave functions
defined in Eqs. (\ref{U-n}) and (\ref{fi}),
and the coefficients $b_n^{1}$ satisfy three-term recurrence
relations.  
So, we have a set of three expansions, one
in series of regular confluent hypergeometric
functions and two in series of irregular
functions. This set corresponds to Leaver's solutions
who have used the definitions (\ref{L-1})
and (\ref{L-2}) for the Coulomb functions. In addition, 
if $U(z)=U(B_{1},B_{2},B_{3}; z_{0},\omega,\eta;z)$
denotes an arbitrary solution of the CHE, we
can find other solutions by means of
the transformations ${T}_{1},\ {T}_{2},\ T_{3}$ and
$T_{4}$  which operate as \cite{lea-1,decarreau1}
\begin{eqnarray}\label{transformacao1}
\begin{array}{l}
{T}_{1}U(z)=z^{1+B_{1}/z_{0}}
U(C_{1},C_{2},C_{3};z_{0},\omega,\eta;z),
\vspace{2mm}\\
{T}_{2}U(z)=(z-z_{0})^{1-B_{2}-B_{1}/z_{0}}U(B_{1},D_{2},D_{3};
z_{0},\omega,\eta;z),
\vspace{2mm}\\
T_{3}U(z)=U(B_{1},B_{2},B_{3}; z_{0},-\omega,-\eta;z),
\vspace{2mm}\\
T_{4}U(z)=
U(-B_{1}-B_{2}z_{0},B_{2},
B_{3}+2\eta\omega z_{0};z_{0},-\omega,
\eta;z_{0}-z),
\end{array}
\end{eqnarray}
where
\begin{eqnarray}\label{constantes-C-D}
\begin{array}{l}
\begin{array}{l}
C_{1}=-B_{1}-2z_{0}, \qquad
C_{2}=2+B_{2}+\frac{2B_{1}}{z_{0}},\qquad C_{3}=B_{3}+
\left(1+\frac{B_{1}}{z_{0}}\right)
\left(B_{2}+\frac{B_{1}}{z_{0}}\right)\end{array},\vspace{2mm}\\
\begin{array}{l}
D_{2}=2-B_{2}-\frac{2B_{1}}{z_{0}},\qquad D_{3}=B_{3}+
\frac{B_{1}}{z_{0}}\left(\frac{B_{1}}{z_{0}}
+B_{2}-1\right).
\end{array}
\end{array}
\end{eqnarray}

These transformations allow constructing a
group with 4 sets of two-sided series 
$\mathbb{U}_i$ ($i=1,\cdots,4$) where 
coefficients $b_n^{i}$ satisfy recurrence
relations having the form
\begin{eqnarray}\label{recursion-1}
\alpha_{n}^{i} b_{n+1}^{i}+\beta_{n}^{i} b_{n}^{i}+
\gamma_{n}^{i} b_{n-1}^{i}=0,
\qquad [-\infty<n<\infty]
\end{eqnarray}
where $\alpha_{n}^{i}$, $\beta_{n}^{i}$ 
and $\gamma_{n}^{i}$ depend on
the parameters of the differential equation as well as on
$\nu$ and $n$. These relations lead to transcendental (characteristic)
equations given as a sum of two infinite continued fractions.
By omitting the superscripts of $\alpha_{n}^{i}$, $\beta_{n}^{i}$ 
and $\gamma_{n}^{i}$, the characteristic equations read
\begin{eqnarray}
\label{recursion-2}
\beta_{0}=\frac{\alpha_{-1}\gamma_{0}}{\beta_{-1}-}\ \frac{\alpha_{-2}
\gamma_{-1}}{\beta_{-2}-}\ \frac{\alpha_{-3}\gamma_{-2}}
{\beta_{-3}-}\cdots+\ \frac{\alpha_{0}\gamma_{1}}{\beta_{1}-}
\ \frac{\alpha_{1}\gamma_{2}}
{\beta_{2}-}\ \frac{\alpha_{2}\gamma_{3}}{\beta_{3}-}\cdots
\end{eqnarray}
which are equivalent to the vanishing of the determinants 
of infinite tridiagonal matrices, as in Eq. (\ref{matriz-tridiagonal}). 
If the CHE has no free parameter, Eq. (\ref{recursion-2})
may be used to find the possible values of $\nu$
(characteristic parameter); if the CHE has 
an arbitrary parameter, Eq. (\ref{recursion-2})
permits to find the values of that parameter
corresponding to suitable values of $\nu$.

To analyze the properties of the solutions
we write explicitly each
of the three solutions, instead of using the abbreviated
form (\ref{forma-de-leaver}). Thus, we denote by
$\mathbb{U}_1=\big(U_{1},U_{1}^{+},
U_{1}^{-}\big)$ the solutions
associated respectively with
the functions $\big(\phi_{n+\nu}$,\
$\psi_{n+\nu}^{+}$,\ $\psi_{n+\nu}^{-}\big)$. This gives
the solutions (\ref{par1-nu}) which,
by the transformations (\ref{transformacao1}),
generate the three sets of solutions that have not
been considered by Leaver. The four sets 
of two-sided solutions are denoted by
\begin{eqnarray}\label{notation-1}
\mathbb{U}_i(z)=\left[U_{i}(z),{U}_{i}^{+}(z),
{U}_{i}^{-}(z)\right], \qquad i=1, \cdots,4,
\end{eqnarray}
if $\eta\neq 0$; if $\eta=0$ the notation is
given in Eq. (\ref{U-eta=0}). From these solutions  
we get eight sets of one-sided solutions which are 
denoted by
\begin{eqnarray}\label{notation-2}
\mathbb{\mathring{U}}_i(z)=\left[\mathring{U}_{i}(z),
\mathring{U}_{i}^{+}(z),
\mathring{U}_{i}^{-}(z)\right], \qquad i=1, \cdots,8,
\end{eqnarray}
and do not depend on $\nu$. In fact they are generated  
by expressing the parameter $\nu$ of each $\mathbb{U}_i$
as two different functions of the parameters of the
CHE. The convergence 
of the solutions $\mathbb{\mathring{U}}_i$ is obtained  
by considering only the limits  $n\to \infty$ in 
the computations given in this section.

\subsubsection{ The four sets of the solutions}
Explicitly  the first set $\mathbb{U}_1$,
given in Eq. (\ref{forma-de-leaver}), reads 
\begin{equation}
\label{par1-nu}
\begin{array}{l}
U_{ 1}(z) =e^{i\omega z}  \displaystyle
\sum _{n} 
\frac{b_{n}^{1}\ [2i\omega z]^{n+\nu+1-\frac{B_2}{2}}}{\Gamma[2n+2\nu+2]}
\Phi\left[n+\nu+1+i\eta,2n+2\nu+2;-2i\omega z\right]
\vspace{2mm}\\
{U}_{ 1}^{\pm}(z) =e^{\pm i\omega z} \displaystyle
\sum _{n} 
\frac{b_{n}^{1}\ 
[-2i\omega z]^{n+\nu+1-\frac{B_2}{2}}}{\Gamma[n+\nu+1\mp i\eta]}
\Psi\left[n+\nu+1\pm i\eta,2n+2\nu+2;\mp 2i\omega z\right], 
\end{array}
\end{equation}
where, in the recurrence relations (\ref{recursion-1}) for
$b_{n}^{1}$, 
\begin{eqnarray}
\label{nu10}
\begin{array}{l}
\alpha_{n}^{1}  =  \frac{2i\omega z_{0}\left[n+\nu+2-\frac{B_{2}}{2}\right]
\left[n+\nu+1-\frac{B_{2}}{2}-\frac{B_{1}}{z_{0}}\right]}
{(2n+2\nu+2)(2n+2\nu+3)},
\vspace{.1cm} \\
\beta_{n}^{1} =  -B_{3}-\eta \omega z_{0}-\left(n+\nu+1-\frac{B_{2}}{2}\right)
\left(n+\nu+\frac{B_{2}}{2}\right)
-\frac{\eta \omega z_{0}\left[B_2-2\right]
\left[B_2+\frac{2B_{1}}{z_{0}}\right]}
{(2n+2\nu)(2n+2\nu+2)},
\vspace{.2cm} \\
\gamma_{n}^{1}  = -\frac{2i\omega z_{0}\left[n+\nu+\frac{B_{2}}{2}-1\right]
\left[n+\nu+\frac{B_{2}}{2}+\frac{B_{1}}{z_{0}}\right]
(n+\nu+i\eta)(n+\nu-i\eta)}
{(2n+2\nu-1)(2n+2\nu)}.
\end{array}
\end{eqnarray}

By applying the transformation ${T}_{3}$ on $\mathbb{U}_1$,
we find the equivalence
%
\begin{eqnarray}\label{T-3}
T_{3}\left[U_{ 1}(z),{U}_{1}^{+}(z),U_{1}^{-}(z)\right]
\quad \Leftrightarrow\quad
\left[U_{ 1}(z),{U}_{1}^{-}(z),U_{1}^{+}(z)\right].
\end{eqnarray}
Precisely, we find $T_{3}\left[U_{ 1},{U}_{1}^{+},U_{1}^{-}\right]
=\left[\bar{U}_{ 1},\bar{U}_{1}^{-},\bar{U}_{1}^{+}\right]$ with
\begin{eqnarray*}
\begin{array}{l}
\bar{U}_{ 1}(z) =e^{i\omega z}  \displaystyle
\sum _{n} 
\frac{\bar{b}_{n}^{1}\ [-2i\omega z]^{n+\nu+1-{B_2}/{2}}}{\Gamma[2n+2\nu+2]}
\Phi\left(n+\nu+1+i\eta,2n+2\nu+2;-2i\omega z\right),\vspace{2mm}\\
\bar{U}_{ 1}^{\pm}(z) =e^{\pm i\omega z} \displaystyle
\sum _{n} 
\frac{\bar{b}_{n}^{1}\ 
[2i\omega z]^{n+\nu+1-{B_2}/{2}}}{\Gamma[n+\nu+1\mp i\eta]}
\Psi\left(n+\nu+1\pm i\eta,2n+2\nu+2;\mp 2i\omega z\right), 
\end{array}
\end{eqnarray*}
where the recurrence relations for
$\bar{b}_{n}^{1}$ are 
\begin{eqnarray*}
-\alpha_{n}^{1} \ \bar{b}_{n+1}^{1}+\beta_{n}^{1} \ \bar{b}_{n}^{1}
-\gamma_{n}^{1} \ \bar{b}_{n-1}^{1}=0.
\end{eqnarray*}
Up to a multiplicative constant independent of $n$,
we can set $\bar{b}_{n}^{1}=(-1)^nb_{n}^{1}$ in
order to establish relation (\ref{T-3}). Thus, the
transformation $T_3$ is ineffective in the present case. 
The remaining transformations allow to form
a group constituted by four sets of solutions,
namely,
\begin{eqnarray}\label{t-1}
\mathbb{U}_{ 1}(z),
\quad \mathbb{U}_{ 2}(z)=T_2\mathbb{U}_{ 1}(z);
\quad \mathbb{U}_{ 3}(z)=T_4\mathbb{U}_{ 1}(z),
\quad \mathbb{U}_{ 4}(z)=T_4\mathbb{U}_{ 2}(z)
=T_1\mathbb{U}_{ 3}(z).
\end{eqnarray}
In fact, from the explicit forms of these sets, 
one verifies that ${T}_{1}$
does not generate new solutions when applied on $\mathbb{U}_{ 1}$
and $\mathbb{U}_{ 2}$;
similarly, $T_{2}$ has no effect on $\mathbb{U}_{ 3}$ and $\mathbb{U}_{ 4}$.

The three sets of solutions generated by the preceding 
transformations are written below.
\begin{eqnarray}\label{par2-nu}
\begin{array}{l}
U_{2}(z)= f_2(z)e^{i\omega z}  \displaystyle
\sum _{n}
\frac{b_{n}^{2}[i\omega z]^{n}}{\Gamma[+2\nu+2]}
\Phi\left[n+\nu+1+i\eta,2n+2\nu+2;-2i\omega z\right]\vspace{2mm}\\
{U}_{ 2}^{\pm}(z) =f_2(z)e^{\pm i\omega z}  \displaystyle
\sum _{n} 
\frac{{b}_{n}^{2}
[-2i\omega z]^{n}}{\Gamma[n+\nu+1\mp i\eta]} 
\Psi\left[+\nu+1\pm i\eta,2n+2\nu+2;\mp 2i\omega z\right] 
\end{array}
\end{eqnarray}\\
where $f_2=f_2(z)=z^{\nu+({B_1}/{z_0})+({B_2}/{2})}
\left(z-z_0\right)^{1-B_2-({B_1}/{z_0})}$.
The coefficients for the recurrence relations are given by
\begin{eqnarray}\label{par2-nu-b}
\begin{array}{l}
\alpha_{n}^{2}  =  \frac{2i\omega z_{0}\left[n+\nu+\frac{B_{2}}{2}\right]
\left[n+\nu+1+\frac{B_{2}}{2}+\frac{B_{1}}{z_{0}}\right]}
{(2n+2\nu+2)(2n+2\nu+3)},
\qquad
\beta_{n}^{2} =  \beta_{n}^{1},
\vspace{.2cm} \\
\gamma_{n}^{2}  = -\frac{2i\omega z_{0}\left[n+\nu+1-\frac{B_{2}}{2}\right]
\left[n+\nu-\frac{B_{2}}{2}-\frac{B_{1}}{z_{0}}\right]
(n+\nu+i\eta)(n+\nu-i\eta)}
{(2n+2\nu-1)(2n+2\nu)}.
\end{array}
\end{eqnarray}
The transformation ${T_4}$ acting on $\mathbb{U}_1$ gives the set $\mathbb{U}_3=\mathbb{U}_3(z)$, namely,
\begin{eqnarray}\label{par3-nu}
\begin{array}{l}
U_{3}= f_3e^{i\omega[z-z_0] }  \displaystyle
\sum _{n} 
\frac{b_{n}^{3}\left[2i\omega(z-z_0)\right]^{n}}{\Gamma[2n+2\nu+2]}
\Phi\left[n+\nu+1+i\eta,2n+2\nu+2;-2i\omega (z-z_0)\right],\vspace{2mm}\\
U_{3}^{\pm} =f_3e^{\pm i\omega [z-z_0]} \displaystyle
\sum _{n} 
\frac{b_{n}^{3}\left[2i\omega(z_0-z)\right]^{n}}{\Gamma[n+\nu+1\mp i\eta]}\times\vspace{2mm}\\
\hspace{5cm}
\Psi\left[n+\nu+1\pm i\eta,2n+2\nu+2;\mp 2i\omega (z-z_0)\right], 
\end{array}
\end{eqnarray}\\
where $f_3=f_3(z)=\left(z-z_0\right)^{\nu+1-({B_2}/{2})}$
and, in the recurrence relations,
\begin{eqnarray}
\begin{array}{l}
\alpha_{n}^{3}  = -\frac{2i\omega z_{0}\left[n+\nu+2-\frac{B_2}{2}\right]
\left[n+\nu+1+\frac{B_2}{2}+\frac{B_1}{z_0}\right]}
{(2n+2\nu+2)(2n+2\nu+3)},
\qquad
\beta_{n}^{3} =  \beta_{n}^{1},
\vspace{.2cm} \\
\gamma_{n}^{3}  = \frac{2i\omega z_0\left[n+\nu-1+\frac{B_2}{2}\right]
\left[n+\nu-\frac{B_2}{2}-\frac{B_1}{z_0}\right]
\left(n+\nu+i\eta\right)\left(n+\nu-i\eta\right)}
{(2n+2\nu-1)(2n+2\nu)}.
\end{array}
\end{eqnarray}
The fourth set, obtained by applying $ {T_1} $ on
$\mathbb{U}_3 $, reads
\begin{eqnarray}\label{par4-nu}
\begin{array}{l}
U_{4}= f_4e^{i\omega [z-z_0]} \displaystyle
\sum _{n} 
\frac{b_{n}^{4}\left[2i\omega(z-z_0)\right]^{n}}{\Gamma[2n+2\nu+2]}
\Phi\left[n+\nu+1+i\eta,2n+2\nu+2;-2i\omega (z-z_0)\right],\vspace{2mm}\\
U_{4}^{\pm} =f_4e^{\pm i\omega [z-z_0]} \displaystyle
\sum _{n} 
\frac{b_{n}^{4}\left[2i\omega(z_0-z)\right]^{n}}
{\Gamma[n+\nu+1\mp i\eta]} \times\vspace{2mm}\\
\hspace{5cm}
\Psi\left[n+\nu+1\pm i\eta,2n+2\nu+2;\mp 2i\omega (z-z_0)\right], 
\end{array}
\end{eqnarray}
where $f_4=f_4(z)=z^{1+(B_1/{z_0})} \left(z-z_0\right)^{\nu-({B_2}/{2})-({B_1}/{z_0})}$ 
and, in recurrence relations for $b_{n}^{4}$,
\begin{eqnarray}
\begin{array}{l}
\alpha_{n}^{4}  = -\frac{2i\omega z_0\left[n+\nu+\frac{B_2}{2}\right]
\left[n+\nu+1-\frac{B_2}{2}-\frac{B_1}{z_0}\right]}
{(2n+2\nu+2)(2n+2\nu+3)},
\qquad
\beta_{n}^{4} =  \beta_{n}^{1},
\vspace{.2cm} \\
\gamma_{n}^{4}  = \frac{2i\omega z_0\left[n+\nu+1-\frac{B_2}{2}\right]
\left[n+\nu+\frac{B_2}{2}+\frac{B_1}{z_0}\right]
(n+\nu+i\eta)(n+\nu-i\eta)}
{(2n+2\nu-1)(2n+2\nu)}.
\end{array}
\end{eqnarray}

If there is no free parameter in the CHE, 
$\nu$ must be determined as solutions 
of a characteristic equation. However, 
by considering the form of the solutions and respective 
recurrence relations for the series coefficients, we find that
\begin{eqnarray}\label{series-0}
2\nu\ \text{ and } \ \nu\mp i\eta \ \text{  cannot be integers }
\end{eqnarray}
for two-sided series.
The restriction $\nu\mp i\eta\neq$integer assures that 
factors $1/\Gamma(n+\nu+1\pm i\eta)$ which appear
in $U_i^{\pm}(z)$ are not zero
for any value of $n$; assures as well that the factors
$(n+\nu+i\eta)(n+\nu-i\eta)$ in $\gamma_{n}^{i}$
do not vanish for any $n$. In fact, such restriction is necessary
to have two-sided infinite series for the three solutions
in each of the four sets $\mathbb{U}_i$. 

The condition $2\nu\neq$integer is necessary in 
order to avoid two terms linearly dependent in
the series of $U_i^{\pm}(z)$. Indeed, suppose that $2\nu=$integer
in the solutions $U_1^{\pm}(z)$. These are 
series expansions in terms of
\[B_n^{\pm}(z)=[-2i\omega z]^{n+\nu+1}\Psi[n+\nu+1\pm i\eta,
2n+2\nu+2;\mp2i\omega z].\]
By setting $n=n_1$ and using (\ref{kummer}), we find
\[B_{n_1}^{\pm}(z)=\pm(-1)^{\nu\mp\nu}[-2i\omega z]^{-n_1-\nu}
\Psi[-n_1-\nu\pm i\eta,
-2n_1-2\nu;\mp 2i\omega z].\]
Hence, $B_{n_1}^{\pm}$ and
$B_{n_2}^{\pm}$ are proportional to each other 
for some $n=n_2$ such that $n_1+n_2+1=-2\nu$.
Similar results are found for the other solutions
$U_i^{\pm}(z)$. On the other side, 
by supposing that $2\nu\neq$integer,
the functions $\Phi(a,c;y)$ which appear in 
$U_i(z)$ are well defined because the parameter
$c=2n+2\nu+2$ cannot be a negative integer. 
Nevertheless, see in the paragraph containing 
Eqs. (\ref{eta-1}) and (\ref{eta-2}) some remarks concerning
the case $\eta=0$.

According to Eqs. (\ref{wronskians}), if the 
conditions (\ref{series-0}) are true, the 
three hypergeometric functions are linearly independent 
and each one can be written as a combination of the others
by means of (\ref{continuation2}). In this case, 
in a common region of validity, we can
write one solution of a given set as a superposition 
of the others. However,  
the three series are really doubly 
infinite ($-\infty<n<\infty$) if, 
in addition to (\ref{series-0}),
$\nu$ satisfies the restrictions
\begin{eqnarray}\label{four-conditions}
\begin{array}{l}
\nu\pm\frac{B_2}{2} \text{  and  } \nu\pm\left(\frac{B_1}{z_0}+\frac{B_2}{2}\right)
\text{  are not integers}.
\end{array}
\end{eqnarray}
These conditions 
assure that $\alpha_{n}^{i}$ and
$\gamma_{n}^{i}$ do not vanish for any value of $n$. In effect, 
if $\alpha_{n}^{i}=0$
for some $n=N_1$, the series should begin at $n=N_1+1$
in order to assure the validity of the theory
of the three-term recurrence relations; by the same reason,
if $\gamma_{n}^{i}=0$ for some $n=N_2$, the series should
terminate at $n=N_2-1$.

Notice that, for two-side solutions, 
\begin{eqnarray}\label{alfa-beta}
\beta_{n}^{i}=\beta_{n}^{1},\qquad
\alpha_{n}^{i}\gamma_{n+1}^{i}=\alpha_{n}^{1}\gamma_{n+1}^{1}, \qquad[i=2,3,4].
\end{eqnarray}
Thence, Eq. (\ref{recursion-2}) implies that all
the solutions $\mathbb{U}_i$ satisfy the same characteristic 
equation and, consequently, the parameter $\nu$ 
takes the same values in all solutions. 
In addition, as noticed by Leaver, the  
characteristic equations are periodic in $\nu$ with period $1$.
In effect, in order to indicate that the coefficients 
depend on $\nu$ we rewrite the recurrence relations as 
$\alpha_{n}^{\nu} b_{n+1}^{\nu}+\beta_{n}^{\nu} b_{n}^{\nu}+
\gamma_{n}^{\nu} b_{n-1}^{\nu}=0$ or as the following
tridiagonal matrix equation:
\begin{eqnarray}\label{matriz-tridiagonal}
\left[
\begin{array}{lllllll}
. & \ . & \ . && & \vspace{1mm}\\
        & \gamma_{n}^{\nu} & \beta_{n}^{\nu} &
\alpha_{n}^{\nu} & \vspace{1mm}\\
%
%
     &  & \gamma_{n+1}^{\nu} &
\beta_{n+1}^{\nu} & \alpha_{n+1}^{\nu} & \vspace{1mm}\\
  &  &  & 
  \gamma_{n+2}^{\nu} &
\beta_{n+2}^{\nu} & \alpha_{n+2}^{\nu} & \vspace{1mm}\\
   &   & &  &  \ .
&  \ .&   \ . 
%
\end{array}
\right]
\left[\begin{array}{l}
\ .\vspace{1mm}\\
%
%
b_{n-1}^{\nu}\vspace{1mm}\\
%
b_{n}^{\nu}\vspace{1mm}\\
b_{n+1}^{\nu}\vspace{1mm}\\
%
%
\ .
\end{array}
\right]=\mathbf{0}
\qquad [-\infty<n<\infty]
\end{eqnarray}
where $\mathbf{0}$ denotes the null column vector.  
Thence, the values for $\nu$ may be determined 
by requiring that the determinant of the above 
matrix vanishes. However, as
\[\gamma_{n}^{\nu+1}=\gamma_{n+1}^{\nu},\qquad 
\beta_{n}^{\nu+1}=\beta_{n+1}^{\nu},\qquad
\alpha_{n}^{\nu+1}=\alpha_{n+1}^{\nu},\qquad \cdots\]
and $-\infty<n<\infty$, the matrix and
its determinant are not modified by the replacement
$\nu\to\nu+1$ (or $\nu\to\nu+N$, where $N$ is any integer).

Further, 
if (\ref{series-0}) and (\ref{four-conditions}) are
fulfilled, all coefficients can be written in terms of  
$b_{n}^{1}$. Up to multiplicative constants independent 
of $n$, we have
\begin{eqnarray}\label{coeff}
&\begin{array}{l}
b_{n}^{2}=\frac{\Gamma\left[n+\nu+2-\frac{B_2}{2}\right]\ 
\Gamma\left[n+\nu+1-\frac{B_1}{z_0}-\frac{B_2}{2}\right]}
{\Gamma\left[n+\nu+\frac{B_2}{2}\right]\ 
\Gamma\left[n+\nu+1+\frac{B_1}{z_0}+\frac{B_2}{2}\right]}\ b_{ n}^{1},
\qquad
b_{n}^{3}=\frac{(-1)^n\ \Gamma\left[n+\nu+1-\frac{B_1}{z_0}-\frac{B_2}{2}\right]}
{\Gamma\left[n+\nu+1+\frac{B_1}{z_0}+\frac{B_2}{2}\right]}\  b_{n}^{1},
\end{array}\nonumber \\ 
&\begin{array}{l}
b_{n}^{4}=\frac{(-1)^n\ \Gamma\left[n+\nu+2-\frac{B_2}{2}\right]}
{\Gamma\left[n+\nu+\frac{B_2}{2}\right]}\ b_{n}^{1}.
\end{array}
\end{eqnarray}
As an example, we consider the solutions $W(u)$ for the WHE (\ref{whe}).
These may be obtained from
the solutions $U(z)$ of the CHE (\ref{gswe}) by taking
\begin{eqnarray}\label{wheasgswe}
\left.
\begin{array}{l}
W(u)=U(z), \quad  z=\cos^{2}(\varsigma u),\quad  [\varsigma=1,i]
\quad\Rightarrow \quad z_{0}=1,
\vspace{2mm}\\
B_{1}=-\frac{1}{2}, \quad B_{2}=1, \quad
B_{3}=\frac{(p+1)\xi-\vartheta}{4}, \quad i\omega=\frac{\xi}{2},
\quad i\eta=\frac{p+1}{2}
\end{array}\right\}
\begin{array}{l}
\mbox{WHE  as}\vspace{2mm}\\
\mbox{a CHE.}
\end{array}
\end{eqnarray}
Thus, the solutions $\mathbb{U}_i
=\left({U}_i,{U}_i^{+},{U}_i^{-}\right)$
lead to four sets of solutions
\letra
\begin{equation}\label{sol-WHill}
\mathbb{W}_i(u)
=\left[{W}_i(u)=U_i(z),\ {W}_i^{+}(u)=U_i^{+}(z),\ {W}_i^{-}(u)=U_i^{-}(z)\right].
\end{equation}
In this case, the coefficients of the recurrence relations for 
$b_{n}^{1}$ simplify to
\begin{eqnarray}\label{examining}
\begin{array}{l}
\alpha_{n}^{1}=\frac{i\omega}{2},\quad 
\beta_{n}^{1}=-B_3-\eta\omega-\left[n+\nu+\frac{1}{2}\right]^2,\quad
\gamma_{n}^{1}=-\frac{i\omega}{2}[n+\nu+i\eta][n+\nu-i\eta],
\end{array}
\end{eqnarray}
whereas Eqs. (\ref{coeff}) reduce to
\begin{eqnarray}
\begin{array}{l}
b_{n}^{2}=\left(n+\nu+\frac{1}{2}\right)b_{\ n}^{1},\qquad 
b_{n}^{3}=(-1)^nb_{n}^{1},\qquad 
b_{n}^{4}=(-1)^n\left(n+\nu+\frac{1}{2}\right)b_{n}^{1}.
\end{array}
\end{eqnarray}
%

\subsubsection{ Convergence and asymptotic behaviour}

The D'Alembert test implies two subgroups
of solutions since $\mathbb{U}_{ 1}$
and $\mathbb{U}_{ 2}$ converge for any finite $z$ such that 
$|z|>|z_0|$, whereas $\mathbb{U}_{ 3}$ and
$\mathbb{U}_{ 4}$ converge for
$|z-z_0|>|z_0|$. However,
by the Raabe test they may converge also at
$|z|=|z_0|$ and $|z-z_0|=|z_0|$ under the conditions
\antiletra
\begin{equation}\label{convergencia1-0}
|z|\geq|z_0|  \text{ if }
\begin{cases}\text{Re}\left[  B_2+\frac{B_1}{z_0}\right] <1 \mbox{ in }
\mathbb{U}_{1},\vspace{2mm} \\
\text{Re}\left[  B_2+\frac{B_1}{z_0}\right] >1\mbox{ in }
\mathbb{U}_{2};
\end{cases}\ \ 
|z-z_0|\geq|z_0| \text{ if }
\begin{cases}\text{Re}\left[ \frac{B_1}{z_0}\right] >{-}1\mbox{ in }
\mathbb{U}_{3},
\vspace{2mm} \\
\text{Re}\left[ \frac{B_1}{z_0}\right] <{-}1\mbox{ in }
\mathbb{U}_{4},
\end{cases}
\end{equation}
where the restrictions on parameters of the equation
are necessary only to assure convergence at $|z|=|z_0|$ or $|z-z_0|=|z_0|$.
In particular, 
for the solutions of the WHEs we find
\begin{eqnarray}\label{conv-WHE-as-CHE}
&&\begin{array}{l}
|\cos(\varsigma u)|\geq 1 \text{ in } \mathbb{W}_1,
\qquad|\cos(\varsigma u)|>1 \text{ in } \mathbb{W}_2,
\end{array}\nonumber
\\
&&\begin{array}{l}
|\sin(\varsigma u)|\geq 1 \text{ in } \mathbb{W}_3,
\qquad|\sin(\varsigma u)|> 1 \text{ in } \mathbb{W}_4.
\end{array}
\end{eqnarray}
Thence the two-sided solutions are useless for the WHE (that is, for $\varsigma=1$ and 
$u$=real), but may be useful for the modified
WHE ($\varsigma=i$, $u$=real). If $\text{Re}[B_2+(B_1/z_0)]=1$ and 
$\text{Re}[B_1/z_0]=-1$ in (\ref{convergencia1-0}), the Raabe test
becomes inconclusive in the sense that the solutions may
converge or diverge at $|z|=|z_0|$ or $|z-z_0|=|z_0|$.

To obtain the conditions (\ref{convergencia1-0}) it
is sufficient to consider the
convergence of the first set of solutions. The results
for the other sets arise from transformations (\ref{transformacao1})
applied in the order given in (\ref{t-1}). Thus, by using the
form (\ref{forma-de-leaver}) for the first set,
the domains of convergence follow from the ratios
\begin{eqnarray*}
\vline\frac{b_{n+1}^{1}
\mathscr{U}_{n+\nu+1}(\eta,\omega z)}
{b_{n}^{1}\mathscr{U}_{n+\nu}(\eta,\omega z)}\vline\text{ when }n\to\infty\quad \text{and} \qquad
\vline\frac{b_{n-1}^{1}
\mathscr{U}_{n+\nu-1}(\eta,\omega z)}
{b_{n}^{1}\mathscr{U}_{n+\nu}(\eta,\omega z)}\vline\text{ when }n\to-\infty.
\end{eqnarray*}
The ratios $b_{n+1}^{1}/b_{n}^{1}$ and $b_{n-1}^{1}/b_{n}^{1}$
come from the relations
$\alpha_{n}^{1}b_{n+1}^{1}+\beta_{n}^{1}b_{n}^{1}+
\gamma_{n}^{1}b_{n-1}^{1}=0$ which, when $n\rightarrow\pm \infty$, yield
\begin{eqnarray}\label{place}
&&\begin{array}{l}
 {i\omega z_0}\left[1-
\frac{1}{n}\left(B_2+\frac{B_1}{z_0}-\frac{1}{2}\right)+
O\left(\frac{1}{n^2}\right)\right]
\frac{{b}_{n+1}^{1}}{{b}_{n}^{1}}-2n\left[ n+2\nu+1+O\left(\frac{1}{n}\right)\right]
\end{array}\nonumber\\
&&\begin{array}{l}
-{i\omega z_0}n
\left[n+2\nu+B_2+\frac{B_1}{z_0}-\frac{1}{2}+O\left(\frac{1}{n}\right)\right]
\frac{{b}_{n-1}^{1}}{{b}_{n}^{1}}=0.
\end{array}
\end{eqnarray}
Hence, the minimal solution for ${{b}_{n+1}^{1}}/{{b}_{n}^{1}}$
when $n\to \infty$ is
\letra
\begin{equation}\label{minimal}
\begin{array}{l}
\frac{b_{n+1}^{1}}{b_{n}^{1}}\sim
\frac{\omega z_0}{2i}\left[1+\frac{1}{n}\left(B_2+\frac{B_1}{z_0}-\frac{3}{2}\right) \right]
\ \Rightarrow
\frac{b_{n-1}^{1}}{b_{n}^{1}}\sim
\frac{2i}{\omega z_0}\left[ 1-\frac{1}{n}\left(B_2+\frac{B_1}{z_0}-
\frac{3}{2}\right)\right],
\end{array}
\end{equation}
and the minimal solution for ${{b}_{n-1}^{1}}/{{b}_{n}^{1}}$
when $n\to -\infty$ is
\begin{equation}\label{minimal-2}
\begin{array}{l}
\frac{b_{n-1}^{1}}{b_{n}^{1}}\sim
\frac{i\omega z_0}{2n^2}\left[1-\frac{1}{n}\left(2\nu+B_2+
\frac{B_1}{z_0}-\frac{3}{2}\right) \right] \Rightarrow 
\frac{b_{n+1}^{1}}{b_{n}^{1}}\sim
\frac{2n^2}{i\omega z_0}\left[ 1+\frac{1}{n}\left(2\nu+B_2+\frac{1}{2}
+\frac{B_1}{z_0}\right)\right].\end{array}
\end{equation}

On the other hand, from relations (\ref{bessel}) and (\ref{hankel})
we find that, for finite values of $z$,
\antiletra
\begin{eqnarray}\label{phi-psi}
\begin{array}{ll}\displaystyle
n\to\infty: &\frac{\phi_{n+\nu+1}}{\phi_{n+\nu}}\sim \frac{i\omega z}{2n^2}
\left[1-\frac{1}{n}\left(2\nu+\frac{5}{2}\right)\right],\qquad
\frac{\psi_{n+\nu+1}^{\pm}}{\psi_{n+\nu}^{\pm}}
\sim \frac{2i}{\omega z}
\left[1-\frac{1}{2n}\right],\vspace{2mm}\\
\displaystyle
n\to -\infty: &\frac{\mathscr{U}_{n+\nu-1}}{\mathscr{U}_{n+\nu}}\sim \frac{2n^2}{i\omega z}
\left[1+\frac{1}{n}\left(2\nu+\frac{1}{2}\right)\right],
\qquad \mathscr{U}_{n+\nu}=\left(\phi_{n+\nu},\psi_{n+\nu}^{(\pm)}\right).
  \end{array}
\end{eqnarray}
Thence, by means of (\ref{minimal}), we find
\letra
\begin{eqnarray}\label{phi}
n\to\infty:\left\{
\begin{array}{l}
\frac{b_{ n+1}^{1} \phi_{n+\nu+1}}{b_{n}^{1}\ \phi_{n+\nu}}
\sim\frac{\omega^2 z_0z}{4n^2}
\left[1+\frac{1}{n}\left(B_2+\frac{B_1}{z_0}-2\nu-4\right)\right],\vspace{2mm}\\
\frac{b_{n+1}^{1} \psi_{n+\nu+1}^{\pm}}
{b_{n}^{1} \psi_{n+\nu}^{\pm}}
\sim\frac{z_0}{z}\left[1+\frac{1}{n}\left({B_2}-2+\frac{B_1}{z_0}\right) \right],
\end{array}\right.
\end{eqnarray}
and, by means of (\ref{minimal-2}),
\begin{eqnarray}\label{psi}\begin{array}{ll}
n\to -\infty:&\frac{b_{n-1}^{1} \mathscr{U}_{n+\nu-1}}
{b_{n}^{1} \mathscr{U}_{n+\nu}}
\sim\frac{z_0}{z}\left[1-\frac{1}{n}\left({B_2}-2+\frac{B_1}{z_0}\right) \right],
\quad \mathscr{U}_{n+\nu}=\left(\phi_{n+\nu},\psi_{n+\nu}^{\pm}\right).
 \end{array}
\end{eqnarray}
From these limits we get
\antiletra
\begin{equation}\label{n+}
\begin{array}{l}
\vline\frac{b_{n+1}^{1}
\phi_{n+\nu+1}}
{b_{n}^{1}\phi_{n+\nu}}\vline\sim\vline\frac{\omega^2 z_0z}{4n^2}\vline \text{ if }n\to\infty,\qquad
\vline\frac{b_{n-1}^{1}
\phi_{n+\nu-1}}
{b_{n}^{1}\phi_{n+\nu}}\vline
\sim
\frac{|z_0|}{|z|}\left[1+\frac{\text{Re}
\left(B_2-2+\frac{B_1}{z_0}\right)}{|n|}\right]\text{ if }n\to-\infty\end{array}
\end{equation}
and
\begin{eqnarray}\label{n-}
&&\begin{array}{l}
\vline\frac{b_{n+1}^{1}
\psi_{n+\nu+1}^{\pm}}
{b_{n}^{1}\psi_{n+\nu}^{\pm}}\vline\sim\frac{|z_0|}{|z|}\left[1+\frac{1}{n}\text{Re}
\left(B_2-2+\frac{B_1}{z_0}\right)\right]\text{ if }n\to\infty\end{array},\nonumber\vspace{2mm}\\
&&\begin{array}{l}
\vline\frac{b_{n-1}^{1}
\psi_{n+\nu-1}^{\pm}}
{b_{n}^{1}\psi_{n+\nu}^{\pm}}\vline
\sim
\frac{|z_0|}{|z|}\left[1+\frac{1}{|n|}\text{Re}
\left(B_2-2+\frac{B_1}{z_0}\right)\right] \text{ if } n\to-\infty.\end{array}
\end{eqnarray}
So, by the D'Alembert test the series converge absolutely for
$|z|>|z_0|$ because the right-hand sides of
(\ref{n+}) and (\ref{n-}) are $<1$. However, if $|z|=|z_0|$,
by the expressions (\ref{Raabe}) for
the Raabe test, the series converge even for $|z|=|z_0|$
provided that the numerators of $|n|$ in
(\ref{n+}) and (\ref{n-}) are  $< -1$, that is,
\begin{eqnarray*}\label{exception-0}
\mbox{if }\ \text{Re}\left[B_2+({B_1}/{z_0})\right]<1,\ \mbox{ the series in }\
\mathbb{U}_1(z) \mbox{ converge for }
|z|\geq|z_0|.
\end{eqnarray*}
If $\text{Re}\left[B_2+({B_1}/{z_0})\right]>1$, the series diverge and,
if $\text{Re}\left[B_2+({B_1}/{z_0})\right]=1$, the test
is inconclusive. The convergence regions (\ref{convergencia1-0})
for the other sets of solutions
are obtained by transforming the parameters and the
variable $z$ of $\mathbb{U}_1$ in accordance with
Eqs. (\ref{t-1}). Only the limit $n\to\infty$ 
is relevant for one-sided series ($n\geq 0$) and, 
then, the solutions $\mathring{U}_i$ converge for any finite 
value $z$ in virtue of the first limit given in (\ref{n+}).
The convergence of $\mathring{U}_i^{\pm }$  is similar to
that of ${U}_i^{\pm }$.

Since the previous regions 
of convergence were derived by 
supposing that $z$ is finite, now we consider the behaviour of the 
solutions at $z=\infty$. By using (\ref{asymptotic-confluent}) 
we find that, when $z\to\infty$,
\begin{equation} \label{bound-2-0}
U_{1}(z)\sim
\begin{array}{l}
e^{i\omega z}[i\omega z]^{-i\eta-\frac{B_2}{2}}
\sum
\frac{b_{n}^{1}}{\Gamma[n+\nu+1-i\eta]} + 
 e^{-i\omega z}[-2i\omega z]^{i\eta-\frac{B_2}{2}}
\sum\frac{(-1)^{n-\nu-1+i\eta}b_{n}^{1}}
{\Gamma[n+\nu+1+i\eta]}.
\end{array} 
\end{equation}
Thus, $U_{1}$ may  
be unbounded by virtue of the exponential factors. This
is consistent with the fact that, if conditions (\ref{series-0}) 
and (\ref{four-conditions}) are satisfied,    
then $U_{1}(z)$ can be written as a linear combination
of ${U}_{1}^{+}(z)$ and  ${U}_{1}^{-}(z)$.
In fact, when $z\to \infty$, Eq. (\ref{asymptotic-confluent1}) gives
\begin{eqnarray}\label{bound-0}
\begin{array}{ll}
{U}_{1}^{+}(z)\sim e^{i\omega z}[-2i\omega z]^{-i\eta-\frac{B_2}{2}}
\sum\frac{b_{n}^{1}}{\Gamma[n+\nu+1-i\eta]},\quad& 
-\frac{3\pi}{2}<\arg(-2i\omega z)<\frac{3\pi}{2};\vspace{2mm}\\
{U}_{1}^{-}(z)\sim e^{-i\omega z}[-2i\omega z]^{i\eta-\frac{B_2}{2}}
\sum\frac{(-1)^{n-\nu-1+i\eta}\ b_{n}^{1}}
{\Gamma[n+\nu+1+i\eta]},\quad& 
-\frac{3\pi}{2}<\arg(2i\omega z)<\frac{3\pi}{2}.
\end{array}
\end{eqnarray}
Thus, the  series in ${U}_{1}^{\pm}$ converge at
$z=\infty$ but one of them may be unbounded depending on
the exponential factors. For instance, 
if $\mathrm{Re}(i\omega z)\to \infty$, ${U}_{1}^{+}\to\infty$
but ${U}_{1}^{-}$ is bounded.

\subsubsection{ The case $\eta= 0$, the spheroidal 
and Mathieu equations}

Taking $\eta=0$ and keeping fixed the
other parameters, the previous solutions are rewritten
in series of Bessel functions
of the first kind, $J_{\kappa}(y)$, and in series of the
first and the second Hankel functions, $H_{\ \kappa}^{(1)}(y)$
and $H_{\ \kappa}^{(2)}(y)$.
We also include the Bessel
functions $Y_{\kappa}(y)$ of the second kind. These four functions are
denoted by $Z_{\ \kappa}^{(j)}(y)$ -- or by $\mathscr{C}_{\ \kappa}^{(j)}(y)$ --
according as \cite{arscott,abramowitz}
\begin{equation}\label{Z2}
Z_{\ \kappa}^{(1)}(y)=J_{\kappa}(y),\quad
Z_{\ \kappa}^{(2)}(y)=Y_{\kappa}(y),\quad
Z_{\ \kappa}^{(3)}(y)=H_{\ \kappa}^{(1)}(y),\quad
Z_{\ \kappa}^{(4)}(y)=H_{\ \kappa}^{(2)}(y).
\end{equation}
There are connections among these functions.
For example, the relation $Y_{\kappa}=[H_{\ \kappa}^{(1)}-
H_{\ \kappa}^{(2)}]/(2i)$
permits to obtain the expansion in series
of $Y_{\kappa}$ as a linear combination of the expansions
in series of Hankel functions. Thus, we get four sets of solutions,
each containing four solutions. These sets are written as
\begin{eqnarray}\label{U-eta=0}
U_{\ i}^{(j)}(z)=\left[U_{\ i}^{(1)}(z),U_{\ i}^{(2)}(z),
U_{\ i}^{(3)}(z),
U_{\ i}^{(4)}(z)\right],\quad i=1,2,3,4,
\end{eqnarray}
where the right-hand side corresponds to the
Bessel functions (\ref{Z2}). For one-sided series
there are eight sets $\mathring{U}_{\ i}^{(j)}$. 
The solutions $\mathbb{U}_1$ lead to $U_{\ 1}^{(j)}$ which, 
in turn, give 
the other sets by means of the
transformations (\ref{t-1}), that is,
\begin{eqnarray}
{U}_{\ 2}^{(j)}(z)=T_2{U}_{\ 1}^{(j)}(z);
\qquad {U}_{\ 3}^{(j)}(z)=T_4{U}_{\ 1}^{(j)}(z),
\qquad {U}_{\ 4}^{(j)}(z)=T_1{U}_{\ 3}^{(j)}(z).
\end{eqnarray}

Thus, we put $\eta=0$ in $\mathbb{U}_1$ given in (\ref{par1-nu}), 
use the relations (\ref{bessel-1})
together with \cite{abramowitz}
\[\Gamma(2z)=2^{2z-1}\Gamma(z)\Gamma\left[z+({1}/{2})\right]
/\sqrt{\pi},\]
and redefine the coefficients as $a_{n}^{1}=i^nb_{n}^{1}/\Gamma(n+\nu+1)$.
So, we find
\letra
\begin{eqnarray}\label{U1}
U_{\ 1}^{(j)}(z)=z^{\frac{1}{2}-\frac{B_2}{2}}\sum_n
a_{ n}^{1}
Z_{n+\nu+\frac{1}{2}}^{(j)}(\omega z),\quad\left[2\nu\neq 0, \pm 1,\pm 2,
\cdots \right]
\end{eqnarray}
In the recurrence relations (\ref{recursion-1}) for $a_{n}^{1}$,
we have
\begin{eqnarray}\label{U1-b}
\begin{array}{l}
\alpha_{n}^{1}=
\frac{\omega z_0\left[ n+\nu+2-\frac{B_2}{2}\right]
\left[n+\nu+1-\frac{B_1}{z_0}-\frac{B_2}{2}\right]}
{\left(2n+2\nu+3\right)},\vspace{.2cm} \\
\beta_{n}^{1} =  -\left(n+\nu+1-\frac{B_{2}}{2}\right)
\left(n+\nu+\frac{B_{2}}{2}\right)-B_{3},
\vspace{.2cm} \\
\gamma_{n}^{1}  = \frac{\omega z_{0}\left[n+\nu+\frac{B_{2}}{2}-1\right]
\left[n+\nu+\frac{B_{2}}{2}+\frac{B_{1}}{z_{0}}\right]}
{(2n+2\nu-1)}.
\end{array}
\end{eqnarray}
The other sets are given by ($\beta_{n}^{i}=\beta_{n}^{1},
\ 2\nu\neq 0, \pm 1, \pm 2,\cdots$):
\antiletra
\begin{eqnarray}\label{U2}
\begin{array}{l}
\displaystyle U_{\ 2}^{(j)}(z)=z^{\frac{B_1}{z_0}+\frac{B_2}{2}-\frac{1}{2}}
[z-z_0]^{1-B_2-\frac{B_1}{z_0}}\sum_n
a_{n}^{2}
Z_{n+\nu+\frac{1}{2}}^{(j)}(\omega z),
\vspace{2mm}\\
 \alpha_{n}^{2}=
\frac{\omega z_0\left[ n+\nu+\frac{B_2}{2}\right]
\left[n+\nu+1+\frac{B_1}{z_0}+\frac{B_2}{2}\right]}
{\left(2n+2\nu+3\right)},\qquad
\gamma_{n}^{2}  = \frac{\omega z_{0}\left[n+\nu+1-\frac{B_{2}}{2}\right]
\left[n+\nu-\frac{B_1}{z_0}-\frac{B_2}{2}\right]}
{(2n+2\nu-1)};\hspace{9,5mm}
\end{array}
\end{eqnarray}
\begin{eqnarray}\begin{array}{l}
\displaystyle U_{\ 3}^{(j)}(z)=
[z-z_0]^{\frac{1}{2}-\frac{B_2}{2}}\sum_n
a_{n}^{3}
Z_{n+\nu+\frac{1}{2}}^{(j)}[\omega (z-z_0)],
\vspace{2mm}\\
\alpha_{n}^{3}=-
\frac{\omega z_0\left[ n+\nu+2-\frac{B_2}{2}\right]
\left[n+\nu+1+\frac{B_1}{z_0}+\frac{B_2}{2}\right]}
{\left(2n+2\nu+3\right)},\qquad
\gamma_{n}^{3}  = -\frac{\omega z_{0}\left[n+\nu-1+\frac{B_{2}}{2}\right]
\left[n+\nu-\frac{B_1}{z_0}-\frac{B_2}{2}\right]}
{(2n+2\nu-1)};
\end{array}
\end{eqnarray}
\begin{eqnarray}\begin{array}{l}
\displaystyle U_{\ 4}^{(j)}(z)=z^{1+\frac{B_1}{z_0}}
[z-z_0]^{-\frac{1}{2}-\frac{B_1}{z_0}-\frac{B_2}{2}}\sum_n
a_{n}^{4}
Z_{n+\nu+\frac{1}{2}}^{(j)}[\omega (z-z_0)],
\vspace{2mm}\\
\alpha_{n}^{4}=-
\frac{\omega z_0\left[ n+\nu+\frac{B_2}{2}\right]
\left[n+\nu+1-\frac{B_1}{z_0}-\frac{B_2}{2}\right]}
{\left(2n+2\nu+3\right)},\qquad
\gamma_{n}^{4}  = -\frac{\omega z_{0}\left[n+\nu+1-\frac{B_{2}}{2}\right]
\left[n+\nu+\frac{B_{1}}{z_{0}}+\frac{B_{2}}{2}\right]}
{(2n+2\nu-1)}.\ \ \
\end{array}\end{eqnarray}
By using the relations \cite{abramowitz}
\begin{eqnarray}\label{abramowitz}
\begin{array}{ll}
J_{\kappa}\left(y e^{i\pi}\right)=e^{i\pi\kappa}J_{\kappa}(y),&
Y_{\kappa}\left(y e^{i\pi}\right)=e^{-i\pi\kappa}Y_{\kappa}(y)+
2i\cos(\pi \kappa)J_{\kappa}(y), \vspace{2mm}\\
H_{\ \kappa}^{(1)}\left(y e^{i\pi}\right)=-e^{-i\pi\kappa}H_{\ \kappa}^{(2)}(y),
\quad&
H_{\ \kappa}^{(2)}\left(y e^{i\pi}\right)=
e^{i\pi\kappa}H_{\ \kappa}^{(1)}(y)+2\cos(\pi\kappa)H_{\ \kappa}^{(2)}(y)
\end{array}
\end{eqnarray}
with $\kappa=n+\nu+(1/2)$, we find that
the change $\omega\to-\omega$ does not
lead to new independent solutions. In this sense, once
more the transformation $T_3$ is ineffective.

Also in the present case ($\eta=0$) conditions
(\ref{four-conditions}), that is,
\letra
\begin{eqnarray} \label{eta-infinita}
\begin{array}{l}
\nu\pm\frac{B_2}{2} \text{  and  } \nu\pm\left(\frac{B_1}{z_0}+\frac{B_2}{2}\right)
\text{  are not integers}, \qquad[\text{ see relations (\ref{four-conditions})}]
\end{array}
\end{eqnarray}
are necessary in order to 
have two-sided infinite series, and 
relations (\ref{coeff}) hold for 
the coefficients $a_n^{1}$. On the other side,
the restrictions (\ref{series-0}) are replaced by 
conditions $2\nu\neq 0,\pm 1,\pm 2\cdots$ which assure the 
independence of the Bessel function:
\begin{eqnarray}\label{eta-independente}
2\nu\neq 0,\pm 1,\pm2,\cdots, \qquad[\text{independence of 
Bessel functions}].
\end{eqnarray}
In fact, it is
necessary that $\nu\neq\pm 1/2,\pm 3/2,\cdots$ in
order to avoid two linearly dependent functions of
integer order, 
like $Z_{\ell}^{(j)}(y)$ and $Z_{-\ell}^{(j)}(y)$
[$Z_{\ell}=(-1)^{\ell}Z_{-\ell}$], where $\ell$ is zero
or positive integer. In addition, $\nu\neq 0,\pm 1,\pm 2,\cdots$ 
assures the independence of the Hankel functions 
in the same series: on the contrary, we would
have functions like $H_{\ell+({1}/{2})}$ and
$H_{-\ell-({1}/{2})}$ which 
are proportional to each other since \cite{nist}
\antiletra
\begin{eqnarray}\label{eta-1}
H_{-\ell-({1}/{2})}^{(1)}(y)= i(-1)^{\ell}H_{\ell+({1}/{2})}^{(1)}(y),
\qquad
H_{-\ell-({1}/{2})}^{(2)}(y)=-i (-1)^{\ell}H_{\ell+({1}/{2})}^{(2)}(y).
\end{eqnarray}
However, for series of Bessel functions
of the first and second kind,  we have
\begin{eqnarray}\label{eta-2}
J_{-\ell-({1}/{2})}(y)=(-1)^{\ell+1}Y_{\ell+({1}/{2})}(y),
\qquad
Y_{-\ell-({1}/{2})}^{(2)}(y)=(-1)^{\ell}J_{\ell+({1}/{2})}^{(2)}(y),
\end{eqnarray}
that is, for $\nu=0,\pm 1, \pm 2,\cdots$ 
the functions $J_{\ell+({1}/{2})}$ 
and $J_{-\ell-({1}/{2})}$ 
(or, $Y_{\ell+({1}/{2})}$ and $Y_{-\ell-({1}/{2})}$) are 
linearly independent. 
In spite of this, by assuming that $\nu\neq0,\pm 1, \pm 2,\cdots$ 
also for $J$ and $Y$ we guarantee that all of the solutions 
(\ref{Meixner-1}) and (\ref{Meixner-2}) for the spheroidal 
equation are two-sided since $\alpha_{n}^{1}$ and 
$\gamma_{n}^{1}$ do not vanish for $-\infty<n<\infty$.

On the other side, 
for two-sided series the domains 
of convergence are again
given by (\ref{convergencia1-0}) 
with  ${U}_{\ i}^{{(j)}}$ substituted for $\mathbb{U}_i$. For 
the one-sided series the solutions $\mathring{U}_{\ i}^{(1)}$,  
in series of Bessel functions of
the first kind, converge for any finite $z$.
The behavior of the solutions at $z=\infty$ can be found
from the fact that, for $\kappa$ fixed and
$|y|\to\infty$ \cite{abramowitz},
\begin{eqnarray}\label{comportamento-infinito}
\begin{array}{l}
J_{\kappa}(y)\sim\sqrt\frac{2}{\pi y}
\cos\left[y-\frac{\kappa\pi}{2}-\frac{\pi}{4}\right],\quad
Y_{\kappa}(y)\sim\sqrt\frac{2}{\pi y}
\sin\left[y-\frac{\kappa\pi}{2}-\frac{\pi}{4}\right]:\ 
|\arg{y}|<\pi;\vspace{2mm}\\
H_{\ \kappa}^{(1)}(y)\sim\sqrt\frac{2}{\pi y}\ 
e^{i\left [y-\frac{\kappa\pi}{2}-\frac{\pi}{4}\right]}:\qquad  -\pi<\arg{y}<2\pi;\vspace{2mm}\\
H_{\ \kappa}^{(2)}(y)\sim\sqrt\frac{2}{\pi y}\ e^{-i\left [y-\frac{\kappa\pi}{2}-\frac{\pi}{4}\right]}:\qquad
-2\pi<\arg{y}<\pi.
\end{array}
\end{eqnarray}


Now we consider the Meixner solutions. The substitutions
\letra
\begin{equation}\label{esferoidal-1}
\begin{array}{l}
y=1-2z,\qquad S(y)=z^{\frac{\mu}{2}}\ \left[z-1\right]^{\frac{\mu}{2}}U(z)
\ \  \Leftrightarrow \ \  S(y)\propto\left[y^2-1\right]^{\frac{\mu}{2}}U\left(z=\frac{1-y}{2}\right)
\end{array}
\end{equation}
transform the spheroidal wave equation (\ref{esferoidal})
into
\begin{eqnarray*}
z(z-1)\frac{d^{2}U}{dz^{2}}+
\left[
-\left(\mu+1\right)+
\left(2\mu+2\right)z
\right]\frac{dU}{dz}+
\left[
\mu
\left(\mu+1\right)-\lambda+4\gamma^{2}z(z-1)
\right]U=0,
\end{eqnarray*}
which is the CHE (\ref{gswe}) with parameters
\begin{equation}\label{esferoidal-2}
\begin{array}{l}
z_0=1,\quad B_1=-\mu-1,\quad B_2=2\mu+2,\quad B_3=\mu(\mu+1)-\lambda,\quad \omega=\pm 2\gamma, \quad \eta=0.
\end{array}
\end{equation}
Instead of $Z_{\ \kappa}^{(j)}(v)$, Meixner used the functions
$\psi_{\ \kappa}^{(j)}(v)$ which are given by \cite{meixner}
\antiletra
\begin{equation}\label{psi-meixner}
\psi_{\ \kappa}^{(j)}(v)=\sqrt{\pi/(2v)}\  Z_{\kappa+(1/2)}^{(j)}(v),
\end{equation}
in analogy with the definitions of the spherical 
Bessel functions ${j}_{\kappa}$,
 ${y}_{\kappa}$, ${h}_{\ \kappa}^{(1)}$ and  ${h}_{\ \kappa}^{(2)}$ \cite{nist}. 
So, by taking $\omega= -2\gamma$ and using this notation, we get
\begin{eqnarray}\label{Meixner-1}
\begin{array}{l}\displaystyle
S_{\ 1}^{(j)}(\mu,y)= \left[ \frac{y+1}{y-1}\right]^{\frac{\mu}{2}}
\sum_n
a_{n}^{1}
\psi_{n+\nu}^{(j)}[ \gamma (y-1)],\qquad S_{\ 2}^{(j)}(\mu,y)=S_{\ 1}^{(j)}(-\mu,y),\vspace{2mm}
\\
\alpha_{n}^{1}=
\frac{ 2 \gamma (n+\nu+1-\mu) (n+\nu+1)}
{\left(2n+2\nu+3\right)},\quad 
\beta_{n}^{1} = (n+\nu)(n+\nu+1)- \lambda ,\quad
%
\gamma_{n}^{1}  = \frac{2 \gamma (n+\nu+\mu) (n+\nu)}
{(2n+2\nu-1)};
\end{array}
\end{eqnarray}
and 
\begin{eqnarray}\label{Meixner-2}
\begin{array}{l}\displaystyle
S_{\ 3}^{(j)}(\mu,y)= \left[ \frac{y-1}{y+1}\right]^{\frac{\mu}{2}}
\sum_n (-1)^n
a_{n}^{1}
\psi_{n+\nu}^{(j)}[ \gamma (y+1)],\qquad S_{4}^{(j)}(\mu,y)=S_{3}^{(j)}(-\mu,y).
\end{array}
\end{eqnarray}
For these solutions, conditions 
(\ref{eta-infinita}) and (\ref{eta-independente}) reduces to
\begin{eqnarray*} 
2\nu\neq 0,\pm1,\pm 2,\cdots,\qquad \nu\pm(\mu+1)\neq\text{integer}.
\end{eqnarray*} 
The Meixner solutions are given by $S_{\ 2}^{(j)}(\mu,y)$ 
and $S_{\ 4}^{(j)}(\mu,y)$. By the D'Alembert test 
$S_{\ i}^{(j)}$ converge for
$|y-1|>2$ (if $i=1,2$) or for $|y+1|>2$ (if $i=3,4$), 
as stated in \cite{meixner,erdelyi3}. 
However, by the Raabe test they may converge at $|y-1|=2$ or $|y+1|=2$
because relations (\ref{convergencia1-0}) 
and (\ref{esferoidal-2}) yield
\begin{equation}\label{Meixner-3}
|y-1|\geq 2 , \text{ if }
\begin{cases}\text{Re}\left(  \mu\right) <0 \mbox{ in }
{S}_{\ 1}^{(j)},\vspace{2mm} \\
\text{Re}\left(  \mu\right) >0\mbox{ in }
{S}_{\ 2}^{(j)};
\end{cases}\qquad
|y+1|\geq 2 \text{ if }
\begin{cases}\text{Re}\left( \mu\right) <0\mbox{ in }
{S}_{\ 3}^{(j)},\vspace{2mm} \\
\text{Re}\left( \mu\right) >0\mbox{ in }
{S}_{\ 4}^{(j)}
\end{cases}\
\end{equation}
(if $\text{Re }(\mu)=0$, the test is inconclusive).

%
On the other side, the solutions
$w(u)$ for the Mathieu equation (\ref{mathieu}), 
considered as a particular case of the CHE, may be 
obtained by setting
\begin{eqnarray}\label{mathieu-as-che}
\left.
\begin{array}{l}
w(u)=U(z), \quad z=\cos^{2}(\frac{\sigma u}{2}),\quad [\sigma=1,i] 
\quad\Rightarrow\quad
z_{0}=1, \vspace{2mm}\\
B_{1}=-{1}/{2}, \quad B_{2}=1, \quad
B_{3}=2k^2-\mathrm{a}, \quad \omega=4 k,\quad
\eta=0
\end{array}\right\}
\begin{array}{l}
\mbox{Mathieu eq. }\vspace{2mm}\\
\mbox{as a CHE,}
\end{array}
\end{eqnarray}
where $U(z)$ are the solutions for CHE with $\eta=0$. Thus, from 
the previous solutions $U_{\ i}^{(j)}(z)$  
we get four sets of two-sided solutions $w_{\ i}^{(j)}(u)$ 
\letra
\begin{eqnarray}\label{?}
\begin{array}{ll}
w_{\ 1}^{(j)}(u) =
\displaystyle \sum_n a_{n} \begin{array}{l}
Z_{n+\nu+\frac{1}{2}}^{(j)}\left[4k\cos^2\frac{\sigma u}{2}\right],\end{array}
\ &\big|\cos\frac{\sigma u}{2}\big|
\geq 1, 
\vspace{2mm}\\
w_{\ 2}^{(j)}(u) =\tan\frac{\sigma u}{2}
\displaystyle \sum_n \begin{array}{l}
\left(n+\nu+\frac{1}{2}\right)
a_{n}
Z_{n+\nu+\frac{1}{2}}^{(j)}\left[4k\cos^2\frac{\sigma u}{2}\right],\end{array}
\ & \big|\cos\frac{\sigma u}{2}\big|
> 1,
\end{array}
\end{eqnarray}
\begin{eqnarray}\label{??}
\begin{array}{ll}
w_{\ 3}^{(j)}(u) =
\displaystyle \sum_n (-1)^na_{n} \begin{array}{l}
Z_{n+\nu+\frac{1}{2}}^{(j)}\left[-4k\sin^2\frac{\sigma u}{2}\right],\end{array}
\quad&\big|\sin\frac{\sigma u}{2}\big|
\geq 1, 
\vspace{2mm}\\
w_{\ 4}^{(j)}(u) =\cot\frac{\sigma u}{2}
\displaystyle \sum_n \begin{array}{l}
(-1)^n\left(n+\nu+\frac{1}{2}\right)
a_{n}
Z_{n+\nu+\frac{1}{2}}^{(j)}\left[-4k\sin^2\frac{\sigma u}{2}\right],\end{array}
\quad& \big|\sin\frac{\sigma u}{2}\big|
> 1,
\end{array}
\end{eqnarray}
where the coefficients $a_n$ satisfy the relations
\begin{eqnarray}\label{???}
\begin{array}{l}
2k(n+\nu+1)a_{n+1}{+}\left[\mathrm{a}-2k^2-\left(n+\nu+\frac{1}{2}\right)^2
\right]a_{n}+2k(n+\nu)a_{n-1}=0.
\end{array}
\end{eqnarray}
For any solutions the behavior when $z=\cos^2(\sigma u/2)\to\infty$
must be determined by using (\ref{comportamento-infinito}).
If $\sigma=1$ and $u=$real (Mathieu equation) the previous 
solutions are useless. Notice, however, that the
one-sided solutions $\mathring{w}_{\ i}^{(j)}$
in series of Bessel functions of the first kind 
are convergent for all finite values of $z$.

\subsection{ Solutions for the Whittaker-Ince limit of the CHE}
To obtain solutions to the Whittaker-Ince limit  of the CHE
(\ref{incegswe}), we apply \cite{erdelyi1}
\antiletra
\begin{eqnarray}\label{J}\begin{array}{l}
\displaystyle
\lim_{a\rightarrow  \infty}
\Phi\left(a,c;-\frac{y}{a}\right)=
\Gamma(c)\ y^{(1-c)/2}J_{c-1}\big(2\sqrt{y}\big),\vspace{2mm}\\
\displaystyle
\lim_{a\rightarrow  \infty}\left[\Gamma(a+1-c)\ \Psi\left(a,c;
-\frac{y}{a}\right)\right]=
\begin{cases}\displaystyle
-i\pi e^{i\pi c}y^{(1-c)/2}H_{c-1}^{(1)}\big(2\sqrt{y}\big),
\quad&\text{Im}\ y>0,
\vspace{2mm}\\
\displaystyle
i\pi e^{-i\pi c}y^{(1-c)/2}H_{c-1}^{(2)}\big(2\sqrt{y}\big),
\quad&\text{Im}\ y<0,
\end{cases}
\end{array}
\end{eqnarray}
on the hypergeometric functions used as basis for
the expansions of the solutions for the CHE. 
For this it is necessary to rewrite the latter solutions 
in a suitable form and keep $n$ fixed. 
 Expansions in series of $Y_{c-1}$ 
are obtained as a linear combination
of the expansions in series of Hankel functions. In this manner,
from  the first set (\ref{par1-nu}), we get 
a set of four solutions for Eq. (\ref{incegswe}). 
These are again denoted by $U_{\ 1}^{(j)}$ ($j=1,\dots,4$).
In fact, we will compute only
the limit of $U_1(z)$: the other solutions follow
from the fact that the four Bessel functions satisfy
the same differential and difference equations.

On the other side, if $U(z)=U(B_{1},B_{2},B_{3};z_{0},q;z)$ 
represents an arbitrary solution for Eq. (\ref{incegswe}), then other
solutions are generated by the transformations
$\mathscr{T}_1$, $\mathscr{T}_2$ and $\mathscr{T}_3$
given by \cite{lea-1}
\begin{eqnarray}\label{Transformacao2}
\begin{array}{l}
\mathscr{T}_{1}
U(z)=z^{1+B_{1}/z_{0}}
U(C_{1},C_{2},C_{3};z_{0},q;z),\vspace{2mm}\\
\mathscr{T}_{2}
U(z)=(z-z_{0})^{1-B_{2}-B_{1}/z_{0}}U(B_{1},D_{2},D_{3};
z_{0},q;z), \vspace{2mm}\\
\mathscr{T}_{3}
U(z)=
U(-B_{1}-B_{2}z_{0},B_{2},
B_{3}-q z_{0};z_{0},-q;z_{0}-z),
\end{array}
\end{eqnarray}
where $C_{i}$ and $D_{i}$ are defined in Eqs. (\ref{constantes-C-D} ).
Thus, it is sufficient to take the limit
of the first set of solutions (\ref{par1-nu}) and of
the coefficients (\ref{nu10}). The other sets are obtained from $U_{\ 1}^{(j)}$ through 
\begin{eqnarray}\label{rules-whitakker}
U_{\ 2}^{(j)}=\mathscr{T}_{2}U_{\ 1}^{(j)},\qquad 
U_{\ 3}^{(j)}=\mathscr{T}_{3}U_{\ 2}^{(j)},\qquad 
U_{\ 4}^{(j)}=\mathscr{T}_{1}U_{\ 3}^{(j)}.
\end{eqnarray}
First we find the four sets of solutions and
use the Raabe test to study their convergence. In the second place, we
write the solution 
for the Mathieu equation. 

\subsubsection{ The four sets of solutions}

To find the limit of $U_1(z)$ given in (\ref{par1-nu}),
we rewrite that solution as
\letra
\begin{eqnarray}\label{truque-1}
\begin{array}{l}\displaystyle
U_{ 1}(z) =e^{i\omega z}\sum_n
\frac{(-1)^nc_n[qz]^{n+\nu+1-({B_2}/{2})}}{\Gamma[2n+2\nu+2]}
\Phi\left[n+\nu+1+i\eta,2n+2\nu+2;-\frac{qz}{i\eta}\right],
\end{array}
\end{eqnarray}
where $c_n=b_{n}^{1}[{-}i\eta]^{-n}$ and $q=-2\eta\omega$.  
From $\alpha_{ n}^{1}b_{n+1}^{1}+
\beta_{n}^{1}b_{n+1}^{1}+\gamma_{n}^{1}b_{n-1}^{1}=0$,
we get 
\begin{eqnarray}\label{truque-2}
-i\eta\ \alpha_{n}^{1}c_{n+1}+
\beta_{n}^{1}c_n+({-}i\eta)^{-1}\gamma_{n}^{1}c_{n-1}=0,
\end{eqnarray}
where $\alpha_{n}^{1}$, $\beta_{n}^{1}$ 
and $\gamma_{n}^{1}$
are given in (\ref{nu10}). By supposing that $n$ is fixed and using (\ref{J}), 
we obtain the solution $U_{\ 1}^{(1)}$ in series of Bessel functions of 
first kind. In fact, we may verify directly that the four
solutions $U_{\ 1}^{(j)}$
given below satisfy Eq. (\ref{incedche}).

Then, the first set is given by
\antiletra\letra
\begin{equation}\label{WIL-ni1a}
U_{\ 1}^{(j)}(z) =z^{(1-B_{2})/2}
\displaystyle \sum_n (-1)^nc_{n}^{1}
Z_{2n+2\nu+1}^{(j)}\left(2\sqrt{qz}\right),
\end{equation}
where, in the recurrence relations $\alpha_{ n}^{1}c_{n+1}^{1}+
\beta_{n}^{1}c_{n+1}^{1}+\gamma_{n}^{1}c_{n-1}^{1}=0$, 
we have
\begin{equation}\label{WIL-ni1b}
\begin{array}{l}
\alpha_{n}^{1} = \frac{q z_{0}\left[n+\nu+2-\frac{B_2}{2}\right]
\left[n+\nu+1-\frac{B_1}{z_0}-\frac{B_2}{2}\right]}
{\left(2n+2\nu+2\right)\left(2n+2\nu+3
\right)},\vspace{2mm}\\
\beta_{n}^{1} =B_{3}{-}\frac{q z_{0}}{2}
{+}\left(n+\nu+1-\frac{B_2}{2}\right)\left(n+\nu+\frac{B_2}{2}\right)
{-}\frac{q z_{0}\left[B_2-2\right]
\left[B_2+\frac{2B_{1}}{z_{0}}\right]}
{2\left(2n+2\nu\right)
\left(2n+2\nu+2\right)},\hspace{1,8cm}\vspace{2mm}\\
\gamma_{n}^{1}=
\frac{ q z_{0}\left[n+\nu-1+\frac{B_2}{2}\right]
\left[n+\nu+\frac{B_{1}}{z_{0}}+\frac{B_2}{2}\right]}
{\left(2n+2\nu-1\right)
\left(2n+2\nu\right)}.
\end{array}
\end{equation}
The transformations (\ref{Transformacao2}), applied
on $U_{\ 1}^{(j)}$ according to
(\ref{rules-whitakker}), generate the other sets, that is,
($\beta_{n}^{i}=\beta_{n}^{1}$)
\antiletra
\begin{eqnarray}\label{WIL-ni2a}\begin{array}{l}
U_{\ 2}^{(j)}(z) =(z-z_0)^{1-B_2-\frac{B_1}{z_0}}
\ z^{\frac{B_1}{z_0}+\frac{B_2}{2}-\frac{1}{2}}
\displaystyle \sum_n (-1)^nc_{ n}^{2}
Z_{2n+2\nu+1}^{(j)}\left(2\sqrt{qz}\right),\vspace{.2cm}\\
\alpha_{n}^{2}  =  \frac{q z_{0}\left[n+\nu+\frac{B_{2}}{2}\right]
\left[n+\nu+1+\frac{B_{2}}{2}+\frac{B_{1}}{z_{0}}\right]}
{(2n+2\nu+2)(2n+2\nu+3)},
\qquad
\gamma_{n}^{2}  = \frac{q z_{0}\left[n+\nu+1-\frac{B_{2}}{2}\right]
\left[n+\nu-\frac{B_{2}}{2}-\frac{B_{1}}{z_{0}}\right]}
{(2n+2\nu-1)(2n+2\nu)};
\end{array}\hspace{9mm}
\end{eqnarray}
\begin{eqnarray}\label{WIL-ni3a}\begin{array}{l}
U_{\ 3}^{(j)}(z) =(z-z_0)^{(1-B_2)/2}
\displaystyle \sum_n (-1)^nc_{n}^{3}
Z_{2n+2\nu+1}^{(j)}\left[2\sqrt{q(z-z_0)}\right],\vspace{.2cm}\\
\alpha_{n}^{3}  = -\frac{q z_0\left[n+\nu+2-\frac{B_2}{2}\right]
\left[n+\nu+1+\frac{B_2}{2}+\frac{B_1}{z_0}\right]}
{(2n+2\nu+2)(2n+2\nu+3)},
\qquad
\gamma_{ n}^{3}  = -\frac{q z_0\left[n+\nu-1+\frac{B_2}{2}\right]
\left[n+\nu-\frac{B_2}{2}-\frac{B_1}{z_0}\right]}
{(2n+2\nu-1)(2n+2\nu)};
\end{array}
\end{eqnarray}
\begin{eqnarray}\label{WIL-ni4a}\begin{array}{l}
U_{\ 4}^{(j)}(z) =z^{1+\frac{B_1}{z_0}}
\ (z-z_0)^{-\frac{1}{2}-\frac{B_1}{z_0}-\frac{B_2}{2}}
\displaystyle \sum_n (-1)^nc_{ n}^{4}
Z_{2n+2\nu+1}^{(j)}\left[2\sqrt{q(z-z_0)}\right],\vspace{.2cm}\\
\alpha_{ n}^{4}  = -\frac{q z_0\left[n+\nu+\frac{B_2}{2}\right]
\left[n+\nu+1-\frac{B_2}{2}-\frac{B_1}{z_0}\right]}
{(2n+2\nu+2)(2n+2\nu+3)},
\qquad
\gamma_{n}^{4}  = -\frac{q z_0\left[n+\nu+1-\frac{B_2}{2}\right]
\left[n+\nu+\frac{B_2}{2}+\frac{B_1}{z_0}\right]}
{(2n+2\nu-1)(2n+2\nu)}.
\end{array}\hspace{3mm}
\end{eqnarray}
For the these four sets of solutions, 
the conditions (\ref{series-0}) 
and (\ref{four-conditions}) are replaced by
\begin{eqnarray}\label{condicao-wil}
\begin{array}{l}
2\nu, \ \ 
\nu\pm\frac{B_2}{2} \ \text{  and  }\  
\nu\pm\left(\frac{B_1}{z_0}+\frac{B_2}{2}\right)
\text{  are not integers},
\end{array}
\end{eqnarray}
while the relations (\ref{coeff}) remain valid for 
the coefficients $c_n^{i}.$

The previous list completes the list given 
in Ref. \cite{eu} where the expansions $U_{\ i}^{(1,2)}$
in series of $J_{\kappa}$ and $Y_{\kappa}$ have
not been taken into account, 
whereas the expansions
$U_{\ i}^{(3,4)}$ 
have been written in terms of the modified Bessel
functions $K_{2n+2\nu+1}[\pm 2i\sqrt{qz}]$
and  $K_{2n+2\nu+1}[\pm 2i\sqrt{q(z-z_0}]$. Moreover,
now the regions of convergence are modified by the
use of the Raabe test.

\subsubsection{ Convergence of the solutions}

As in the case of the CHE, by the D'Alembert test
the two-sided expansions ${U}_{\ 1}^{(j)} $
and ${U}_{\ 2}^{(j)} $ converge absolutely for $|z|>|z_0|$
while ${U}_{\ 3}^{(j)} $ and ${U}_{\ 4}^{(j)} $ converge
for $|z-z_0|>|z_0|$. However, by the Raabe test the
solutions also converge at $|z|=|z_0|$ and $|z-z_0|=|z_0|$
under the conditions similar (\ref{convergencia1-0}),
that is,
\begin{equation}\label{convergencia-WI-limit}
|z|\geq|z_0|  \text{ if}
\begin{cases}\text{Re}\left[  B_2+\frac{B_1}{z_0}\right] <1 \mbox{ in }
{U}_{\ 1}^{(j)},\vspace{2mm} \\
\text{Re}\left[  B_2+\frac{B_1}{z_0}\right] >1\mbox{ in }
{U}_{\ 2}^{(j)},
\end{cases}
|z-z_0|\geq|z_0| \text{ if}
\begin{cases}\text{Re}\left[ \frac{B_1}{z_0}\right] >-1\mbox{ in }
{U}_{\ 3}^{(j)},\vspace{2mm} \\
\text{Re}\left[ \frac{B_1}{z_0}\right] <-1\mbox{ in }
{U}_{\ 4}^{(j)}.
\end{cases}
\end{equation}
The test does not assure convergence at $z=\infty$
and, so, the behavior at $z=\infty$
again deserves special attention.
We will find that the one-sided infinite series 
$\mathring{U}_{\ i}^{(1)}$ in series of Bessel functions of first kind 
converge for any finite value of $z$.

Relations (\ref{convergencia-WI-limit}) correspond
to (\ref{convergencia1-0}) with the replacements $U_{\ i}^{(j)}
\leftrightarrow \mathbb{U}_i$. In fact, a few modifications
in the previous demonstration lead to (\ref{convergencia-WI-limit}).   
Thus, if $n\rightarrow\pm\infty$, we find
\begin{eqnarray*}\begin{array}{l}
 qz_0\left[1-\frac{1}{n}\left(B_2+\frac{B_1}{z_0}-
 \frac{1}{2}\right)\right]\frac{c_{n+1}^{1}}{c_{n}^{1}}+
\big[ 4n(n+2\nu+1)\big]+ \vspace{2mm}\\
qz_0\left[1+\frac{1}{n}\left(B_2+\frac{B_1}{z_0}-
\frac{1}{2}\right)\right]
\frac{c_{n-1}^{1}}{{c}_{ n}^{1}}=0.
\end{array}
\end{eqnarray*}
When $n\to \infty$ the minimal solution for $c_{n+1}/c_{n}$  is
%
\begin{eqnarray}
\begin{array}{ll}
n\to\infty: & \frac{c_{n+1}^{1}}{c_{ n}^{1}}\sim
-\frac{q z_0}{4n^2}\left[1-\frac{1}{n}\left(2\nu-B_2-\frac{B_1}{z_0}+
\frac{7}{2}\right) \right]\Rightarrow
\end{array}\vspace{2mm} \\
\begin{array}{l}
 \frac{c_{n-1}^{1}}{c_{n}^{1}}\sim-
\frac{4n^2}{q z_0}\left[ 1+\frac{1}{n}\left(2\nu-B_2-\frac{B_1}{z_0}+
\frac{3}{2}\right)\right], \nonumber
\end{array}
\end{eqnarray}
while the minimal solution for $c_{n-1}/c_{n}$, when $n\to -\infty$, is
\begin{eqnarray}
\begin{array}{ll}
n\to-\infty:&\frac{c_{n-1}^{1}}{c_{n}^{1}}\sim-
\frac{qz_0}{4n^2}\left[1-\frac{1}{n}\left(2\nu+B_2+
\frac{B_1}{z_0}
-\frac{3}{2}\right) \right] \Rightarrow
\end{array}
\vspace{2mm} \\
\begin{array}{l}
\frac{c_{n+1}^{1}}{c_{n}^{1}}\sim-
\frac{4n^2}{qz_0}\left[ 1+\frac{1}{n}\left(2\nu+B_2
+\frac{B_1}{z_0}+\frac{1}{2}\right)\right].\nonumber
\end{array}
\end{eqnarray}
On the other side,  the behaviors (\ref{Z})
and (\ref{Z-2}) for the Bessel functions
lead to 
\begin{eqnarray}\label{rediscutir}
\begin{array}{l}n\to\infty: \;\;
\frac{J_{2n+2\nu+3}}{J_{2n+2\nu+1}}=
\frac{qz}{4n^2}\left[1-\frac{1}{n}
\left(2\nu+\frac{5}{2}\right)\right],\qquad 
%
\frac{Z_{2n+2\nu+3}^{(2,3,4)}}{Z_{2n+2\nu+1}^{(2,3,4)}}
=\frac{4n^2}{qz}\left[1+\frac{1}{n}
\left(2\nu+\frac{3}{2}\right)\right];\vspace{2mm}\\
%
n\to-\infty:\;\;\frac{Z_{2n+2\nu-1}^{(j)}}{Z_{2n+2\nu+1}^{(j)}}=
{\frac{4n^2}{qz}}\left[1+\frac{1}{n}
\left(2\nu+\frac{1}{2}\right)\right],
\qquad [j=1,2,3, 4].
\end{array}
\end{eqnarray}
Thus, when $n\to+\infty$,
\begin{eqnarray}\label{following-0}
\begin{array}{l}
\frac{c_{n+1}^{1}J_{2n+2\nu+3}}{c_{n}^{1}J_{2n+2\nu+1}}\sim
- \frac{q^2z_0z}{16n^4}\left[1-\frac{1}{n}
\left(4\nu+6-B_2-\frac{B_1}{z_0}\right)\right],\vspace{2mm}\\
\frac{c_{n+1}^{1}Z_{2n+2\nu+3}^{(j)}}{c_{n}^{1}
Z_{2n+2\nu+1}^{(j)}}
\sim-\frac{z_0}{z}\left[1-\frac{1}{n}
\left(2-B_2-\frac{B_1}{z_0}\right)\right],\qquad[ j=2,3,4]
\end{array}
\end{eqnarray}
and, when $n\to-\infty$,
\begin{eqnarray}\label{following}
\begin{array}{l}
\frac{c_{n-1}^{1}
Z_{2n+2\nu-1}^{(j)}}{c_{n}^{1}Z_{2n+2\nu+1}^{(j)}}
\sim -\frac{z_0}{z}\left[1+\frac{1}{n}
\left(2-B_2-\frac{B_1}{z_0}\right)\right],\qquad[ j=1,2,3,4].
\end{array}
\end{eqnarray}
Hence, by the Raabe test the expansions
$U_{\ 1}^{(j)}$ are convergent for $|z|\geq|z_0|$
as indicated in (\ref{convergencia-WI-limit}). 
For the other sets of solutions the domains
of convergence follow from the transformations
(\ref{Transformacao2}) applied on $U_{\ 1}^{(j)}$
according to (\ref{rules-whitakker}). Moreover,
from the first limit given in (\ref{following-0})
we see that the one-sided infinite series 
$\mathring{U}_{\ i}^{(1)}$ converge for all $z$ excepting probably
the point $z=\infty$. The behavior 
any solution when $z\to\infty$
must be determined by using (\ref{comportamento-infinito}).

\subsubsection{ Heine's solutions for the Mathieu equation}

From the previous solutions we recover 
the usual solutions in series of Bessel functions for Mathieu's equation
(called Heine's solutions \cite{erdelyi-2}) by means of the 
substitutions
\begin{equation}\label{mathieu-as-wil}
\left.
\begin{array}{l}
w(u)=U(z), \quad z=\cos^{2}(\sigma u)\quad\Rightarrow
\quad z_{0}=1,\vspace{2mm}\\
B_{1}=-\frac{1}{2}, \quad B_{2}=1, \quad
B_{3}=\frac{k^2}{2}-\frac{\mathrm{a}}{4}, \quad q=k^2
\end{array}
\right\}
\begin{array}{l}
\mbox{  Mathieu eq. as Whittaker-}\\
\mbox{  Ince limit of the CHE.}
\end{array}
\end{equation}
However, the regions of convergence for some solutions
turn out to be improved by the Raabe test. This fact
is useful for some applications, as we will see in Sec. 4.

Relations ({\ref{coeff}), with $b_{n}^{i}$ replaced by 
$c_{ n}^{i}$, yield
\begin{eqnarray*}
\begin{array}{l}
c_{n}^{2}=\left(n+\nu+\frac{1}{2}\right)c_{n}^{1},
\qquad c_{n}^{3}=(-1)^nc_{n}^{1},\qquad c_{ n}^{4}=(-1)^n\left(n+\nu+\frac{1}{2}\right)c_{ n}^{1}.
\end{array}
\end{eqnarray*}
Then, by writing $c_n=c_{n}^{1}$, 
$w_{\ i}^{(j)}(u)=U_{\ i}^{(j)}(z)$
and setting $\sqrt{k^2}=k$ we find
\letra
\begin{eqnarray}\label{Heine-1}
\begin{array}{ll}
w_{\ 1}^{(j)}(u) =
\displaystyle \sum_n 
\begin{array}{l}
(-1)^nc_{n}
Z_{2n+2\nu+1}^{(j)}\left[2k\cos(\sigma u)\right],
\end{array}
&|\cos(\sigma u)|
\geq 1,
\vspace{2mm}\\
w_{\ 2}^{(j)}(u) =\tan[\sigma u]
\displaystyle \sum_n 
\begin{array}{l}
(-1)^n
\left(n+\nu+\frac{1}{2}\right)
c_{n}
Z_{2n+2\nu+1}^{(j)}\left[2k\cos(\sigma u)\right],\end{array}
& |\cos(\sigma u)|
> 1,
\end{array}
\end{eqnarray}
\begin{eqnarray}\label{Heine-2}
\begin{array}{ll}
w_{\ 3}^{(j)}(u) =
\displaystyle \sum_n 
\begin{array}{l}
c_{n}
Z_{2n+2\nu+1}^{(j)}\left[2ki\sin(\sigma u)\right],\end{array}
&|\sin(\sigma u)|\geq 1\vspace{2mm}\\
w_{\ 4}^{(j)}(u) =\cot(\sigma u)
\displaystyle \sum_n 
\begin{array}{l}
\left(n+\nu+\frac{1}{2}\right)c_{n}
Z_{2n+2\nu+1}^{(j)}\left[2ki\sin(\sigma u)\right],\end{array}
&|\sin(\sigma u)|> 1,
\end{array}
\end{eqnarray}
where the coefficients $c_n$ satisfy the relations
\begin{eqnarray}\label{Heine-3}
k^2c_{n+1}{+}\left[\left(2n+2\nu+1\right)^2-
\mathrm{a}\right]c_{n}+k^2c_{n-1}=0.
\end{eqnarray}
As in the case of the two-sided solutions 
(\ref{?}) and (\ref{??}), obtained
from the CHE, the above solutions are useless
for $\sigma=1$ and $u=$real (Mathieu equation).

The conditions (\ref{condicao-wil}) 
reduce to $2\nu\notin \mathbb{Z}$ and
assure the linear independence of the terms in a given series.
In addition, the above notation for solutions of the Mathieu equation
is similar to the one used by Erd\'{e}lyi \cite{erdelyi-2}. However,
in Ref. \cite{abramowitz,nist} the 
coefficients ($c_{n+1}, c_n, c_{n-1}$) are 
replaced by ($c_{2n+2}, c_{2n}, c_{2n-2}$)
and the Bessel functions $Z_{2n+2\nu+1}$ written as
 $Z_{2n+\bar{\nu}}$: this is equivalent to put 
 $2\nu+1=\bar{\nu}$ with $\bar{\nu}\notin \mathbb{Z}$.
The above domains of convergence may be compared
 with the ones of solutions 28.23.2-28.23.5 of \cite{nist}.

By the Raabe test, the 
the two-sided solutions $w_{\ 1}^{(j)}(u)$
and $w_{\ 3}^{(j)}(u)$ are absolutely convergent also at
$|\cos(\sigma u)|=1$ and $|\sin(\sigma u)|=1$, respectively.
In Refs. \cite{arscott,abramowitz,erdelyi3,meixner2} these
points are not included in the domains
of convergence due to the use of the D'Alembert test. 
The one-sided solutions $\mathring{w}_i^{(1)}$ 
converge for any finite $u$; in Ref. \cite{nist}
it is stated that this property holds also for
two-sided solutions (a misprint, for certain).

\subsection{ Solutions for the double-confluent Heun equation}

For $z_0\rightarrow 0$ the limits of the
solutions of the sets (\ref{par1-nu}) and (\ref{par2-nu}) 
are equivalent, respectively, to the limits of (\ref{par3-nu}) 
and (\ref{par4-nu}). This gives a subgroup constituted by 
two sets of solutions. A second subgroup, with different 
regions of convergence, results from the transformations 
of the DCHE and cannot be obtained
as limits of the expansions in series of
Coulomb wave functions. In effect, 
if $U(z)=U(B_{1},B_{2},B_{3};\omega,\eta;z)$
denotes one solution for the DCHE, new solutions can 
be generated by using the transformations \cite{lea-1}
\antiletra\letra
\begin{eqnarray}\label{transformacao-dche}
\begin{array}{l}
t_{1}U(z)=e^{\frac{B_1}{z}}z^{2-B_{2}}U(-B_{1},4-B_{2},
B_{3}+2-B_{2}; \omega,\eta;z),
\vspace{2mm}\\
t_{2}U(z)=e^{i\omega z+\frac{B_{1}}{2z}}z^{-i\eta-\frac{B_{2}}{2}}
U\left( \bar{B}_{1},\bar{B}_{2},\bar{B}_{3};\bar{\omega}=1,\bar{\eta};
\frac{iB_{1}}{2z}\right),\vspace{2mm}\\
t_{3}U(z)=U(B_{1},B_{2},B_{3};-\omega,-\eta;z),
\end{array}
\end{eqnarray}
where, on the right-hand side of the second relation, we have
\begin{eqnarray}\begin{array}{l}
\bar{B}_{1}=\omega B_{1},  \quad
\bar{B}_{2}=2+2i\eta,  \quad
\bar{B}_{3}=B_{3}-\left[\frac{B_{2}}{2}+i\eta\right]
\left[\frac{B_{2}}{2}-i\eta-1\right], 
\quad
i\bar{\eta}=\frac{B_{2}}{2}-1.
\end{array}
\end{eqnarray}
The transformation $t_2$ is the one responsible
for the second subgroup of solutions.

In the following we use the notation
given in (\ref{notation-1}) and (\ref{notation-2})
also for solutions of the DCHE. We find that 
the Raabe test is useless for the DCHE.

\subsubsection{The two subgroups of solutions}
%
%
%
The limit $z_0\to 0$ does not change the form
of (\ref{par1-nu}) but modifies the series coefficients.
Thus we have
\antiletra\letra
\begin{eqnarray}\label{dche-1a}
\begin{array}{l}
U_{ 1}(z) =e^{i\omega z}\displaystyle\sum_n
\frac{b_{n}^{1}[2i\omega z]^{n+\nu+1-\frac{B_2}{2}}}{\Gamma[2n+2\nu+2]}
\Phi\left(n+\nu+1+i\eta,2n+2\nu+2;-2i\omega z\right),\vspace{2mm}\\
{U}_{1}^{\pm}(z) =e^{\pm i\omega z}\displaystyle
\sum_n
\frac{b_{ n}^{1}
[-2i\omega z]^{n+\nu+1-\frac{B_2}{2}}}{\Gamma[n+\nu+1\mp i\eta]}
\Psi\left(n+\nu+1\pm i\eta,2n+2\nu+2;\mp 2i\omega z\right), 
\end{array}
\end{eqnarray}
whereas coefficients (\ref{nu10}) when $z_0\to 0$
give
\begin{eqnarray}\label{dche-1b}
\begin{array}{l}
\alpha_{n}^{1}  =  \frac{2i\omega B_1\left[n+\nu+2-\frac{B_{2}}{2}\right]}
{(2n+2\nu+2)(2n+2\nu+3)},
\vspace{.2cm} \\
\beta_{n}^{1} =  B_{3}+\left(n+\nu+1-\frac{B_{2}}{2}\right)
\left(n+\nu+\frac{B_{2}}{2}\right)
+\frac{2\eta\omega B_1\left(B_2-2\right)}
{(2n+2\nu)(2n+2\nu+2)},
\vspace{.3cm} \\
\gamma_{n}^{1}  = \frac{2i\omega B_1\left[n+\nu+\frac{B_{2}}{2}-1\right]
(n+\nu+i\eta)(n+\nu-i\eta)}
{(2n+2\nu-1)(2n+2\nu)}.
\end{array}
\end{eqnarray}
On the other side, since
\antiletra\letra
\begin{eqnarray*}
\lim_{z_0\to 0}f_2(z)=\lim_{z_0\to 0}\left[z^{\nu+\frac{B_{2}}{2}}
(z-z_0)^{1-B_2}\left(1-\frac{z_0}{z}\right)^{-{B_1}/{z_{0}}}\right]=
z^{1+\nu-\frac{B_2}{2}\ }e^{\frac{B_1}{z}},
\end{eqnarray*}
the solutions obtained from (\ref{par2-nu})
and (\ref{par2-nu-b}) are
\begin{eqnarray}\label{dche-2a}
\begin{array}{l}
U_{2}(z)= e^{i\omega z+\frac{B_1}{z}} \displaystyle
\sum_n
\frac{b_{n}^{2}[2i\omega z]^{n+\nu+1-\frac{B_2}{2}}}{\Gamma[2n+2\nu+2]}
\Phi\left[n+\nu+1+i\eta,2n+2\nu+2;-2i\omega z\right],\vspace{2mm}\\
{U}_{2}^{\pm}(z) =e^{\pm i\omega z+\frac{B_1}{z}} \displaystyle
\sum_n\frac{b_{ n}^{2}
[-2i\omega z]^{n+\nu+1-\frac{B_2}{2}}}{\Gamma[n+\nu+1\mp i\eta]}
\Psi\left[n+\nu+1\pm i\eta,2n+2\nu+2;\mp 2i\omega z\right], 
\end{array}
\end{eqnarray}
where the coefficients for the recurrence relations are given by
\big($\beta_{n}^{2} =  \beta_{ n}^{1}$\big)
\begin{eqnarray}\label{dche-2b}
\begin{array}{l}
\alpha_{n}^{2}  =  \frac{-2i\omega B_1\left[n+\nu+\frac{B_{2}}{2}\right]}
{(2n+2\nu+2)(2n+2\nu+3)},
\qquad
\gamma_{n}^{2}  = -\frac{2i\omega B_1\left[n+\nu+1-\frac{B_{2}}{2}\right]
(n+\nu+i\eta)(n+\nu-i\eta)}
{(2n+2\nu-1)(2n+2\nu)}.
\end{array}
\end{eqnarray}

Notice that $\mathbb{U}_2=t_1\mathbb{U}_1$.
We can check that $t_3$ does not generate new
independent solutions when applied on
$\mathbb{U}_1$ and $\mathbb{U}_2$.
Furthermore, by using
$t_2$ we construct a second subgroup having two
sets of solutions. Thus, we form a group of four
sets of solutions $\mathbb{U}_i$, given by
\antiletra
\begin{eqnarray}\label{hipergeometricas}
\mathbb{U}_1(z),\qquad
\mathbb{U}_2(z)=t_1\mathbb{U}_1(z);\qquad
\mathbb{U}_3(z)=t_2\mathbb{U}_1(z),\qquad
\mathbb{U}_4(z)=t_2\mathbb{U}_2(z)=t_3\mathbb{U}_3(z)
\end{eqnarray}
where $\mathbb{U}_i=(U_i,U_i^{+}, U_i^{-})$ as in Eq. (\ref{notation-1}). 

The solutions of the third set are
\letra
\begin{eqnarray}\label{dche-3a}
\begin{array}{l}
U_{3}(z)  =e^{i\omega z} \displaystyle
\sum_n
\frac{b_{n}^{3}\left[-{B_1}/{z}\right]^{n+\nu+\frac{B_2}{2}}
}{\Gamma[2n+2\nu+2]}
\Phi\left[n+\nu+\frac{B_{2}}{2},2n+2\nu+2;
\frac{B_{1}}{z}\right],
\vspace{0.2cm}\\
{U}_{3}^{\pm}(z) =e^{i\omega z+(1\mp 1)\frac{B_1}{2z}}
\displaystyle
\sum_n
\frac{b_{n}^{3}\left[{B_1}/{z}\right]^{n+\nu+\frac{B_2}{2}}}
{\Gamma\left[n+\nu+1\mp\frac{B_2-2}{2}\right]}\times
\vspace{0.2cm}\\
\hspace{4.5cm}\Psi\left[n+\nu+1\pm\frac{B_2-2}{2},2n+2\nu+2;
\pm\frac{B_{1}}{z}\right],
\end{array}
\end{eqnarray}
where, in the recurrence relations for $b_n^{3}$ ($\beta_{n}^{3} 
=\beta_{n}^{1}$)
\begin{eqnarray}\label{dche-3b}
\begin{array}{l}
\alpha_{n}^{3}  =\frac{2i\omega B_1(n+\nu+1-i\eta)}
{(2n+2\nu+2)(2n+2\nu+3)}  ,
\qquad
%
\gamma_{n}^{3}=\frac{2i\omega B_1\left[n+\nu-1+\frac{B_2}{2}\right]
\left[n+\nu+1-\frac{B_2}{2}\right][n+\nu+i\eta]}{(2n+2\nu-1)
(2n+2\nu)}.
\end{array}
\end{eqnarray}
Finally, since the fourth set is given  by
\antiletra
\begin{eqnarray}\label{dche-4a}
\begin{array}{l}
\left[U_4(z),U_4^{\pm}(z)\right]=\left[t_2U_2(z),t_2U_2^{\pm}(z)\right]=
\left[t_3U_3(z),t_3U_3^{\pm}(z)\right],
\end{array}
\end{eqnarray}
it is obtained by replacing ($\eta,\omega$)
by ($-\eta,-\omega$) in ($U_3, U_3^{\mp}$) and 
respective recurrence relations.

For these solutions of the DCHE,
the restrictions (\ref{series-0}) are replaced
by 
\begin{eqnarray}\label{series-1}
\begin{array}{l}
2\nu \text{ and } \nu\mp i \eta\text{ cannot be integers in
$\mathbb{U}_1$ and $\mathbb{U}_2$ },\vspace{2mm}\\
2\nu \text{ and }  \nu\mp\frac{B_2}{2}\text{ cannot be integers in
$\mathbb{U}_3$ and $\mathbb{U}_4$ }.
\end{array}
\end{eqnarray}
Under these conditions, the 
three hypergeometric functions of a given set are 
linearly independent 
and, consequently, each solution can be written as a 
combination of the others
by means of (\ref{continuation2}). However, 
to have two-sided infinite series, in addition 
(\ref{series-1}), we must require that
\begin{eqnarray}
\begin{array}{l}
\nu\mp \frac{B_2}{2}\text{ cannot be integer in
$\mathbb{U}_1$ and $\mathbb{U}_2$ },\ 
\nu\mp i\eta\text{ cannot be integer in
$\mathbb{U}_3$ and $\mathbb{U}_4$ }
\end{array}
\end{eqnarray}
These conditions take the place of (\ref{four-conditions}).
Eqs. (\ref{alfa-beta}) hold also for the present solutions,
but Eqs. (\ref{coeff}) are replaced by
\begin{eqnarray}\label{coeff-2}
\begin{array}{l}
b_{ n}^{2}=\frac{[-1]^n\Gamma\left[n+\nu+2-\frac{B_2}{2}\right] }
{\Gamma\left[n+\nu+\frac{B_2}{2}\right]} b_{ n}^{1},\ \ \ 
b_{n}^{3}=
\frac{\Gamma\left[n+\nu+2-\frac{B_2}{2}\right]}
{\Gamma\left[n+\nu+1-i\eta\right]} b_{ n}^{1} ,\ \ \ 
b_{n}^{4}=\frac{[-1]^n \Gamma\left[n+\nu+2-\frac{B_2}{2}\right]}
{\Gamma\left[n+\nu+1+i\eta\right]}  b_{n}^{1}.
\end{array}
\end{eqnarray}

\subsubsection{Convergence, Whittaker-Hill and Mathieu equations}
The two subgroups present different regions of convergence:
\begin{itemize}
\itemsep-3pt
\item The solutions $\mathbb{U}_1$
and $\mathbb{U}_2$ converge absolutely for $|z|>0$
but may be unbounded at $z=\infty$.
The one-sided series ($n\geq 0$ ),
$\mathring{U}_1$
and $\mathring{U}_2$, converge also at $z=0$.
\item The solutions $\mathbb{U}_3$
and $\mathbb{U}_4$ converge for $|z|<\infty$
but may be unbounded at $z=0$.
The one-sided series,
$\mathring{U}_3$
and $\mathring{U}_4$, converge also at $|z|=\infty$.
\end{itemize}

Once more we take the solutions (\ref{dche-1a}) in the
form (\ref{forma-de-leaver}).
Then, the ratios $b_{n+1}^{1}/b_{ n}^{1}$ and $b_{n-1}^{1}/b_{ n}^{1}$
are obtained from
\begin{eqnarray*}
\displaystyle
\frac{i\omega B_1}{n}\
\frac{{b}_{n+1}^{1}}{{b}_{ n}^{1}}+2n\Big[ n+2\nu+1\Big]
+i\omega B_1 n \
\frac{{b}_{n-1}^{1}}{{b}_{n}^{1}}=0,
\end{eqnarray*}
which takes the place of Eq. (\ref{place}).
The minimal solutions are 
\begin{eqnarray*}
\begin{array}{lll}
n\to \infty: &\frac{b_{n+1}^{1}}{b_{ n}^{1}}\sim
\frac{\omega B_1}{2in}\left[1-\frac{2\nu+2}{n} \right]
\quad& \Rightarrow\quad
\frac{b_{n-1}^{1}}{b_{n}^{1}}\sim
-\frac{2(n+2\nu+1)}{i\omega B_1},\qquad\vspace{2mm}\\
n\to -\infty:&\frac{b_{n-1}^{1}}{b_{ n}^{1}}\sim
-\frac{i\omega B_1}{2n^3}\left[1-\frac{2\nu+2}{n} \right]&
\Rightarrow\quad
\frac{b_{n+1}^{1}}{b_{ n}^{1}}\sim-
\frac{2n^2(n+2\nu+1)}{i\omega B_1}.
\end{array}
\end{eqnarray*}
Combining these limits with (\ref{phi-psi}), we find
the relations
%
\begin{eqnarray}\label{n+dche}
\begin{array}{l}
n\to\infty:\;\;
\vline\frac{b_{n+1}^{1}
\phi_{n+\nu+1}(\eta,\omega z)}
{b_{ n}^{1}\phi_{n+\nu}(\eta,\omega z)}\vline\sim\
\vline\frac{\omega^2 B_1 z}{4n^3}\vline\ ,\qquad\
 \vline\frac{b_{n+1}^{1}
\psi_{n+\nu+1}^{\pm}(\eta,\omega z)}
{b_{n}^{1}\psi_{n+\nu}^{(\pm)}(\eta,\omega z)}\vline\sim\
\vline\frac{B_1}{zn}\vline\ ;\vspace{2mm}\\
%
%
n\to-\infty: \; \; 
\vline\frac{b_{n+1}^{1}
\mathscr{U}_{n+\nu+1}(\eta,\omega z)}
{b_{n}^{1}\mathscr{U}_{n+\nu}(\eta,\omega z)}\vline
\displaystyle\sim\
\vline\frac{B_1}{zn}\vline\ ,
\qquad \mathscr{U}_{n+\nu}=\left(\phi_{n+\nu},\psi_{n+\nu}^{\pm}\right),\end{array}
\hspace{2.2cm}
\end{eqnarray}
which follow as well from (\ref{phi})
and (\ref{psi}) as $z_0\to 0$. Thence, by the ratio
test, the solutions $\mathbb{U}_1$ converge for $|z|>|B_1/n|$,
that is, for $|z|>0$.
The same holds for the second set of solutions,
$\mathbb{U}_2=t_1\mathbb{U}_1$.
For $z\to\infty$ the
solutions may be unbounded insofar as their behaviors
are formally given by (\ref{bound-2-0})
and (\ref{bound-0}). The corresponding one-sided solutions 
in series of regular confluent hypergeometric
functions converges for $|z|\geq 0$ because
the limits when $n\to-\infty$  become irrelevant.

On the other side, by applying $t_2$ on (\ref{n+dche}) we find
\begin{eqnarray}\label{n+dche-b}
\begin{array}{lll}
n\to\infty:\;\;& \vline\frac{b_{n+1}^{3}
\phi_{n+\nu+1}\left(\bar{\eta},\frac{iB_1}{2z}\right)}
{b_{ n}^{3}\phi_{n+\nu}\left(\bar{\eta},\frac{iB_1}{2z}\right)}
\vline\sim\ \vline\frac{\omega B_1^2 }{8n^3z}\vline,\qquad&
 \vline\frac{b_{n+1}^{3}
\psi_{n+\nu+1}^{\pm}\left(\bar{\eta},\frac{iB_1}{2z}\right)}
{b_{ n}^{3}\psi_{n+\nu}^{\pm}\left(\bar{\eta},\frac{iB_1}{2z}\right)}
\vline\sim\
\vline\frac{2\omega z}{n}\vline\ ;\vspace{2mm}\\
%
%
 n\to-\infty:& \vline\frac{b_{n+1}^{3}
\mathscr{U}_{n+\nu+1}\left(\bar{\eta},\frac{iB_1}{2z}\right)}
{b_{ n}^{3}\mathscr{U}_{n+\nu}\left(\bar{\eta},\frac{iB_1}{2z}\right)}
\vline\displaystyle\sim\
\vline\frac{2\omega z}{n}\vline\ ,
& \mathscr{U}_{n+\nu}=
\left(\phi_{n+\nu},\psi_{n+\nu}^{\pm}\right). 
\end{array}
\end{eqnarray}
Then, the solutions $\mathbb{U}_3$ converge for finite values of $z$; 
the same is true for $\mathbb{U}_3$ 
(if $n\geq 0$, $\mathring{U}_3$ and $\mathring{U}_4$
converge at $z=\infty$ as well).
However, we must determine the behavior of each 
solution at $z=0$. Thus, from relations (\ref{asymptotic-confluent}) 
%
\begin{eqnarray} \label{bound-dche}
{z\to 0}:\;\;U_{3}(z)\sim
\begin{array}{l}
\displaystyle
\sum_n
\frac{{b}_{ n}^{3}}{\Gamma[n+\nu+2-B_2/2]} +
\left(\frac{z}{B_1}\right)^{2-{B_2}}\ e^{B_1/z}
\sum_n \frac{(-1)^{n+\nu+B_2/2}\ b_{ n}^{3}}{\Gamma[n+\nu+B_2/2]}
\end{array}
\end{eqnarray}
while from Eq. (\ref{asymptotic-confluent1}) 
\begin{eqnarray} \label{bounded-dche}
\begin{array}{ll} \displaystyle
{z\to 0}:\;\;{U}_{3}^{+}(z)\sim
\sum_n
\frac{{b}_{ n}^{3}}{\Gamma[n+\nu+2-B_2/2]},&\ 
-\frac{3\pi}{2}<\arg\left(\frac{B_1}{z}\right)<\frac{3\pi}{2},
\vspace{2mm}\\
\displaystyle
{z\to 0}:\;\;{U}_{3}^{-}(z)\sim \left(-{z}/{B_1}\right)^{2-{B_2}}\ e^{B_1/z}
\sum_n
\frac{(-1)^{n+\nu+B_2/2}\ b_{ n}^{3}}{\Gamma[n+\nu+B_2/2]},&\ 
-\frac{3\pi}{2}<\arg
\left(-\frac{B_1}{z}\right)<\frac{3\pi}{2}. 
\end{array}
\end{eqnarray}
Therefore, $U_3$ and $U_3^{-}$ may be unbounded at $z=0$ 
as occur, for example, if $\mathrm{Re}(B_1/z)\to \infty$.

Now we consider the properties of the solutions
for the  Whittaker-Hill and Mathieu equations 
obtained from the solutions of 
the DCHE. The Whittaker-Hill equation (\ref{whe}) (WHE)  
is transformed into the DCHE (\ref{dche}) by
the substitutions
%
\begin{eqnarray}\label{whe2}
\left.\begin{array}{l}
z=e^{2i\varsigma u},\quad W(u)=z^{1+(p/2)}e^{\xi/(8z)}U(z)
\quad\Rightarrow\quad
B_{1}=-\frac{\xi}{4},
\vspace{2mm}\\
B_{2}=p+3, \ \
B_{3}=\left(\frac{p}{2}+1\right)^{2}+\frac{\xi^2}{32}-
\frac{\vartheta}{4}, \ \  i\omega=\frac{\xi}{8},
\ \  i\eta=\frac{1}{2}(p+1)
\end{array}\right\}
\begin{array}{l}
\mbox{WHE  as}\\
\mbox{a DCHE.}
\end{array}
\end{eqnarray}
So, by inserting the above expressions into the solutions
$\mathbb{U}_{i}(z)=({U}_{i}(z),{U}_{i}^{\pm}(z))$ for the DCHE, we get
the solutions $\mathbb{W}_{i}(u)=({W}_{i}(u),{W}_{i}^{\pm}(u))$ for the
Whittaker-Hill equation. Thus
\begin{eqnarray}
\begin{array}{l}
W_1=e^{i\varsigma[2\nu+1]u}f^{+}\displaystyle\sum_{n}
\begin{array}{l}\frac{b_n^1\left[\xi e^{2i\varsigma u}/4\right]^n}{\Gamma[2n+2\nu+2]}
\Phi\left[n+\nu+\frac{p+3}{2},2n+2\nu+2;-\frac{\xi}{4}
e^{2i\varsigma u}\right],\end{array}\vspace{2mm}\\
W_1^{\pm}=e^{i\varsigma[2\nu+1]u}f^{\pm}\displaystyle\sum_{n}
\begin{array}{l}
\frac{b_n^1\left[-\xi e^{2i\varsigma u}/4\right]^n}{\Gamma[n+\nu+1\mp\frac{1}{2}(p+1)]}
\Psi\left[n+\nu+1\pm\frac{p+1}{2},2n+2\nu+2;\mp\frac{\xi}{4}
e^{2i\varsigma u}\right],\end{array}
\end{array}
\end{eqnarray}
where
\[f^{+}=f^{+}(u)=e^{(\xi/4)\cos(2\varsigma u)}, 
\qquad f^{-}=f^{-}(u)=
e^{i(\xi/4)\sin(2\varsigma u)},\]
and
\begin{eqnarray}
\begin{array}{l}
W_2=e^{i\varsigma[2\nu+1]u}g^{+}\displaystyle\sum_{n}
\begin{array}{l}\frac{b_n^2\left[\xi e^{2i\varsigma u}/4\right]^n}{\Gamma[2n+2\nu+2]}
\Phi\left[n+\nu+\frac{p+3}{2},2n+2\nu+2;-\frac{\xi}{4}
e^{2i\varsigma u}\right],\end{array}\vspace{2mm}\\
W_2^{\pm}=e^{i\varsigma[2\nu+1]u}g^{\pm}\displaystyle\sum_{n}
\begin{array}{l}
\frac{b_n^2\left[-\xi e^{2i\varsigma u}/4\right]^n}{\Gamma[n+\nu+1\mp\frac{1}{2}(p+1)]}
\Psi\left[n+\nu+1\pm\frac{p+1}{2},2n+2\nu+2;\mp\frac{\xi}{4}
e^{2i\varsigma u}\right],\end{array}
\end{array}
\end{eqnarray}
where
\[g^{+}=g^{+}(u)=e^{i(\xi/4)\sin(2\varsigma u)}, 
\qquad g^{-}=g^{-}(u)=
e^{-(\xi/4)\cos(2\varsigma u)}.\]
The other sets are given by
\begin{equation} 
\mathbb{W}_3(u)=\mathbb{W}_1(-u),
\qquad
\mathbb{W}_4(u)=\mathbb{W}_2(-u).
\end{equation}
In the recurrence relations (\ref{recursion-1}) for $b_n^1$ 
and $b_n^2$, 
the coefficients are obtained by replacing the 
parameters (\ref{whe2}) into (\ref{dche-1b}) 
and (\ref{dche-2b}) respectively. 
The convergence of the two-sided infinite series is given by
\begin{equation} 
\big|e^{2i\varsigma u}\big|>0 \text{ for }\mathbb{W}_1(u) \text{ and } \mathbb{W}_2(u),
\qquad
\big|e^{-2i\varsigma u}\big|<\infty \text{ for }\mathbb{W}_3(u) \text{ and } \mathbb{W}_4(u).
\end{equation}
Then, if $\varsigma=i$ and $u$ is real (WHE, not 
modified WHE) the four 
sets of solutions are convergent for all values of
$z=\exp{(2iu)}$, in opposition
to the solutions obtained from the solutions for the
the CHE -- see Eqs (\ref{conv-WHE-as-CHE}).

On the other side, the Mathieu equation (\ref{mathieu}) 
is converted into a DCHE
by the substitutions
\begin{eqnarray}\label{mathieu3}
\left.\begin{array}{l}
z=e^{i\sigma u},\quad w(u)=z^{1/2}e^{ik/z}U(z)
\qquad\Rightarrow
\vspace{2mm}\\
B_{1}=-2ik,\quad B_{2}=2, \quad
B_{3}=(1/4)-\mathrm{a}, \quad \omega=k,
\quad \eta=0
\end{array}\right\}
\begin{array}{l}
\mbox{Mathieu equation}\\
\mbox{as a DCHE.}
\end{array}
\end{eqnarray}
For these parameters the preceding solutions
$\mathbb{U}_i(z)$ for the DCHE are
expressed in series of Bessel functions $J_{\kappa}$,
$H_{\ \kappa}^{(1)}$ and $H_{\ \kappa}^{(2 )}$ by using
the relation (\ref{bessel-1}). We can as well include
the function $Y_{\kappa}$ since these are linear combinations
of the Hankel functions. Thus, from $\mathbb{U}_i(z)$
we obtain the sets of solutions, 
$w_{\ i}^{(j)}(u)$,
\letra
\begin{eqnarray} \label{mathieu-dche}
\begin{array}{ll}
\displaystyle
w_{\ 1}^{(j)}(u)=e^{ik[\cos(\sigma u)-i\sin(\sigma u)]}
\sum_{n}i^n a_nZ_{n+\nu+\frac{1}{2}}^{(j)}\left[ke^{i\sigma u}\right],
& w_{\ 3}^{(j)}(u)=w_{\ 1}^{(j)}(-u),
\vspace{2mm}\\
\displaystyle
w_{\ 2}^{(j)}(u)=e^{-ik[\cos(\sigma u)-i\sin(\sigma u)]}
\sum_{n}(-i)^n a_nZ_{n+\nu+\frac{1}{2}}^{(j)}\left[ke^{i\sigma u}\right],
& w_{\ 4}^{(j)}(u)=w_{\ 2}^{(j)}(-u)
\end{array}
\end{eqnarray}
where the recurrence relations are
\begin{equation}\begin{array}{l}
2k^2\left[\frac{n+\nu+1}{2n+2\nu+3}\right]a_{n+1}+
\left[\frac{1}{4}-\mathrm{a}+(n+\nu)(n+\nu+1)\right]a_n+
2k^2\left[\frac{n+\nu}{2n+2\nu-1}\right]a_{n-1}=0.
\end{array}
\end{equation}
The convergence of the above series requires that
\begin{equation} \label{mathieu-curved}
\big|e^{i\sigma u}\big|>0 \text{ for }{w}_{\ 1}^{(j)}(u) 
\text{ and } {w}_{\ 2}^{(j)}(u),
\qquad
\big|e^{-i\sigma u}\big|<\infty \text{ for }{w}_{\ 3}^{(j)}(u) \text{ and } {w}_{\ 4}^{(j)}(u).
\end{equation}
%
%
%


\subsection{Solutions for the Whittaker-Ince limit of the DCHE}
For the limit (\ref{incedche}) of the DCHE
we find only three sets of solutions. Two of them are obtained
by applying the Leaver limit ($z_0\to 0$) on the solutions
in series of Bessel functions (\ref{WIL-ni1a}) and (\ref{WIL-ni2a})
for the Whitakker-Ince limit of the CHE; the other set is given by series of
confluent hypergeometric functions and is obtained by applying
the Whittaker-Ince limit on the solutions (\ref{dche-3a})
or (\ref{dche-4a}) for the DCHE. 
We have found no transformation of variables 
connecting the solutions in series of Bessel functions
with the solutions in series of confluent hypergeometric functions. 

\subsubsection{The two types of solutions}
When $z_0\to 0$ the set (\ref{WIL-ni1a}) retains its form, that is,
\antiletra
\letra
\begin{eqnarray}
U_{\ 1}^{(j)}(z) =z^{(1-B_{2})/2}
\sum_n (-1)^n c_{n}^{1}
Z_{2n+2\nu+1}^{(j)}\left(2\sqrt{qz}\right)
\end{eqnarray}
but the coefficients (\ref{WIL-ni1b}) for the recurrence relations
reduce to
\begin{eqnarray}
\begin{array}{l}
\alpha_{n}^{1}  =  - \frac{qB_{1}\left[n+\nu+2-\frac{B_{2}}{2}\right]}
{(2n+2\nu+2)\left(2n+2\nu+3\right)},
\vspace{.2cm} \\
\beta_{ n}^{1}  =  B_{3}+\left(n+\nu+1-\frac{B_{2}}{2}\right)
\left(n+\nu+\frac{B_{2}}{2}\right)
-\frac{q B_{1}\left(B_{2}-2\right)}
{(2n+2\nu)(2n+2\nu+2)},
\vspace{.2cm} \\
\gamma_{n}^{1} =  \frac{qB_{1}\left[n+\nu+\frac{B_{2}}{2}-1\right]}
{\left(2n+2\nu-1\right)(2n+2\nu)}.
\end{array}
\end{eqnarray}
This set arises also from (\ref{WIL-ni3a}) when $z_0\to 0$.
It can  as well be obtained by taking the Whittaker-Ince limit
of solutions (\ref{dche-1a}) provided we add the solution
in series of $Y_{2n+2\nu+1}$. On the other hand, when $z_0\to 0$
the sets (\ref{WIL-ni2a}) and (\ref{WIL-ni4a}) give
%

\antiletra\letra
\begin{eqnarray}
U_{\ 2}^{(j)}(z)=e^{{B_{1}}/{z}}z^{(1-B_{2})/2}
\displaystyle \sum_n (-1)^n c_{n}^{2}
Z_{2n+2\nu+1}^{(j)}\left(2\sqrt{qz}\right),
\end{eqnarray}
with
\begin{eqnarray}
\begin{array}{l}
\alpha_{ n}^{2}=  \frac{qB_{1}\left(n+\nu+\frac{B_{2}}{2}\right)}
{(2n+2\nu+2)\left(2n+2\nu+3\right)}, \qquad
\beta_{ n}^{2}= \beta_{ n}^{1},\qquad
\gamma_{ n}^{2}= -\frac{qB_{1}\left(n+\nu+1-\frac{B_{2}}{2}\right)}
{\left(2n+2\nu-1\right)(2n+2\nu)},
\end{array}
\end{eqnarray}
in the recurrence relations for $c_{\ n}^{(2)}$.
This set can also be obtained by taking the Whittaker-Ince limit
of solutions (\ref{dche-2a}). On the other side,
solutions (\ref{dche-3a}) and (\ref{dche-4a}) yield the
same Whittaker-Ince limits in series of confluent hypergeometric
functions, namely,
%
%
\antiletra\letra
\begin{eqnarray}\label{incedche-3}
\begin{array}{l}
U_{3}(z)  = \displaystyle
\sum_n\begin{array}{l}
\frac{c_{ n}^{3}\left[-{B_1}/{z}\right]^{n+\nu+\frac{B_2}{2}}
}{\Gamma[2n+2\nu+2]}
\Phi\left[n+\nu+\frac{B_{2}}{2},2n+2\nu+2;
\frac{B_{1}}{z}\right],\end{array}
\vspace{0.2cm}\\
{U}_{3}^{\pm}(z) =e^{(1\mp 1)\frac{B_1}{2z}}
\displaystyle
\sum_n \begin{array}{l}
\frac{c_{ n}^{3}\left[{B_1}/{z}\right]^{n+\nu+\frac{B_2}{2}}}
{\Gamma\left[n+\nu+1\mp({B_2-2})/{2}\right]} 
%
%
\Psi\left[n+\nu+1\pm\frac{B_2-2}{2},2n+2\nu+2;
\pm\frac{B_{1}}{z}\right],\end{array}
\end{array}
\end{eqnarray}
where, in the recurrence relations for $c_{\ n}^{(3)}$,
\begin{eqnarray}
\begin{array}{l}
\alpha_{ n}^{3}  =\frac{-qB_1}
{(2n+2\nu+2)(2n+2\nu+3)}  ,
\qquad
\beta_{ n}^{3}  =\beta_{ n}^{1} ,\qquad
%
\gamma_{ n}^{3}=\frac{q B_1\left[n+\nu-1+\frac{B_2}{2}\right]
\left[n+\nu+1-\frac{B_2}{2}\right]}{(2n+2\nu-1)
(2n+2\nu)}.
\end{array}
\end{eqnarray}
This set yields expansions in series of Bessel functions if $B_2=2$. 

The three sets of solutions are given by two-sided
infinite series if
\antiletra
\begin{eqnarray} 
\begin{array}{l}
2\nu \ \text{  and  }\ 
\nu\pm\frac{B_2}{2}  
\text{  are not integers}.
\end{array}
\end{eqnarray}
The condition $2\nu\neq$integer avoids vanishing denominators
in the recurrence relations and assures the independence of 
the Bessel and hypergeometric functions in each series.
Besides this, if $\nu\pm({B_2}/{2})$ is not integer, the coefficients 
are connected by
\begin{eqnarray}
\begin{array}{l}
c_{ n}^{2}=(-1)^n\frac{\Gamma\left[n+\nu+2-(B_2/2)\right]}
{\Gamma\left[n+\nu+(B_2/2)\right]}\ c_{ n}^{1},\qquad
c_{ n}^{3}=\Gamma\left[n+\nu+2-\frac{B_2}{2}\right]c_{ n}^{1}.
\end{array}
\end{eqnarray}

For completeness we mention that, if $U(z)=U(B_1,B_2,B_3;z)$
is an arbitrary solution
for the Whittaker-Ince limit of the DCHE,
then a new solution may be generated by
the transformation $T$ whose effect on
$U(z)$ is
\begin{eqnarray}\label{transformationT}
T U(z)=e^{{B_{1}}/{z}}z^{2-B_{2}}U(-{B}_{1},4-{B}_{2},
{B}_{3}+2-B_2;z).
\end{eqnarray}
We can check that $TU_1^{(j)}=U_2^{(j)}$ while 
$T(U_3,U_3^{\pm})$ does not produce new solutions.

The solutions in series of the Bessel $J_{2n+2\nu+1}$
and $Y_{2n+2\nu+1}$, as well as the solution $U_3(z)$ in series of
regular confluent hypergeometric functions, are
new, whereas the other solutions have already appeared in \cite{eu}.

\subsubsection{Convergence of the solutions}
As in the case of the solutions for the DCHE, 
we find that
\begin{itemize}
\itemsep-3pt
\item The solutions ${U}_{\ 1}^{(j)}$
and ${U}_{\ 2}^{(j)}$ converge absolutely for $|z|>0$
but may be unbounded at $z=\infty$.
The one-sided series,
$\mathring{U}_{\ 1}^{(1)}$
and $\mathring{U}_{\ 2}^{(1)}$, converge also at $|z|=0$.
\item The solutions $({U}_3,U_3^{\pm} )$
converge for $|z|<\infty$
but may be unbounded at $z=0$.
The one-sided series 
$\mathring{U}_3$ converges also at $|z|=\infty$.
\end{itemize}
In effect, if $n\rightarrow\pm\infty$ the  recurrence relations
for $c_{ n}^{1}$ give
\begin{eqnarray*}\begin{array}{l}
\frac{qB_1}{n}\frac{c_{n+1}^{1}}{c_{ n}^{1}}-
\big[ 4n(n+2\nu+1)\big]-
\frac{qB_1}{n}
\frac{c_{n-1}^{1}}{{c}_{ n}^{1}}=0
\end{array}
\end{eqnarray*}
Then, the minimal solutions are 
\begin{eqnarray*}\begin{array}{lll}
n\to\infty:&\frac{c_{n+1}^{1}}{c_{ n}^{1}}\sim
\frac{-qB_1}{4(n+1)^2(n+2\nu+2)}=
-\frac{qB_1}{4n^3}\left[1+O\left(\frac{1}{n}\right)\right]
&\Rightarrow
\frac{c_{n-1}^{1}}{c_{ n}^{1}}\sim
\frac{4n^2(n+2\nu+1)}{-q B_1},\vspace{2mm}\\
n\to -\infty:& \frac{c_{n-1}^{1}}{c_{ n}^{1}}\sim\frac{qB_1}{4(n-1)^2(n+2\nu)} =
\frac{qB_1}{4n^3}\left[1+O\left(\frac{1}{n}\right)\right]\
&\Rightarrow
\frac{c_{n+1}^{1}}{c_{ n}^{1}}\sim
\frac{4n^2(n+2\nu+1)}{qB_1}.
\end{array}
\end{eqnarray*}
Thus, by using the ratios (\ref{rediscutir}) we find
\begin{eqnarray}
&&\begin{array}{l}
{n\to\infty}:\;\;
\frac{c_{n+1}^{1}J_{2n+2\nu+3}}{c_{ n}^{1}J_{2n+2\nu+1}}\sim
- \frac{q^2B_1z}{16n^5},\qquad
%
\frac{c_{n+1}^{1}Z_{2n+2\nu+3}^{(2,3,4)}}{c_{n}^{1}
Z_{2n+2\nu+1}^{(2,3,4)}}
\sim-\frac{B_1}{nz},\vspace{2mm}\\
\end{array}\\
&&\begin{array}{l}
{n\to-\infty}:\;\;\frac{c_{n-1}^{1}
Z_{2n+2\nu-1}^{(j)}}{c_{ n}^{1}Z_{2n+2\nu+1}^{(j)}}
\sim\frac{B_1}{nz},\ \qquad\quad [j=1,2,3,4]
\end{array}
\end{eqnarray}
[these expressions also result as limits of (\ref{following-0})
and (\ref{following}) when $z_0\to 0 $]. Therefore,
by the D'Alembert test the two-sided series solutions $U_{\ 1}^{(j)}$
converge absolutely for $|z|>0$; for one-sided series,
$\mathring{U}_{\ 1}^{(1)}$ converges also at $|z|=0$ because
only the limit $n\to +\infty$ is relevant. However, 
the behavior of each solution at $z=\infty$
must be analyzed by using the relations (\ref{comportamento-infinito}).
%
%
The convergence of the set $U_{\ 2}^{(j)}$ is inferred from 
$U_{\ 2}^{(j)}=TU_{\ 1}^{(j)}$; since
the transformation $T$ replaces $B_1$ by $-B_1$, it does not
change the convergence properties for the series in $U_{\ 2}^{(j)}$ but
can change the behaviors at $z=\infty$ -- see Eq. (\ref{transformationT}).

To get the convergence of the third set, 
we write the solutions (\ref{incedche-3}) as 
\begin{eqnarray*} 
\mathbb{U}_3(z)=z^{1-\frac{B_2}{2}}e^{B_1/(2z)}\sum_n c_{ n}^{3}
\mathscr{U}_{n+\nu}\left[\frac{i(2-B_2)}{2},\frac{iB_1}{2z}\right],
\qquad\mathscr{U}_{n+\nu}=(\varphi_{n+\nu},\psi_{n+\nu}^{\pm})
\end{eqnarray*}
where we have used the definitions (\ref{fi}) for $\mathscr{U}_{n+\nu}$.
When $n\to \pm\infty$, $c_{ n}^{3}$ obeys
\begin{eqnarray*}\begin{array}{l}
\frac{qB_1}{n^2}\frac{c_{n+1}^{3}}{c_{ n}^{3}}-
\big[ 4n(n+2\nu+1)\big]-
qB_1
\frac{c_{n-1}^{3}}{c_{ n}^{3}}=0
\end{array}
\end{eqnarray*}
whose minimal solutions are
\begin{eqnarray}\begin{array}{l}
\frac{c_{n+1}^{3}}{c_{ n}^{3}}\sim
\frac{-qB_1}{4(n+1)(n+2\nu+2)}=
-\frac{qB_1}{4n^2}\left[1+O\left(\frac{1}{n}\right)\right]
\ \Leftrightarrow\ 
\frac{c_{n-1}^{3}}{c_{ n}^{3}}\sim-
\frac{4n(n+2\nu+1)}{q B_1}, \quad[n\to\infty]\vspace{2mm}\\
%
\frac{c_{n-1}^{3}}{c_{ n}^{3}}\sim\frac{qB_1}{4(n-1)^3(n+2\nu)} =
\frac{qB_1}{4n^4}\left[1+O\left(\frac{1}{n}\right)\right]\
\Leftrightarrow\
\frac{c_{n+1}^{3}}{c_{n}^{3}}\sim
\frac{4n^3(n+2\nu+1)}{qB_1},\qquad [n\to -\infty].
\end{array}
\end{eqnarray}
On the other hand, for finite values of $y=iB_1/(2z)$
relations (\ref{bessel}) and (\ref{hankel}) give
\begin{eqnarray}
\begin{array}{lll} 
\frac{\phi_{n+\nu+1}}{\phi_{n+\nu}}\sim \frac{-B_1}{4n^2 z}
\left[1-\frac{1}{n}\left(2\nu+\frac{5}{2}\right)\right],\qquad&
\frac{\psi_{n+\nu+1}^{\pm}}{\psi_{n+\nu}^{\pm}}
\sim \frac{4z}{B_1}
\left[1-\frac{1}{2n}\right] \quad & [n\to\infty]\vspace{2mm}\\
%
\frac{\mathscr{U}_{n+\nu-1}}{\mathscr{U}_{n+\nu}}\sim -\frac{4n^2z}{B_1}
\left[1+\frac{1}{n}\left(2\nu+\frac{1}{2}\right)\right],
& \mathscr{U}_{n+\nu}=\left(\phi_{n+\nu},\psi_{n+\nu}^{(\pm)}\right)
& [n\to -\infty].
  \end{array}
\end{eqnarray}
Thence
\begin{eqnarray}\begin{array}{l}
\frac{c_{ n+1}^{3} \phi_{n+\nu+1}}{c_{ n}^{3}\ \phi_{n+\nu}}
\sim\frac{qB_1^2}{16n^4 z},\qquad
\frac{c_{n+1}^{3} \psi_{n+\nu+1}^{\pm}}
{c_{ n}^{3} \psi_{n+\nu}^{\pm}}
\sim-\frac{qz}{n^2},
\quad [n\to\infty]\vspace{2mm}\\
\frac{c_{n-1}^{3} \mathscr{U}_{n+\nu-1}}
{c_{n}^{3} \mathscr{U}_{n+\nu}}
\sim-\frac{qz}{n^2},\quad\mathscr{U}_{n+\nu}=
\left(\phi_{n+\nu},\psi_{n+\nu}^{(\pm)}\right)
\quad[n\to -\infty].
 \end{array}
\end{eqnarray}
Thus, the two-sided series solutions $\mathbb{U}_3(z)$
converges for finite values of $z$, excepting 
possibly the point $z=0$.  
Since the limit $n\to\infty$ is the only one applicable
for one-sided series, the solution $\mathring{U}_3$ 
converge also at $z=\infty$.
In order to verify if the solutions are bounded or not
$z=0$, we must find the behavior of such solutions at $z=0$. 
The results are the same given in Eqs. (\ref{bound-dche}) 
and (\ref{bounded-dche}) for the DCHE.
%

\section{One-sided Series Expansions}

In  Ref. \cite{eu2} one-sided solutions in series
of Coulomb wave functions have been obtained from
two-sided solutions. In this section this problem is reconsidered 
because the redefinitions of the Coulomb functions modify 
the recurrence relations for the series coefficients, 
while the Raabe test modifies the regions of convergence for the
solutions of the CHE. We examine also 
the Whittaker-Ince limits
of the CHE and DCHE and the conditions for having finite-series solutions.

In fact, it is possible to restrict 
the summation of the two-sided series
to $n\geq 0$ by writing the parameter $\nu$ as function of the 
parameters of the CHE. For each of the four 
sets $\mathbb{U}_i(z)$ we get
two expressions for $\nu$ and, accordingly, 
eight sets $\mathring{\mathbb{U}}_i(z)$ of one-sided series solutions. 
Thus, for $\mathbb{U}_1(z)$ 
we obtain $\nu=(B_2/2)-1$ and $\nu=(B_2/2)+(B_1/z_0)$
-- see Appendix B. We consider only the set $\mathring{\mathbb{U}}_1(z)$
obtained by setting $\nu=(B_2/2)-1$ in $\mathbb{U}_1(z)$,  
and generate the other sets by applying the transformations 
(\ref{transformacao1}) in the order given in Eqs. (\ref{transf}). 
If $\eta\neq0$, there are three
possible types of recurrence relations for the series coefficients; 
if $\eta=0$, only two types.

Solutions for the DCHE, as well as for the 
Whittaker-Ince limit of the CHE and DCHE, are obtained
by the procedures used in section II. 
We will see that in a fixed set 
for the CHE and DCHE, it is possible to have one solution given
by infinite and one given by finite series. 
On the other side, the one-sided expansions in 
series of regular hypergeometric functions $\Phi(a,c;y)$,
their particular and limiting cases, are convergent 
also in the vicinity of $z=0$, as we have already 
seen in section II. 

First we regard the solutions for the CHE and DCHE; 
after that, the solutions
for the Whittaker-Ince limit for the CHE and DCHE.

\subsection{Solutions for the confluent Heun equation}

The one-sided solutions $\mathring{\mathbb{U}}_1(z)$,
obtained in Appendix B,
are 
\begin{eqnarray}\label{forma-de-leaver-truncada}
\mathring{\mathbb{U}}_1(z) =z^{-\frac{B_2}{2}}\displaystyle\sum_{n=0}^{\infty} \mathring{b}_{n}^{1}
\mathscr{U}_{n-1+\frac{B_2}{2}}\left(\eta,\omega{z}\right),
\qquad 
\mathscr{U}_{n-1+\frac{B_2}{2}}=
\left(\phi_{n-1+\frac{B_2}{2}},\psi_{n-1+\frac{B_2}{2}}^{\ \pm}\right),
\end{eqnarray}
which can be rewritten as in Eqs. (\ref{par1}). 
Then, the transformations (\ref{transformacao1}) 
give the sets:
\begin{eqnarray}\label{transf}
\begin{array}{lllll}
\mathring{\mathbb{U}}_1=\mathbb{U}_1\big|_{\nu=\nu_1},\quad &\mathring{\mathbb{U}}_2=T_1\mathring{\mathbb{U}}_1,\quad &
\mathring{\mathbb{U}}_3=T_2\mathring{\mathbb{U}}_2,
\quad &\mathring{\mathbb{U}}_4=T_1\mathring{\mathbb{U}}_3:
\quad & \text{first subgroup};
\vspace{2mm}\\
\mathring{\mathbb{U}}_5=T_4\mathring{\mathbb{U}}_1,
&\mathring{\mathbb{U}}_6=T_1\mathring{\mathbb{U}}_5,&
\mathring{\mathbb{U}}_7=T_2\mathring{\mathbb{U}}_6,&
\mathring{\mathbb{U}}_8=T_1\mathring{\mathbb{U}}_7:&
\text{sec. subgroup}.
\end{array}
\end{eqnarray}
We can check that these solutions are obtained 
by setting $\nu=\nu_i$ in the two-sided solutions $\mathbb{U}_i $
(and also in $\alpha_n^{i}$, $\beta_n^{i}$ and $\gamma_n^{i}$) 
according to
\begin{eqnarray}\label{combination-0}
\begin{array}{llll}
\mathring{\mathbb{U}}_1=\mathbb{U}_1\big|_{ \nu_1},\qquad &
\mathring{\mathbb{U}}_2=\mathbb{U}_1\big|_{\nu_2},
\qquad & 
\mathring{\mathbb{U}}_3=\mathbb{U}_2\big|_{\nu_3},\qquad &
\mathring{\mathbb{U}}_4=\mathbb{U}_2\big|_{\nu_4};
\vspace{2mm}\\
\mathring{\mathbb{U}}_5=\mathbb{U}_3\big|_{ \nu_5},&
\mathring{\mathbb{U}}_6=\mathbb{U}_4\big|_{\nu_6},
&
\mathring{\mathbb{U}}_7=\mathbb{U}_4\big|_{\nu_7},&
\mathring{\mathbb{U}}_8=\mathbb{U}_3\big|_{\nu_8}
\end{array}
\end{eqnarray}
with 
\begin{equation}\label{combination}
\begin{array}{l}
\nu_1=\frac{B_{2}}{2}-1,\quad\nu_2=\frac{B_{1}}{z_0}+\frac{B_2}{2},
\quad \nu_3=1-\frac{B_{2}}{2},\quad\nu_4=-\frac{B_{1}}{z_0}-\frac{B_2}{2},
\quad \nu_{i+4}=\nu_i,
\end{array}
\end{equation}
where $i=1,2,3,4$. Thus, $\nu$ is given by the same expression in
the solutions $\mathring{\mathbb{U}}_i$ and $\mathring{\mathbb{U}}_{i+4}$.

First we write the general form of 
the recurrence relations 
for the series coefficients and state the convergence of
the infinite series. Then, we examine the conditions
for having finite series for the cases
$\eta\neq 0$ and $\eta=0$. Actually, here we write only
the sets $\mathring{\mathbb{U}}_1$ and 
$\mathring{\mathbb{U}}_5$; the other sets are
given in Appendix C.

\subsubsection{Recurrence relations, transformations and convergence}

In the following $\mathring{\alpha}_n$, $\mathring{\beta}_n$,
 $\mathring{\gamma}_n$ and $\mathring{b}_n$ mean 
$\mathring{\alpha}_n^{i}$, $\mathring{\beta}_n^{i}$, 
$\mathring{\gamma}_n^{i}$ and $\mathring{b}_n^{i}$.  
Then, by Appendix B the forms of the recurrence 
and characteristic relations are
\begin{eqnarray}
\left.
\begin{array}{l}
\mathring{\alpha}_{0}\mathring{b}_{1}+
\mathring{\beta}_{0}\mathring{b}_{0}=0,
\vspace{.2cm} \\
\mathring{\alpha}_{n}\mathring{b}_{n+1}+
\mathring{\beta}_{n}\mathring{b}_{n}+
\mathring{\gamma}_{n}\mathring{b}_{n-1}=0,\ n\geq1
\end{array}\right\} \Rightarrow
\mathring{\beta}_{0}=\frac{\mathring{\alpha}_{0}\mathring{\gamma}_{1}}
{\mathring{\beta}_{1}-}\ \frac{\mathring{\alpha}_{1}
\mathring{\gamma}_{2}}
{\mathring{\beta}_{2}-}\ \frac{\mathring{\alpha}_{2}\mathring{\gamma}_{3}}
{\mathring{\beta}_{3}-}\cdots,
\label{r1a}
\end{eqnarray}
\begin{eqnarray}
\left.
\begin{array}{l}
\mathring{\alpha}_{0}\mathring{b}_{1}+\mathring{\beta}_{0}\mathring{b}_{0}=0,
\qquad\qquad \mathring{\alpha}_{1}\mathring{b}_{2}+ \vspace{.2cm} \\
\mathring{\beta}_{1}\mathring{b}_{1}+\left[
\mathring{\gamma}_{1}-\left(\eta^2+\frac{1}{4}\right)\mathring{\alpha}_{-1}
\right]\mathring{b}_{0}=0,
\vspace{.2cm} \\
\mathring{\alpha}_{n}\mathring{b}_{n+1}+\mathring{\beta}_{n}
\mathring{b}_{n}+\mathring{\gamma}_{n}
\mathring{b}_{n-1}=0,\ n\geq2
\end{array}\right\}\Rightarrow
%
\mathring{\beta}_{0}=
\frac{\mathring{\alpha}_{0}\left[\mathring{\gamma}_{1}-
\left(\eta^2+\frac{1}{4}\right)\mathring{\alpha}_{-1}
\right]}
{\mathring{\beta}_{1}-}
 \frac{\mathring{\alpha}_{1}\mathring{\gamma}_{2}}
{\mathring{\beta}_{2}-} 
\cdots, 
\label{r2a}
\end{eqnarray}
\begin{eqnarray}
\left.
\begin{array}{l}
\mathring{\alpha}_{0}\mathring{b}_{1}+
\left[\mathring{\beta}_{0}-i\eta\mathring{\alpha}_{-1}
\right]\mathring{b}_{0}=0,
\vspace{.2cm} \\
\mathring{\alpha}_{n}\mathring{b}_{n+1}+
\mathring{\beta}_{n}\mathring{b}_{n}
+\mathring{\gamma}_{n}\mathring{b}_{n-1}=0,\ n\geq1
\end{array}\right\}\Rightarrow
\mathring{\beta}_{0}-i\eta\mathring{\alpha}_{-1}=
\frac{\mathring{\alpha}_{0}\mathring{\gamma}_{1}}
{\mathring{\beta}_{1}-}\ \frac{\mathring{\alpha}_{1}
\mathring{\gamma}_{2}}{\mathring{\beta}_{2}-}
\ \frac{\mathring{\alpha}_{2}\mathring{\gamma}_{3}}
{\mathring{\beta}_{3}-}\cdots ,
\label{r3a}
\end{eqnarray}
where $\mathring{\alpha}_n$,  $\mathring{\beta}_n$ and  
$\mathring{\gamma}_n$ depend
on the parameters of the equation. Notice that
$n\geq -1$ for $\mathring{\alpha}_n$, $n\geq 0$ for $\mathring{\beta}_n$,
and $n\geq 1$ for $\mathring{\gamma}_n$.


Besides this, relations (\ref{r1a}-\ref{r3a})
stand if the series really begin at $n=0$, that is, if 
$\mathring{b}_0\neq 0 $. If the summation begins
at $n=N\geq 1$, the recurrence relations are given
by (\ref{N>1}). 
%
%
For series beginning at $n=0$, the recurrence 
relations for $\eta\neq 0$ and $\eta=0$ are given 
by 
\begin{equation}\label{rel}
\begin{array}{ll}
\eta\neq0:\text{ (\ref{r1a})} \text{ if }2\nu_i+2\neq 1,2;
\quad  \text{ (\ref{r2a})} \text{ if }2\nu_i+2=1;
\quad  \text{ (\ref{r3a})} \text{ if }2\nu_i+2=2,
\vspace{2mm}\\
\eta=0:\text{ (\ref{r1a})} \text{ if }2\nu_i+2\neq 1;
\quad  \text{ (\ref{r2a})} \text{ if }2\nu_i+2=1
\end{array}
\end{equation}
where $\nu_i$ are the parameters given in (\ref{combination}).
Thus, $\mathring{\mathbb{U}}_i$ and 
$\mathring{\mathbb{U}}_{i+4}$ ($i=1,\cdots,4$) have 
the same type of recurrence relations. 
On the other side, to assure the linear independence 
of each term of the series we demand that
\begin{equation}\label{L-IND}
2\nu_i+2\neq 0,-1,-2,\cdots\text{ for series beginning at } n=0.
\end{equation}
These restrictions also avoid
vanishing denominators in the recurrence
relations. 
Explicitly the initial set reads
\begin{eqnarray}
\label{par1}
\begin{array}{l}
\mathring{U}_{ 1}(z) =e^{i\omega z}\displaystyle\sum_{n=0}^{\infty}
\frac{\mathring{b}_{n}^{1}\ [2i\omega z]^{n}}{\Gamma(2n+B_2)}
\Phi\left(n+i\eta+\frac{B_{2}}{2},2n+B_{2};-2i\omega z\right),\vspace{2mm}\\
\mathring{U}_{1}^{\pm}(z) =e^{\pm i\omega z}\displaystyle\sum_{n=0}^{\infty}
\frac{\mathring{b}_{n}^{1}\ [-2i\omega z]^{n}}{\Gamma[n\mp i\eta+(B_2/2)]}
\Psi\left(n\pm i\eta+\frac{B_{2}}{2},2n+B_{2};\mp 2i\omega z\right), 
\end{array} 
%
\end{eqnarray}
where the recurrence relations for $\mathring{b}_n^{1}$ are given
by (\ref{s-che-1}) with
\begin{eqnarray}\label{coef-3'}
\begin{array}{l}\displaystyle
\mathring{\alpha}_{n}^{1}=
\frac{2i\omega z_0\left[ n+1\right]
\left[n-\frac{B_1}{z_0}\right]}
{\left(2n+B_2\right)\left(2n+B_2+1\right)},\vspace{2mm}\\
\mathring{\beta}_{n}^{1}=-n\left(n+B_2-1\right) -
\omega \eta z_0 - B_3-
\frac{\omega \eta z_0\left[B_2+\frac{2B_1}{z_0}\right]\left[B_2-2\right]}
{\left(2n+B_2-2\right)\left(2n+B_2\right)},\vspace{2mm}\\
\displaystyle
\mathring{\gamma}_{n}^{1}=-\frac{2i\omega z_0
\left[n+B_2-2\right]
\left[ n+B_2+\frac{B_1}{z_0}-1\right]\left[ n-1+i\eta+\frac{B_2}{2}\right]
\left[ n-1-i\eta +\frac{B_2}{2}\right]}
{\left(2n+B_2-3\right)\left(2n+B_2-2\right)}.
\end{array}
\end{eqnarray}
The fifth set, $\mathring{\mathbb{U}}_{5}(z)=T_4\mathring{\mathbb{U}}_{1}(z)$, is 
\begin{eqnarray}
\label{par5}
\begin{array}{l}
\mathring{U}_{5}(z) =e^{i\omega z}\displaystyle\sum_{n=0}^{\infty}
\frac{\mathring{b}_{n}^{5}\ \left[2i\omega (z-z_0)\right]^{n}}{\Gamma[2n+B_2]}
\Phi\left[n+i\eta+\frac{B_{2}}{2},2n+B_{2};-2i\omega (z-z_0)\right],
\vspace{2mm}\\
U_{5}^{\pm}(z) =e^{\pm i\omega z}\displaystyle\sum_{n=0}^{\infty}
\frac{\mathring{b}_{n}^{5}\ \left[-2i\omega (z-z_0)\right]^{n}}
{\Gamma[n\mp i\eta+(B_2/2)]}
\Psi\left[n\pm i\eta+\frac{B_{2}}{2},2n+B_{2};\mp 2i\omega (z-z_0)\right]
\end{array}
\end{eqnarray}
with 
%
\begin{eqnarray}\label{coef-par5}
\begin{array}{l}\displaystyle
\mathring{\alpha}_{n}^{5}=-
\frac{2i\omega z_0\left[ n+1\right]
\left[n+B_2+\frac{B_1}{z_0}\right]}
{\left(2n+B_2\right)\left(2n+B_2+1\right)},\quad \quad
\mathring{\beta}_{n}^{5}=\mathring{\beta}_{n}^{1},\vspace{2mm} \\
\displaystyle
\mathring{\gamma}_{n}^{5}=\frac{2i\omega z_0
\left[n+B_2-2\right]
\left[ n-1-\frac{B_1}{z_0}\right]\left[ n-1+i\eta+\frac{B_2}{2}\right]
\left[ n-1-i\eta +\frac{B_2}{2}\right]}
{\left(2n+B_2-3\right)\left(2n+B_2-2\right)}.
\end{array}
\end{eqnarray}
The recurrence relations for $\mathring{b}_{n}^{1}$
and $\mathring{b}_{n}^{5}$ are given by
\begin{eqnarray}\label{s-che-1}
\begin{array}{ll}
\eta\neq 0:&\text{Eqs. (\ref{r1a})} \text{ if } B_2\neq 1,2; \quad
\text{Eqs. (\ref{r2a})} \text{ if } B_2=1; \quad
\text{Eqs. (\ref{r3a})} \text{ if } B_2=2.\vspace{2mm}\\
\eta=0 :&\text{Eqs. (\ref{r1a})} \text{ if } B_2\neq 1; \quad
\text{Eqs. (\ref{r2a})} \text{ if } B_2=1,
\end{array}
\end{eqnarray}
while 
for both sets conditions (\ref{L-IND})
become
\begin{equation}\label{U1-U5}
B_2\neq 0,-1,-2,\cdots\text{ for } \mathring{\mathbb{U}}_1
\text{ and } \mathring{\mathbb{U}}_5.
\end{equation}

Although the recurrence relations (\ref{r2a}) 
and (\ref{r3a}) hold only for
special cases, their relevance becomes 
apparent by regarding the Whittaker-Hill equation as a CHE. By 
Eqs. (\ref{wheasgswe}), for this case $z_0=1$, $B_1=-1/2$ and $B_2=1$. 
Then, examining also the solutions
written in Appendix C, we find:  
Eqs. (\ref{r2a}) for $\mathring{\mathbb{U}}_{1}$ and 
$\mathring{\mathbb{U}}_{5}$; Eqs. (\ref{r1a}) 
for $\mathring{\mathbb{U}}_{3}$ and $\mathring{\mathbb{U}}_{7}$; Eqs. 
(\ref{r3a}) for the other solutions. 

The relations (\ref{r1a}-\ref{r3a})
and (\ref{L-IND}) hold if the series begin at $n=0$ and extend 
to infinite. However, in some cases the series truncate on 
the right-hand side leading to finite series 
(see page 146 of \cite{arscott}): for these solutions the 
ratio tests are meaningless. In other cases the series are 
truncated on the left-hand side and, then,
the series begin at $n>0$ (page 171 of \cite{arscott}, Ex. 2).

In effect, if $\mathring{\gamma}_{n}=0$ 
for some $n=\text{\small N}\geq 1$, 
we have three-term recurrence relations only if the series
terminate at $n=\text{\small N}-1$, that is, if 
$\mathring{b}_{\text{\tiny  N}}=\mathring{b}_{\text{\tiny N}+1}=\cdots=0$. 
In particular, for $\text{\small N}=1$ the series presents
just the first term except when the recurrence relations
are given by (\ref{r2a}): in this case the series
presents only the first term if 
$\mathring{\gamma}_1-[\eta^2+(1/4)]\mathring{\alpha}_{-1}=0$.

On the other hand, if $\mathring{\alpha}_{n}=0$ for some 
$n=\text{\small N}-1\geq 0$, then
the theory of three-term recurrence relations
remains consistent only for series beginning
at $n=\text{\small N}$, that is, if $\mathring{b}_{0}=\cdots=
\mathring{b}_{\text{\tiny N}-1}=0$.
In this case, the recurrence relations and the
corresponding characteristic equation (in terms of
continued fractions) are
\letra
\begin{eqnarray}\label{N>1}
\left.
\begin{array}{l}
\mathring{\alpha}_{\text{\tiny N}}\mathring{b}_{\text{\tiny N}+1}+
\mathring{\beta}_{\text{\tiny N}}\mathring{b}_{\text{\tiny N}}=0,
\vspace{.2cm} \\
\mathring{\alpha}_{n}\mathring{b}_{n+1}+\mathring{\beta}_{n}\mathring{b}_{n}+
\mathring{\gamma}_{n}\mathring{b}_{n-1}=0\ (n\geq \text{\small N}+1)
\end{array}\right\} \Rightarrow
\mathring{\beta}_{\text{\tiny N}}=
\frac{\mathring{\alpha}_{\text{\tiny N}}\mathring{\gamma}_{\text{\tiny N}+1}}
{\mathring{\beta}_{\text{\tiny N}+1}-}\ 
\frac{\mathring{\alpha}_{\text{\tiny N}+1}
\mathring{\gamma}_{\text{\tiny N}+2}}
{\mathring{\beta}_{\text{\tiny N}+2}-}\cdots. 
\end{eqnarray}
In this case, conditions (\ref{L-IND}) are replaced by
\begin{equation}\label{N2>1}
2\nu_i+2 \text{\small N}+2\neq 0,-1,-2,\cdots\text{ for series beginning at }n=N.
\end{equation}

The convergence of the one-sided infinite series 
is obtained from the convergence of the two-sided
infinite series. We find that
\begin{itemize}
\itemsep-3pt
\item The expansions $\mathring{U}_i(z)$ in series 
of regular hypergeometric
functions converge for finite values of $z$ in the 
complex plane, as explained in sec. II. 
\item By the D'Alembert test the expansions $\mathring{U}_i^{\pm}(z)$ in 
series of irregular hypergeometric functions 
converge for finite values of $z$ such that $|z|>|z_0|$ 
(if $i=1,\cdots,4$) or $|z-z_0|>|z_0|$ (if $i=5,\cdots,8$); 
by the Raabe test, they converge also at
$|z|=|z_0|$ and $|z-z_0|=|z_0|$ if the conditions given in 
(\ref{convergencia1-nu=0}) are fulfilled.
\item The behaviour of the solutions
at $z=\infty$ must be determined 
by using (\ref{asymptotic-confluent1}) and (\ref{asymptotic-confluent}).
\end{itemize}
The regions of convergence given by the Raabe test  
result from the combination 
of relations (\ref{convergencia1-0}) with (\ref{combination}).
So, we find
\antiletra
\begin{equation}\label{convergencia1-nu=0}
|z|\geq|z_0|  \text{ if }
\begin{cases}\text{Re}\left[  B_2+\frac{B_1}{z_0}\right] <1: \
\mathring{U}_{1,2}^{\pm};\vspace{2mm} \\
\text{Re}\left[  B_2+\frac{B_1}{z_0}\right] >1:\ \mathring{U}_{3,4}^{\pm};
\end{cases} 
|z-z_0|\geq|z_0| \text{ if }
\begin{cases}\text{Re}\left[ \frac{B_1}{z_0}\right] >{-}1: \ 
\mathring{U}_{5,8}^{\pm};
\vspace{2mm} \\
\text{Re}\left[ \frac{B_1}{z_0}\right] <{-}1: \ 
\mathring{U}_{6,7}^{\pm}, 
\end{cases}
\end{equation}
where $\mathring{U}_{m,n}^{\pm}$ means: $\mathring{U}_{\ m}^{\pm}$
and $\mathring{U}_{\ n}^{\pm}$. 
The test is inconclusive if $\text{Re}[B_2+(B_1/z_0)]=1$ and 
$\text{Re}[B_1/z_0]=-1$ in first and second cases, respectively.
The Raabe test is not valid if the above restrictions 
are incompatible with the restrictions which
assure the independence of the hypergeomertic functions
in each series; in addition,
it does not assure that the solutions
are bounded at the singular points.

\subsubsection{Finite-series solutions 
for $\eta\neq 0$ }

If $\eta\neq 0$, in the same set it 
is possible to have solutions in which the series
truncate on the left-hand side (this gives infinite
series solutions) and on the right-hand side (finite series). 
In this case, the finite and infinite
series present different series coefficients, despite
the fact that these are formally denoted by the same symbol
$\mathring{b}_n ^i$.

We consider only the first set but the following
considerations can be extended for the other sets. Thus,
if
\begin{eqnarray}\label{serieinfinita-1}
2n+B_2 \quad\text{and}
\quad n\pm i\eta+\frac{B_{2}}{2}\quad \text{are not zero or negative integers
in } \mathring{\mathbb{U}}_1,
\end{eqnarray}
for any admissible $n$, 
each of the solutions (\ref{par1})} 
can be expressed as a linear
combination of the others by means of Eq. (\ref{continuation2}). 
For series beginning at $n=0$, the above conditions are 
satisfied if $B_2$, $i\eta+{B_{2}}/{2}$
and $-i\eta+{B_{2}}/{2}$ are not zero or negative integers. 
However, when the second or the third condition is not valid,
the solutions can be interpreted as ($\eta\neq 0$):
\begin{itemize}
\itemsep-3pt
\item If $i\eta+{B_{2}}/{2}=-\ell$,
where $\ell$ is zero or a natural
number, then $\mathring{U}_{1}$ and $\mathring{U}_{1}^{+}$
are solutions linearly dependent given by a
finite series with $0\leq n\leq \ell$ because
$\mathring{\gamma}_{n=\ell+1}^{1}=0$, 
whereas  the series in $\mathring{U}_{1}^{-}$ begins at $n=\ell+1$
because $1/\Gamma(n-\ell)=0$ for $n\leq \ell$: 
in this case the solutions cannot be related by 
(\ref{continuation2}). 
    \item If $-i\eta+{B_{2}}/{2}=-\ell$,
where $\ell$ is zero or a natural
number, then $\mathring{U}_{1}\propto\mathring{U}_{1}^{-}$ 
is given by a finite series with $0\leq n \leq \ell$, whereas the 
series in $\mathring{U}_{1}^{+}$ begins at $n=\ell+1$.
\end{itemize}
The linear dependence follows from the generalized Laguerre 
polynomials of degree $l$ \cite{erdelyi1} 
\begin{eqnarray}\label{laguerre-1}
\begin{array}{l}
L_l^{\alpha}(y)=\frac{\Gamma(\alpha+l+1)}{l !\ \Gamma(\alpha+1)}
\ \Phi(-l,\alpha+1;y)=\frac{(-1)^{l}}{l !}\ \Psi(-l,\alpha+1;y),
\qquad [l=0,1,2,\cdots]
\end{array}
\end{eqnarray}
which yield
\letra
\begin{equation}\label{laguerre-2}
\mathring{U}_1\propto
\mathring{U}_1^{+}=e^{i\omega z}\displaystyle \sum_{n=0}^{\ell}
\frac{(\ell-n)! \ \mathring{b}_n^{1}}{\Gamma(n+\ell+B_2)}[2i\omega z]^n
L_{\ell-n}^{2n+B_2-1}(-2i\omega z),\text{ if } i\eta+\frac{B_2}{2}=-\ell
\end{equation}
and
\begin{equation}\label{laguerre-3}
\mathring{U}_1\propto
\mathring{U}_1^{-}=e^{-i\omega z}\displaystyle \sum_{n=0}^{\ell}
\frac{(\ell-n)! \ \mathring{b}_n^{1}}{\Gamma(n+\ell+B_2)}[2i\omega z]^nL_{\ell-n}^{2n+B_2-1}(2i\omega z), 
\text{ if }-i\eta+\frac{B_2}{2}=-\ell.
\end{equation}
%

%
%
The above types of finite series occur under
the same conditions [that is, $\pm i\eta+(B_2/2)=-\ell$]
as the ones obtained from the Baber-Hass\'e expansions 
in ascending power series of ($z-z_0$) \cite{baber,lea-1,leaver}.
However, the power series are more general, since the
expansions in confluent hypergeometric functions 
require restrictions on the parameters
of the CHE. On the other side, 
finite and infinite series solutions concerning  
the same set present different multiplicative
factors, $\exp(i\omega z)$ or $\exp(-i\omega z)$, 
and so it is possible that only one of these solutions 
is bounded at $z=\infty$.

The expansions in Coulomb functions afford
a second type of finite series. For instance,
\antiletra
\begin{eqnarray}\label{second-type}
\text{if } B_2\neq 0,-1,-2,\cdots\  \text { and } \ 
B_2+(B_1/z_0)= -\ell\  \text{ in  }\ \mathring{\mathbb{U}}_1, 
\qquad [ \ell=0,\ 1,
\  2,\ \cdots]\end{eqnarray}
then the coefficient $\mathring{\gamma}_{ n}^{1}$ 
of $\mathring{b}_{n-1}^{1}$ is zero for 
$n=\ell+1$ and so the three solutions $\mathring{\mathbb{U}}_1$ 
are given by finite series with $0\leq n\leq \ell$. 
For this second type of finite series, the hypergeometric 
functions do not reduce to polynomials and, in general, the 
solutions are not bounded for all values of $z$. Indeed, 
$\mathring{U}_1$ is bounded for finite values of $z$, 
but may be unbounded at $z=\infty$ because the limit 
(\ref{asymptotic-confluent}) for $\Phi$ contains
both the factors $\exp{( i\omega z)}$ and $\exp{(-i\omega z)}$.
On the other side, the solutions $\mathring{U}_1^{\pm}$
can be unbounded at $z=0$ because in general the 
functions $\Psi$ present logarithmic terms at $z=0$. 

The second type of finite series does not occur  
in solutions of the DCHE ($z_0\to 0$) but is the only type 
which may be valid for the CHE with $\eta=0$ and also in the 
Whittaker-Ince limit (\ref{incegswe}) of the CHE. This 
fact becomes relevant because these equations do  
not admit finite series in ascending powers of $z$
or $z-z_0$.

\subsubsection{Solutions for $\eta=0$ and the spheroidal equation}
As in the case of two-sided solutions, if $\eta=0$ we
write the confluent hypergeometric functions of the first set
in terms of Bessel functions, redefine the series coefficients
as 
%
\begin{eqnarray}\label{redefinition}
\begin{array}{l}
\mathring{b}_{n}^{1}=i^{-n}\Gamma\left(n+\frac{B_2}{2}\right)
\mathring{a}_{n}^{1}
\end{array}\end{eqnarray}
and add the expansions in series of Bessel functions
of the second kind, $Y_{\kappa}$. Thus, from the first
set (\ref{par1}) we find the set $\mathring{U}_{\ 1}^{(j)}$
given in Eq. (\ref{Uj}). The full group is generated by 
the transformations (\ref{transformacao1}) applied  
as in Eqs. (\ref{transf}), that is,
\begin{eqnarray*}\label{transf-eta-zero}
\begin{array}{lllll}
\mathring{U}_{\ 1}^{(j)},\quad 
&\mathring{U}_{\ 2}^{(j)}=T_1\mathring{U}_{\ 1}^{(j)},\quad 
&\mathring{U}_{\ 3}^{(j)}=T_2\mathring{U}_{\ 2}^{(j)},
\quad 
&\mathring{U}_{\ 4}^{(j)}=T_1\mathring{U}_{\ 3}^{(j)}:
\  
& \text{first subgroup};
\vspace{2mm}\\
\mathring{U}_{\ 5}^{(j)}=T_4\mathring{U}_{\ 1}^{(j)},\ 
&\mathring{U}_{\ 6}^{(j)}=T_1\mathring{U}_{\ 5}^{(j)},\quad 
&\mathring{U}_{\ 7}^{(j)}=T_2\mathring{U}_{\ 6}^{(j)},
\  
&\mathring{U}_{\ 8}^{(j)}=T_1\mathring{U}_{\ 7}^{(j)}:
&\text{sec. subgroup}.
\end{array}
\end{eqnarray*}

Due to the redefinition (\ref{redefinition}), 
relations (\ref{r1a}) and (\ref{r2a}) give
\begin{eqnarray}
\left.
\begin{array}{l}
\mathring{\alpha}_{0}\mathring{a}_{1}+
\mathring{\beta}_{0}\mathring{a}_{0}=0,
\vspace{.2cm} \\
\mathring{\alpha}_{n}\mathring{a}_{n+1}+
\mathring{\beta}_{n}\mathring{a}_{n}+
\mathring{\gamma}_{n}\mathring{a}_{n-1}=0,\ [n\geq1]
\end{array}\right\} \Rightarrow
\mathring{\beta}_{0}=\frac{\mathring{\alpha}_{0}\mathring{\gamma}_{1}}
{\mathring{\beta}_{1}-}\ \frac{\mathring{\alpha}_{1}
\mathring{\gamma}_{2}}
{\mathring{\beta}_{2}-}\ \frac{\mathring{\alpha}_{2}\mathring{\gamma}_{3}}
{\mathring{\beta}_{3}-}\cdots,
\label{R1a}
\end{eqnarray}
\begin{eqnarray}
\left.
\begin{array}{l}
\mathring{\alpha}_{0}\mathring{a}_{1}+\mathring{\beta}_{0}\mathring{a}_{0}=0,
 \vspace{.2cm} \\
\mathring{\alpha}_{1}\mathring{a}_{2}+ \mathring{\beta}_{1}\mathring{a}_{1}+\left[
\mathring{\gamma}_{1}-\mathring{\alpha}_{-1}
\right]\mathring{a}_{0}=0,
\vspace{.2cm} \\
\mathring{\alpha}_{n}\mathring{a}_{n+1}+\mathring{\beta}_{n}
\mathring{a}_{n}+\mathring{\gamma}_{n}
\mathring{a}_{n-1}=0,\ [n\geq2]
\end{array}\right\}\Rightarrow
\mathring{\beta}_{0}=
\frac{\mathring{\alpha}_{0}\left[\mathring{\gamma}_{1}-
\mathring{\alpha}_{-1}
\right]}
{\mathring{\beta}_{1}-}
 \frac{\mathring{\alpha}_{1}\mathring{\gamma}_{2}}
{\mathring{\beta}_{2}-} 
\cdots, 
\label{R2a}
\end{eqnarray}
where the coefficients $\mathring{\alpha}_{n}$ and 
$\mathring{\gamma}_{n}$ have also been redefined.
For example, the redefinition (\ref{redefinition})
leads to the redefinitions $\mathring{\alpha}_n^1\mapsto-i(n+B_2/2)
\mathring{\alpha}_n^1$ and 
$\mathring{\gamma}_n^1\mapsto i\mathring{\gamma}_n^1/(n-1+B_2/2)$.
This explains the difference between relations (\ref{r2a}) and 
(\ref{R2a}) when $\eta=0$ and $B_2=1$.

Now we write first and the fifth sets of solutions. The first set
reads
\letra
\begin{eqnarray}\label{Uj}
\mathring{U}_{\ 1}^{(j)}(z)=z^{\frac{1}{2}-\frac{B_2}{2}}
\sum_{n=0}^{\infty}\mathring{a}_{n}^{1}
Z_{n+\frac{B_2}{2}-\frac{1}{2}}^{(j)}(\omega z),
\end{eqnarray}
%
%
\begin{equation}
\begin{array}{l}
\mathring{\alpha}_{n}^{1}=
\frac{\omega z_0\left[ n+1\right]
\left[n-\frac{B_1}{z_0}\right]}
{\left(2n+B_2+1\right)},\quad
\mathring{\beta}_{n}^{1}=-n\left[n+B_2-1\right] - B_3,\quad
%
\mathring{\gamma}_{n}^{1}=\frac{\omega z_0
\left[n+B_2-2\right]
\left[ n+B_2-1+\frac{B_1}{z_0}\right]}
{\left(2n+B_2-3\right)}.
\end{array}
\end{equation}
For the fifth set, we find
\antiletra\letra
\begin{eqnarray}
\mathring{U}_{\ 5}^{(j)}(z)=(z-z_0)^{\frac{1}{2}-\frac{B_2}{2}}
\sum_{n=0}^{\infty}\mathring{a}_{ n}^{5}
Z_{n+\frac{B_2}{2}-\frac{1}{2}}^{(j)}\left[\omega (z-z_0)\right],
\end{eqnarray}
%
%
\begin{equation}
\begin{array}{l}
\mathring{\alpha}_{ n}^{5}=\frac{-\omega z_0\left[ n+1\right]\left[n+B_2+\frac{B_1}{z_0}\right]}
{\left(2n+B_2+1\right)},\qquad
\mathring{\beta}_{ n}^{5}=\mathring{\beta}_{n}^{1},\qquad
\mathring{\gamma}_{n}^{5}=\frac{-\omega z_0\left[n+B_2-2\right]\left[ n-1-\frac{B_1}{z_0}\right]}
{\left(2n+B_2-3\right)}.
\end{array}
\end{equation}

In both sets the independence of the Bessel functions requires that
\antiletra\letra
\begin{eqnarray}\label{particular-1}
\begin{array}{l}
B_2\neq 0,-1,-2,\cdots \text{ if the series begin at $n=0$ in } 
\mathring{U}_{\ 1}^{(j)} \text {and } \mathring{U}_{\ 5}^{(j)},
\end{array}
\end{eqnarray}
while the recurrence relations are given by 
 \begin{eqnarray}\label{particular-2}
 \text{Eqs. (\ref{R1a})} \text{ if } B_2\neq 1, \qquad
 \text{Eqs. (\ref{R2a})} \text{ if } B_2=1
 \text { for } \mathring{U}_{\ 1}^{(j)} \text {and } \mathring{U}_{\ 5}^{(j)}.
  \end{eqnarray}
On the other side, for the full group of solutions (Appendix C), 
we find that
\antiletra
\begin{eqnarray}\label{geral-1}
2\nu_i+2\neq 0,-1,-2,\cdots, \quad\left[\text{independence of
$Z_{\ \kappa}^{(j)}$ in } \mathring{U}_{\ i}^{(j)} \text {and } \mathring{U}_{i+4}^{(j)}\right]
\end{eqnarray}
for series beginning at $n=0$, while the recurrence relations are given by
 \begin{eqnarray}\label{geral-2}
 \text{Eqs. (\ref{R1a})} \text{ if } 2\nu_i+2\neq 1, \qquad
 \text{Eqs. (\ref{R2a})} \text{ if } 2\nu_i+2=1,
  \end{eqnarray}
the parameters $\nu_i$ being defined in (\ref{combination}). 
Relations (\ref{particular-1}) and (\ref{particular-2})
stand for $\nu_i=\nu_1$ in (\ref{geral-1}) and (\ref{geral-2}).

For the present case ($\eta=0$), finite and infinite
series are not possible in solutions belonging to the same set, unlike 
the case $\eta\neq 0$. Further,   
finite-series solutions are of the type given in (\ref{second-type}) 
which may be unbounded at
some of the singular points. This holds also for solutions of the spheroidal equation.%


In fact, in the literature we have 
found no finite-series solutions for the spheroidal 
equation.  The preceding solutions give 
four sets of solutions which do not 
admit finite series, and four sets 
which admit finite series,
namely,
\begin{eqnarray}\label{esferoidal-finitas}
\begin{array}{l}
\text{only  infinite  series: }\qquad  \mathring{S}_{\ 1}^{(j)}, \  
\mathring{S}_{\ 3}^{(j)}, \
\mathring{S}_{\ 5}^{(j)}, \ 
\mathring{S}_{\ 7}^{(j)}; \vspace{2mm}\\
\text{finite or infinite series: } \ \mathring{S}_{\ 2}^{(j)}, \  
\mathring{S}_{\ 4}^{(j)}, \
\mathring{S}_{\ 6}^{(j)}, \ 
\mathring{S}_{\ 8}^{(j)}. 
\end{array}
\end{eqnarray} 
As examples we write the solutions $\mathring{S}_{\ 1}^{(j)}$ 
and $\mathring{S}_{\ 2}^{(j)}$. Thus, by setting $\omega=-2\gamma$ in 
Eqs. (\ref{esferoidal-2}) 
the solutions (\ref{esferoidal-1}) corresponding to
$\mathring{U}_{\ 1}^{(j)}$ read
\letra
\begin{eqnarray}\label{Meixner-1-1}
\begin{array}{l}
\displaystyle
\mathring{S}_{\ 1}^{(j)}(\mu,y)= \left[ \frac{y+1}{y-1}\right]^{\frac{\mu}{2}}
\sum_{n=0}^{\infty}\mathring{a}_{n}^{1}
\psi_{n+\mu}^{(j)}[ \gamma (y-1)],\qquad [\mu\neq-1,-{3}/{2},-2,\cdots]
\end{array}
\end{eqnarray}
where the recurrence relations for $\mathring{a}_{n}^{1}$
are given by
 \begin{eqnarray}
 \text{Eqs. (\ref{R1a})} \text{ if } \mu\neq -1/2, \qquad
 \text{Eqs. (\ref{R2a})} \text{ if } \mu=-1/2
\end{eqnarray}
with
\begin{equation}
\begin{array}{l}
\mathring{\alpha}_{n}^{1}=
\frac{ 2 \gamma (n+1) (n+\mu+1)}
{\left(2n+2\mu+3\right)},\qquad 
\mathring{\beta}_{n}^{1} = (n+\mu)(n+\mu+1)- \lambda ,\qquad
\mathring{\gamma}_{n}^{1}  = \frac{2 \gamma (n+2\mu) (n+\mu)}
{(2n+2\mu-1)}.
\end{array}
\end{equation}
Due to the restrictions $\mu\neq-1,-{3}/{2},-2,\cdots$
which assure the independence of each term of the series, 
it is impossible to have $\mathring{\gamma}_{n}^{1}=0$ and, 
consequently, these solutions do not admit finite series. Similarly,
the other solutions which do not permit finite series
are expansions in series of $\psi_{n\pm\mu}^{(j)}$
and require restrictions on $\mu$.

The solutions admitting finite series
are expansions in series of $\psi_{\ n}^{(j)}$ which
are all independent, regardless the values of $\mu$.
In particular, from $\mathring{U}_{\ 2}^{(j)}$ given in Appendix C,  
we find
\antiletra\letra
\begin{eqnarray} 
\begin{array}{l}
\displaystyle
\mathring{S}_{\ 2}^{(j)}(\mu,y)= \left[ \frac{y+1}{y-1}\right]^{\frac{\mu}{2}}
\sum_{n=0}^{\infty}\mathring{a}_{n}^{2}\ 
\psi_{n}^{(j)}[ \gamma (y-1)],
\end{array}
\end{eqnarray}
where the recurrence relations for $\mathring{a}_{n}^{2}$
are given by Eqs. (\ref{R1a}) if the series begin 
at $n=0$ and by Eqs. (\ref{N>1}) if the series begin 
at $n=N\geq1$. The coefficients 
for $\mathring{a}_{n}^{2}$
are
\begin{equation}
\begin{array}{l}
\mathring{\alpha}_{n}^{2}=
\frac{  2\gamma (n+1) (n+1-\mu)}
{\left(2n+3\right)},\qquad 
\mathring{\beta}_{n}^{2} = n(n+1)-\lambda ,\qquad
%
\mathring{\gamma}_{n}^{2}  = \frac{2 \gamma n (n+\mu)}
{(2n-1)}.
\end{array}
\end{equation}
So, if $\mu=-\ell-1=-1,-2.-3,\cdots$,
then $\mathring{\gamma}_{\ell+1}^{2}=0$ 
implies finite series with $0\leq n\leq \ell$.

\subsection{ Solutions for the double-confluent Heun equation}

Only the solutions $\mathring{\mathbb{U}}_1$, $\mathring{\mathbb{U}}_3$, 
$\mathring{\mathbb{U}}_5$ and  $\mathring{\mathbb{U}}_7$ of the CHE admit 
the limit $z_0\to 0$. Besides this,
the limits of $\mathring{\mathbb{U}}_1$ and 
$\mathring{\mathbb{U}}_3$ are equivalent to the 
limits of $\mathring{\mathbb{U}}_5$ and 
$\mathring{\mathbb{U}}_7$.
Thus, we obtain a subgroup of two sets of one-sided 
solutions for the DCHE. A second subgroup 
is generated by using the transformation $t_2$. In fact, denoting
by $\mathring{\mathbb{U}}_1$ the first set for the DCHE  
and using the transformations (\ref{transformacao-dche}), 
the full group is given as in Eqs. (\ref{hipergeometricas}),
that is,
\antiletra
\begin{eqnarray}\label{hipergeometricas-DCHE}
\mathring{\mathbb{U}}_1(z),\quad
\mathring{\mathbb{U}}_2(z)=t_1\mathring{\mathbb{U}}_1(z);\quad
\mathring{\mathbb{U}}_3(z)=t_2\mathring{\mathbb{U}}_1(z),\quad
\mathring{\mathbb{U}}_4(z)=t_2\mathring{\mathbb{U}}_2(z)
=t_3\mathring{\mathbb{U}}_3(z).
\end{eqnarray}
Conditions for connecting solutions of a given set may
be found as in the case of the CHE. 
In the following we write the four sets of solutions 
and discuss the possibility of finite-series solutions (Heun polynomials).

%
The limit $z_0\to 0$ modifies the expressions for 
coefficients (\ref{coef-3'}) but leaves the solutions (\ref{par1})
unaltered. So, we have 
\begin{eqnarray}\label{forma-de-leaver-truncada-DCHE}
\mathring{\mathbb{U}}_1(z) =z^{-\frac{B_2}{2}}\displaystyle\sum_{n=0}^{\infty} \mathring{b}_{n}^{1}
\mathscr{U}_{n-1+\frac{B_2}{2}}\left(\eta,\omega{z}\right)
\  \Leftrightarrow\   
\text{ Eq. (\ref{par1})}\qquad [B_2\neq 0,-1.-2.\cdots]
\end{eqnarray}
or, explicitly,
\begin{eqnarray}
\begin{array}{l}
\mathring{U}_{ 1}(z) =e^{i\omega z}\displaystyle\sum_{n=0}^{\infty}
\frac{\mathring{b}_{n}^{1}\ [2i\omega z]^{n}}{\Gamma(2n+B_2)}
\Phi\left(n+i\eta+\frac{B_{2}}{2},2n+B_{2};-2i\omega z\right),\vspace{2mm}\\
\mathring{U}_{1}^{\pm}(z) =e^{\pm i\omega z}\displaystyle\sum_{n=0}^{\infty}
\frac{\mathring{b}_{n}^{1}\ [-2i\omega z]^{n}}{\Gamma[n\mp i\eta+(B_2/2)]}
\Psi\left(n\pm i\eta+\frac{B_{2}}{2},2n+B_{2};\mp 2i\omega z\right), 
\end{array} 
%
\end{eqnarray}
where in the recurrence relations for $\mathring{b}_n^{1}$ 
\begin{eqnarray}\label{coef-1-DHCE}
\begin{array}{l}
\mathring{\alpha}_{n}^{1}=
\frac{2i\omega B_1\left( n+1\right)}
{\left(2n+B_2\right)\left(2n+B_2+1\right)},\vspace{2mm}\\
\mathring{\beta}_{n}^{1}=n\left(n+B_2-1\right) + B_3+
\frac{2\omega \eta B_1\left(B_2-2\right)}
{\left(2n+B_2-2\right)\left(2n+B_2\right)},\vspace{2mm}\\
%
\mathring{\gamma}_{n}^{1}=\frac{2i\omega B_1
\left[n+B_2-2\right]\left[ n-1+i\eta+\frac{B_2}{2}\right]
\left[ n-1-i\eta +\frac{B_2}{2}\right]}
{\left(2n+B_2-3\right)\left(2n+B_2-2\right)}.
\end{array}
\end{eqnarray}
Thence, the recurrence relations 
are again given by (\ref{s-che-1}), that is,
\begin{eqnarray}
\begin{array}{ll}
\eta\neq 0:&\text{Eqs. (\ref{r1a})} \text{ if } B_2\neq 1,2; \quad
\text{Eqs. (\ref{r2a})} \text{ if } B_2=1; \quad
\text{Eqs. (\ref{r3a})} \text{ if } B_2=2.\vspace{2mm}\\
\eta=0 :&\text{Eqs. (\ref{r1a})} \text{ if } B_2\neq 1; \quad
\text{Eqs. (\ref{r2a})} \text{ if } B_2=1.
\end{array}
\end{eqnarray}
The second set of solutions, corresponding to the 
solutions $\mathring{\mathbb{U}}_3$
of the CHE, is
\begin{eqnarray*}
\mathring{\mathbb{U}}_2(z)=t_1 \mathring{\mathbb{U}}_1(z)=
e^{\frac{B_1}{z}} z^{-\frac{B_2}{2}}
\displaystyle\sum_{n=0}^{\infty} \mathring{b}_{n}^{2}
\mathscr{U}_{n+1-\frac{B_2}{2}}\left(\eta,\omega{z}\right),
\qquad[B_2\neq 4,5,6,\cdots]
\end{eqnarray*}
which, explicitly, reads
\begin{eqnarray} 
\begin{array}{l}
\mathring{U}_{2}= 
e^{i\omega z+\frac{B_1}{z}}z^{2-B_2} \displaystyle
\sum_{n=0}^{\infty}\begin{array}{l}
\frac{\mathring{b}_{n}^{2}\ [2i\omega z]^{n}}{\Gamma[2n+4-B_2]}
\Phi\left[n+2+i\eta-\frac{B_2}{2},2n+4-B_2;-2i\omega z\right]\end{array},\vspace{2mm}\\
\mathring{U}_{2}^{\pm} =e^{\pm i\omega z+\frac{B_1}{z}} 
z^{2-B_2}\displaystyle
\sum_{n=0}^{\infty} \begin{array}{l}
\frac{\mathring{b}_{ n}^{2}\ 
[-2i\omega z]^n}{\Gamma\left[n+2\mp i\eta-\frac{B_2}{2}\right]}
\Psi\left[n+2\pm i\eta-\frac{B_2}{2},2n+4-B_2;\mp 2i\omega z\right]
\end{array}, 
\end{array} 
\end{eqnarray}
with
\begin{eqnarray}
\begin{array}{l}
\mathring{\alpha}_{n}^{2} =  \frac{2i\omega B_1(n+1)}
{\left(2n+4-B_{2}\right)\left(2n+5-
B_{2}\right)},
\vspace{.2cm} \\
%
\mathring{\beta}_{ n}^{2}  =  -(n+1)(n+2-B_{2})-B_{3}
-\frac{2\eta \omega B_1\left(B_{2}-2\right)}
{\left(2n+2-B_{2}\right)\left(2n+4-
B_{2}\right)},
\vspace{.3cm} \\
%
\mathring{\gamma}_{n}^{2}  = \frac{2i\omega B_1\left[n+2-B_{2}\right]
\left[n+1+i\eta-\frac{B_{2}}{2}\right]\left[n+1-i\eta -\frac{B_{2}}{2}\right]}
{\left(2n+1-B_{2}\right)\left(2n+2-B_{2}\right)}
\end{array}
\end{eqnarray}
in the following recurrence relations
\begin{eqnarray}\begin{array}{l}
\eta\neq0: \ \text{Eqs. (\ref{r1a})} \text{ if } B_2 \neq  2,3; \qquad
 \text{(\ref{r2a})} \text{ if } B_2=3; \qquad
 \text{(\ref{r3a})} \text{ if } B_2=2;\vspace{2mm}\\
 \eta=0: \ \text{Eqs. (\ref{r1a})} \text{ if } B_2 \neq 3, \qquad
  \text{(\ref{r2a})} \text{ if } B_2=3.
\end{array}
\end{eqnarray}

If $\eta=0$, $\mathring{\mathbb{U}}_1$ and  
$\mathring{\mathbb{U}}_2$ do not admit finite-series expansions
($\mathring{\gamma}_n\neq0$) and can be rewritten in 
terms of Bessel functions.  
Finite series are possible if $\eta \neq 0$  
because, by setting $\nu_1=(B_2/2)-1$ and $\nu_2=1-(B_2/2)$,
%
%
we can interpret the
solutions as in the two items following relations 
(\ref{serieinfinita-1}). 

For the second subgroup ($\mathring{\mathbb{U}}_3$ and  
$\mathring{\mathbb{U}}_4$) 
the transformation $t_2$ changes the form of the recurrence 
relations since $i\eta$ is transformed into $(B_2/2)-1$. 
Precisely, the relations (\ref{r1a}), (\ref{r2a}) and (\ref{r3a})
are replaced by
\letra
\begin{eqnarray}
\begin{array}{l}
\mathring{\alpha}_{0}\mathring{b}_{1}+
\mathring{\beta}_{0}\mathring{b}_{0}=0,\qquad
\mathring{\alpha}_{n}\mathring{b}_{n+1}+
\mathring{\beta}_{n}\mathring{b}_{n}+
\mathring{\gamma}_{n}\mathring{b}_{n-1}=0,\quad [n\geq1]\ \Leftrightarrow
\ \text{(\ref{r1a})}.
\end{array}
\label{r1A}
\end{eqnarray}
\begin{eqnarray}
\begin{array}{l}
\mathring{\alpha}_{0}\mathring{b}_{1}+\mathring{\beta}_{0}\mathring{b}_{0}=0,
 \vspace{.2cm} \\
\mathring{\alpha}_{1}\mathring{b}_{2}+
\mathring{\beta}_{1}\mathring{b}_{1}+\left\{
\mathring{\gamma}_{1}+\left[\left(\frac{B_2}{2}-1\right)^2-
\frac{1}{4}\right]\mathring{\alpha}_{-1}
\right\}\mathring{b}_{0}=0,
\vspace{.2cm} \\
\mathring{\alpha}_{n}\mathring{b}_{n+1}+\mathring{\beta}_{n}
\mathring{b}_{n}+\mathring{\gamma}_{n}
\mathring{b}_{n-1}=0,\quad [n\geq2].
\end{array}
\label{r2A}
\end{eqnarray}
\begin{eqnarray}
\begin{array}{l}
\mathring{\alpha}_{0}\mathring{b}_{1}+
\left[\mathring{\beta}_{0}-
\left(\frac{B_2}{2}-1\right)\mathring{\alpha}_{-1}
\right]\mathring{b}_{0}=0,
\vspace{.2cm} \\
\mathring{\alpha}_{n}\mathring{b}_{n+1}+
\mathring{\beta}_{n}\mathring{b}_{n}
+\mathring{\gamma}_{n}\mathring{b}_{n-1}=0,\quad [n\geq1].
\end{array}
\label{r3A}
\end{eqnarray}
If $B_2=2$, relations (\ref{r3A}) reduce to (\ref{r1A}). 

Up to a constant factor, the third set is
\begin{eqnarray*}
\mathring{\mathbb{U}}_3(z) =
e^{i\omega z+ \frac{B_1}{2z}} z^{1-\frac{B_2}{2}}\displaystyle\sum_{n=0}^{\infty} \mathring{b}_{n}^{3}
\mathscr{U}_{n+i\eta}\left[i\left(1-\frac{B_2}{2}\right),
\frac{iB_1}{2z}\right],\qquad [2i\eta\neq -2,-3,-4,.\cdots]
\end{eqnarray*}
where the conditions $2i\eta\neq 0,-1.-2.\cdots$ 
assure the independence of each term if
the series begin at $n=0$. Explicitly,
\antiletra
\begin{eqnarray}
\begin{array}{l}
\mathring{U}_{3}(z)  =e^{i\omega z} \displaystyle
\sum_{n=0}^{\infty}\begin{array}{l}
\frac{\mathring{b}_{n}^{3}\left[-{B_1}/{z}\right]^{n+i\eta+\frac{B_2}{2}}
}{\Gamma[2n+2i\eta+2]}
\Phi\left[n+i\eta+\frac{B_{2}}{2},2n+2i\eta+2;
\frac{B_{1}}{z}\right],\end{array}
\vspace{0.2cm}\\
\mathring{U}_{3}^{\pm}(z) =e^{i\omega z+(1\mp 1)\frac{B_1}{2z}}
\displaystyle
\sum_{n=0}^{\infty}\begin{array}{l}
\frac{\mathring{b}_{n}^{3}\left[{B_1}/{z}\right]^{n+i\eta+\frac{B_2}{2}}}
{\Gamma\left[n+i\eta+1\mp\frac{B_2-2}{2}\right]}\times\end{array}
\vspace{0.2cm}\\
\hspace{4.5cm}
\begin{array}{l}\Psi\left[n+i\eta+1\pm\frac{B_2-2}{2},2n+2i\eta+2;
\pm\frac{B_{1}}{z}\right],\end{array}
\end{array}
\end{eqnarray}
which correspond to solutions (\ref{dche-3a}) with $\nu=i\eta$. 
The recurrence relations for $\mathring{b}_n^{3}$ 
are
\begin{eqnarray}
\begin{array}{ll}
B_2\neq 2:&\text{Eqs. (\ref{r1A})} \text{ if } 2i\eta\neq -1,0, \quad
\text{ (\ref{r2A})} \text{ if } 2i\eta=-1; \quad
\text{ (\ref{r3A})} \text{ if } 2i\eta=0;\vspace{2mm}\\
B_2=2 :&\text{Eqs. (\ref{r1A})} \text{ if } 2i\eta\neq -1; \quad
\text{ (\ref{r2A})} \text{ if } 2i\eta=-1
\end{array}
\end{eqnarray}
with
\begin{eqnarray}
\begin{array}{l}
\mathring{\alpha}_{n}^{3}  =\frac{2i\omega B_1(n+1)}
{(2n+2i\eta+2)(2n+2i\eta+3)}  ,
\qquad
\vspace{.2cm} \\
\mathring{\beta}_{n}^{3} =  B_{3}+\left(n+i\eta+1-\frac{B_{2}}{2}\right)
\left(n+i\eta+\frac{B_{2}}{2}\right)
+\frac{2\eta\omega B_1\left(B_2-2\right)}
{(2n+2i\eta)(2n+2i\eta+2)},
\vspace{2mm}\\
\mathring{\gamma}_{n}^{3}=\frac{2i\omega B_1\left[n+i\eta-1+\frac{B_2}{2}\right]
\left[n+i\eta+1-\frac{B_2}{2}\right][n+2i\eta]}{(2n+2i\eta-1)
(2n+2i\eta)}.
\end{array}
\end{eqnarray}
As in the case of two-sided solutions, the 
fourth set is obtained by replacing ($\eta,\omega$)
by ($-\eta,-\omega$) in ($\mathring{U}_3, \mathring{U}_3^{\mp}$) 
and respective recurrence relations, that is, 
%
\begin{eqnarray}
\begin{array}{l}
\left[\mathring{U}_4(z),\mathring{U}_4^{\pm}(z)\right]=
\left[t_3\mathring{U}_3(z),t_3\mathring{U}_3^{\pm}(z)\right],
\qquad [2i\eta\neq 2,3,4.\cdots].
\end{array}
\end{eqnarray}
These correspond to solutions (\ref{dche-4a}) with $\nu=-i\eta$.

If $B_2=2$, $\mathring{\mathbb{U}}_3$ and  
$\mathring{\mathbb{U}}_4$ do not admit 
finite-series expansions and can be expressed in
terms of Bessel functions. Finite series are possible 
when $B_2 \neq 2$ since (for $\ell$ equals to zero or a natural
number) 
\begin{itemize}
\itemsep-3pt
\item If $i\eta+(B_2/2)=-\ell$,  
$\mathring{U}_{3}$ and $\mathring{U}_{3}^{+}$
are solutions linearly dependent and lead to 
finite series with $0\leq n\leq \ell$,  
whereas  the series in $\mathring{U}_{3}^{-}$ begins at $n=\ell+1$.
If $2+i\eta-(B_2/2)=-\ell$, $\mathring{U}_{3}$ and $\mathring{U}_{3}^{-}$
are multiple of each other and give 
finite series with $0\leq n\leq \ell$,  
whereas  the series in $\mathring{U}_{3}^{+}$ begins at $n=\ell+1$.
    \item If $-i\eta+(B_2/2)=-\ell$,
$\mathring{U}_{4}$  and $\mathring{U}_{4}^{+}$ 
give finite series with $0\leq n \leq \ell$, whereas the 
series in $\mathring{U}_{4}^{-}$ begins at $n=\ell+1$.
If $2-i\eta-(B_2/2)=-\ell$, $\mathring{U}_{4}$ and $\mathring{U}_{4}^{-}$
are multiple of each other and give 
finite series with $0\leq n\leq \ell$,  
whereas  the series in $\mathring{U}_{4}^{+}$ begins at $n=\ell+1$.
\end{itemize}
By Eq. (\ref{laguerre-1}), the finite series  
may be written in terms of generalized Laguerre
polynomials.

The convergence of 
the one-sided infinite-series 
solutions follows from the convergence of two-sided solutions
(see section 2.C):
\begin{itemize}
\itemsep-3pt
\item $\mathring{U}_1^{\pm}$
and $\mathring{U}_2^{\pm}$ converge absolutely for $|z|>0$
but may be unbounded at $z=\infty$; $\mathring{U}_3^{\pm}$
and $\mathring{U}_4^{\pm}$ converge for $|z|<\infty$
but may be unbounded at $z=0$. 
\item By the ratio test, the series which appear in $\mathring{U}_i$ 
converge for any $z$. Nevertheless, it is necessary to
compute the limits of $\mathring{U}_1$ as
$z\to 0$ and $z\to\infty$ in order to get the
behaviour of these solutions at the singular
points $z=0$ and $z=\infty$.
\end{itemize}

Solutions for the Whitaker-Hill and Mathieu equations,
regarded as DCHEs, may be obtained by using (\ref{whe2}) 
and (\ref{mathieu3}). On the other side,
from the expressions for $\mathring{\beta}_{n}^{3}$
and $\mathring{\beta}_{n}^{4}$, we see that  
$\mathring{\mathbb{U}}_{3}$ and $\mathring{\mathbb{U}}_{4}$ 
do not admit the Whittaker-Ince limit (\ref{ince}). Thus, 
in this limit we will find only the two sets written in Eqs. 
(\ref{primeiro}) and (\ref{segundo}).

\subsection{ Solutions for the Whittaker-Ince limits of the CHE}

The Raabe test has been used in 
a previous paper \cite{lea-2} in the study of one-sided series
solutions for the Whittaker-Ince limit of the CHE. Such solutions
are exactly the ones obtained by applying the Whittaker-Ince
limit (\ref{ince}) to the preceding one-sided solutions for the
CHE. Thus, here we restrict ourselves to the form
of the recurrence relations, write the first set of
solutions and indicate how to obtain the other sets.
We show as well how the domains
of convergence of some solutions for the Mathieu
equation are affected by the Raabe test.

For series beginning at $n=0$, instead of the forms 
(\ref{r1a}-\ref{r3a}) for the recurrence relations we find,
respectively,
\begin{eqnarray}
\left.
\begin{array}{l}
\mathring{\alpha}_{0}\mathring{c}_{1}+
\mathring{\beta}_{0}\mathring{c}_{0}=0,
\vspace{.2cm} \\
\mathring{\alpha}_{n}\mathring{c}_{n+1}+
\mathring{\beta}_{n}\mathring{c}_{n}+
\mathring{\gamma}_{n}\mathring{c}_{n-1}=0,\ [n\geq1]
\end{array}\right\} \Rightarrow
\mathring{\beta}_{0}=\frac{\mathring{\alpha}_{0}\mathring{\gamma}_{1}}
{\mathring{\beta}_{1}-}\ \frac{\mathring{\alpha}_{1}
\mathring{\gamma}_{2}}
{\mathring{\beta}_{2}-}\ \frac{\mathring{\alpha}_{2}\mathring{\gamma}_{3}}
{\mathring{\beta}_{3}-}\cdots,
\label{rIa}
\end{eqnarray}
\begin{eqnarray}
\left.
\begin{array}{l}
\mathring{\alpha}_{0}\mathring{c}_{1}+\mathring{\beta}_{0}\mathring{c}_{0}=0,
 \vspace{.2cm} \\
\mathring{\alpha}_{1}\mathring{c}_{2}+
\mathring{\beta}_{1}\mathring{c}_{1}+\left[
\mathring{\gamma}_{1}+\mathring{\alpha}_{-1}
\right]\mathring{c}_{0}=0,
\vspace{.2cm} \\
\mathring{\alpha}_{n}\mathring{c}_{n+1}+\mathring{\beta}_{n}
\mathring{c}_{n}+\mathring{\gamma}_{n}
\mathring{c}_{n-1}=0,\ [n\geq2]
\end{array}\right\}\Rightarrow
%
\mathring{\beta}_{0}=
\frac{\mathring{\alpha}_{0}\left[\mathring{\gamma}_{1}+\mathring{\alpha}_{-1}
\right]}
{\mathring{\beta}_{1}-}
 \frac{\mathring{\alpha}_{1}\mathring{\gamma}_{2}}
{\mathring{\beta}_{2}-}  \frac{\mathring{\alpha}_{2}
\mathring{\gamma}_{3}}
{\mathring{\beta}_{3}-}
\cdots, 
\label{rIIa}
\end{eqnarray}
\begin{eqnarray}
\left.
\begin{array}{l}
\mathring{\alpha}_{0}\mathring{c}_{1}+
\left[\mathring{\beta}_{0}+\mathring{\alpha}_{-1}
\right]\mathring{c}_{0}=0,
\vspace{.2cm} \\
\mathring{\alpha}_{n}\mathring{c}_{n+1}+
\mathring{\beta}_{n}\mathring{c}_{n}
+\mathring{\gamma}_{n}\mathring{c}_{n-1}=0,\ [n\geq1]
\end{array}\right\}\Rightarrow
\mathring{\beta}_{0}+\mathring{\alpha}_{-1}=
\frac{\mathring{\alpha}_{0}\mathring{\gamma}_{1}}
{\mathring{\beta}_{1}-}\ \frac{\mathring{\alpha}_{1}
\mathring{\gamma}_{2}}{\mathring{\beta}_{2}-}
\ \frac{\mathring{\alpha}_{2}\mathring{\gamma}_{3}}
{\mathring{\beta}_{3}-}\cdots ,
\label{rIIIa}
\end{eqnarray}
Once more, $n\geq -1$ for $\mathring{\alpha}_n$; 
$n\geq 0$ for $\mathring{\beta}_n$;  
and $n\geq 1$ for $\mathring{\gamma}_n$. If the series begin
at $n=N\geq 1$ (that is, if $\mathring{\alpha}_{\text{\tiny N}-1}=0$), 
the recurrence relations have the form (\ref{N>1}).

Observe that the relations (\ref{rIa}-\ref{rIIIa})
are not obtained by applying directly the 
Whittaker-Ince limit on (\ref{r1a}-\ref{r3a}). For example, before
taking the limits of the solution
$\mathring{U}_1$ given in (\ref{par1}), we redefine 
the series coefficients as -- see Eqs. (\ref{truque-1}) and (\ref{truque-2}) --
\[\mathring{b}_n^1=[-i\eta]^n\mathring{c}_n^1\qquad\Rightarrow\quad
\mathring{\alpha}_n^1\mapsto\bar{\alpha}_n^1=[-i\eta]\mathring{\alpha}_n^1,
\qquad
\mathring{\gamma}_n^1\mapsto\bar{\gamma}_n^1=\mathring{\gamma}_n^1/[-i\eta].
\]
Then, we insert these expressions into (\ref{r1a}-\ref{r3a})
and perform the limits.

The first set, corresponding to solutions (\ref{par1}), 
is \cite{lea-2}
\letra
\begin{eqnarray}\label{WIL-1a}
\mathring{U}_{\ 1}^{(j)}(z) =z^{(1-B_{2})/2}\displaystyle\sum_{n=0}^{\infty}(-1)^n
\mathring{c}_{n}^{1}
Z_{2n+B_2-1}^{(j)}\left(2\sqrt{qz}\right),\qquad [B_{2}\neq 0,-1,-2,\cdots]
\end{eqnarray}
where recurrence relations for $ \mathring{c}_{n}^{1}$,
if the series begin at $n=0$, are given by
\begin{eqnarray}\label{bessel1d}
 \text{ Eqs. (\ref{rIa}) if }
B_{2}\neq 1,2; \quad \text{   Eqs. (\ref{rIIa}) if } B_{2}=1;\qquad
\text{   Eqs. (\ref{rIIa}) if }B_{2}=2,
\end{eqnarray}
with
\begin{eqnarray}\label{WIL-1b}
&\begin{array}{l}
\mathring{\alpha}_{n}^{1} = \frac{q z_0(n+1)\left[n-\frac{B_1}{z_0}\right]}
{(2n+B_2)(2n+B_2+1)},\qquad 
%
\mathring{\beta}_{n}^{1} =n(n+B_2-1)- \frac{q z_0}{2}+B_3-\frac{q z_0 (B_2-2)\left[B_2+\frac{2B_{1}}{z_{0}}\right]}
{2(2n+B_2-2)(2n+B_2)},\end{array}\nonumber\\
%
&\begin{array}{l}
\mathring{\gamma}_{n}^{1}= \frac{ q z_0(n+B_2-2)\left[n+B_2+\frac{B_1}{z_0}-1\right]}
{(2n+B_2-3)(2n+B_2-2)}.
\end{array}
\end{eqnarray}
From $\mathring{U}_{\ 1}^{(j)}(z)$, the two subgroups are obtained 
by using the transformations (\ref{Transformacao2}) as 
\antiletra
\begin{eqnarray}
\begin{array}{lllll}
\text{First}:&\mathring{U}_{\ 1}^{(j)},\quad& 
\mathring{U}_{\ 2 }^{(j)}=\mathscr{T}_1\mathring{U}_{\ 1 }^{(j)},\quad&
\mathring{U}_{\ 3 }^{(j)}=\mathscr{T}_2\mathring{U}_{\ 3 }^{(j)},\quad&
\mathring{U}_{\ 4 }^{(j)}=\mathscr{T}_1\mathring{U}_{\ 3 }^{(j)};\vspace{2mm}\\
\text{Second}:&\mathring{U}_{\ 5}^{(j)}=\mathscr{T}_{3}\mathring{U}_{\ 1}^{(j)},\quad& 
\mathring{U}_{\ 6 }^{(j)}=\mathscr{T}_1\mathring{U}_{\ 5 }^{(j)},\quad&
\mathring{U}_{\ 7 }^{(j)}=\mathscr{T}_2\mathring{U}_{\ 6 }^{(j)},\quad&
\mathring{U}_{\ 8 }^{(j)}=\mathscr{T}_1\mathring{U}_{\ 7 }^{(j)}.
\end{array}
\end{eqnarray}
The convergence of the one-sided infinite series
now results from the 
convergence (\ref{convergencia-WI-limit})
for the two-sided series, the point
$z=\infty$ requiring special attention. Explicitly: (i) the solutions 
$\mathring{U}_{\ i}^{(1)}$ in series of Bessel functions of first kind 
converge for finite values of $z$; (ii) for $j=2,3,4$
the solutions $\mathring{U}_{\ i}^{(j)}$ converge
for $|z|>|z_0|$ in the first subgroup and for $|z-z_0|>|z_0|$ in 
the second subgroup but, according to the Raabe test, they
may converge also at $|z|=|z_0|$ and $|z-z_0|=|z_0|$ under
conditions similar to (\ref{convergencia1-nu=0}), namely,
\begin{equation} \label{convergencia1-W}
|z|\geq |z_0| \text{ if }
\begin{cases}\text{Re}\left[  B_2+\frac{B_1}{z_0}\right] <1\text{ in }
\mathring{U}_{1,2}^{(j)}, \vspace{2mm} \\
\text{Re}\left[  B_2+\frac{B_1}{z_0}\right] >1\text{ in }
\mathring{U}_{3,4}^{(j)};
\end{cases}
\ 
|z-z_0|\geq|z_0| \text{ if }
\begin{cases}\text{Re}\left[ \frac{B_1}{z_0}\right] >-1\text{ in }
\mathring{U}_{5,8}^{(j)},\vspace{2mm} \\
\text{Re}\left[ \frac{B_1}{z_0}\right] <-1\text{ in }
\mathring{U}_{6,7}^{(j)},
\end{cases}
\end{equation}
%
%
%
%
where $\mathring{U}_{m,n}^{(j)}$ means: $\mathring{U}_{\ m}^{(j)}$
and $\mathring{U}_{\ n}^{(j)}$. 
The behavior of the solutions
at $z=\infty$ must be determined by using the
limits (\ref{comportamento-infinito}). In virtue of
the  multiplicative factors it is also necessary
to examine the behavior of the solutions at
the other singular points lying inside the
regions afforded by the ratio test.

Now we consider 
the convergence of the solutions $\mathring{w}_{\ i}^{(j)}(u)$
for the Mathieu equation. According to (\ref{mathieu-as-wil}),
$\mathring{w}_{\ i}^{(j)}(u)=\mathring{U}_{\ i}^{(j)}(z)$
where $\mathring{U}_{\ i}^{(j)}(z)$ are the solutions
of the Whittaker-Ince limit of the CHE with $z_0=1$, $B_1=-1/2$, $B_2=1$ 
and $z=\cos^{2}(\sigma u)$. Thence, the expansions 
of $\mathring{w}_{\ i}^{(1)}(u)$ in series Bessel functions
of the first kind converge for all $u$ but may be unbounded
at the singular points, as state above. In comparison with
the usual literature, the modifications 
implied by the Raabe test consist in the equal sign in
the following domains of convergence ($j=2,3,4$):
\letra
\begin{equation} \label{181}
|\cos(\sigma u)|\geq 1 \text{ for }
\mathring{w}_{\ 1}^{(j)} \text{ and }
\mathring{w}_{\ 2}^{(j)},\qquad|\sin(\sigma u)|\geq 1 \text{ for }
\mathring{w}_{\ 5}^{(j)} \text{ and }
\mathring{w}_{\ 8}^{(j)}.
\end{equation}
In effect, by taking $\sigma=i$ and writing explicitly the 
solutions, we find that $\mathring{w}_{\ 1}^{(j)}$,
$\mathring{w}_{\ 2}^{(j)}$, $\mathring{w}_{\ 5}^{(j)}$ and 
$\mathring{w}_{\ 8}^{(j)}$ are equivalent, 
respectively, to the solutions 28.23.6, 28.23.8, 28.23.7 
and 28.23.11 of Ref. \cite{nist}. 
For the other solutions, the conditions given in (\ref{convergencia1-W}) 
are not satisfied and, then,  
\begin{equation} 
|\cos(\sigma u)|> 1 \text{ for }
\mathring{w}_{\ 3}^{(j)} \text{ and }
\mathring{w}_{\ 4}^{(j)},\qquad|\sin(\sigma u)|> 1 \text{ for }
\mathring{w}_{\ 6}^{(j)} \text{ and }
\mathring{w}_{\ 7}^{(j)},
\end{equation}
as usual.

\subsection{ Solutions for the Whittaker-Ince limit of the DCHE}

The Whittaker-Ince limit of the one-sided solutions for the DCHE
yields two sets of solutions for Eq. (\ref{incedche}), as
mentioned at the end of section III.B. The first set can
be obtained as limits of Eqs. (\ref{WIL-1a}-\ref{WIL-1b}) 
when $z_0\to 0$. Thus,
\antiletra\letra
\begin{eqnarray}\label{primeiro}
\mathring{U}_{\ 1}^{(j)}(z) =z^{(1-B_{2})/2}\displaystyle\sum_{n=0}^{\infty}(-1)^n
\mathring{c}_{n}^{1}
Z_{2n+B_2-1}^{(j)}\left(2\sqrt{qz}\right),\qquad [B_{2}\neq 0,-1,-2,\cdots]
\end{eqnarray}
where recurrence relations for $ \mathring{c}_{n}^{1}$ are given by
\begin{eqnarray}
 \text{ Eqs. (\ref{rIa}) if }
B_{2}\neq 1,2; \quad \text{  (\ref{rIIa}) if } B_{2}=1;\qquad
\text{  (\ref{rIIa}) if }B_{2}=2,
\end{eqnarray}
with
\begin{eqnarray}
&\begin{array}{l}
\mathring{\alpha}_{n}^{1} = -\frac{q B_1(n+1)}
{(2n+B_2)(2n+B_2+1)},\qquad
\mathring{\beta}_{n}^{1} =n(n+B_2-1)+B_3-\frac{q B_1 (B_2-2)}
{(2n+B_2-2)(2n+B_2)},\end{array}\nonumber\\
%
%
&\mathring{\gamma}_{n}^{1}= \frac{ q B_1(n+B_2-2)}
{(2n+B_2-3)(2n+B_2-2)}.
\end{eqnarray}
The second set is obtained as 
$\mathring{U}_{\ 2}^{(j)}(z)=T\mathring{U}_{\ 1}^{(j)}(z)$,
where $T$ is the transformation written in Eq. 
(\ref{transformationT}). Hence
\antiletra\letra
\begin{eqnarray}\label{segundo}
\mathring{U}_{\ 2}^{(j)}(z) =e^{B_1/z}\ z^{(1-B_{2})/2}
\displaystyle \sum_{n=0}^{\infty}(-1)^{n}\mathring{c}_{n}^{2}
Z_{2n+3-B_{2}}^{(j)}\left(2\sqrt{qz}\right),\qquad
\left[B_{2}\neq 4,5,6,\cdots \right]
\end{eqnarray}
where the coefficients $\mathring{c}_{n}^{2}$
satisfy the recurrence relations given in 
\begin{eqnarray}
\text{Eqs. (\ref{rIa}) if }
B_{2}\neq 2,3;\qquad \text{ (\ref{rIIa}) if } B_{2}=3;\qquad 
\text{(\ref{rIIIa}) if } B_{2}=2.
\end{eqnarray}
with 
\begin{eqnarray}
&\begin{array}{l}
\mathring{\alpha}_{n}^{2} = \frac{qB_1(n+1)}
{\left(2n+4-B_{2}\right)\left(2n+5-
B_{2}\right)},\qquad
\mathring{\beta}_{n}^{2} = (n+1)(n+2-B_{2})+B_{3}
 -\frac{q B_1\left(B_{2}-2\right)}
{\left(2n+2-B_{2}\right)\left(2n+4-
B_{2}\right)},\end{array}\nonumber\\
&\mathring{\gamma}_{n}^{2} =-
\frac{q B_1\left(n+2-B_{2}\right)}
{\left(2n+1-B_{2}\right)
\left(2n+2-B_{2}\right)}.
\end{eqnarray}
By examining the coefficients $\mathring{\alpha}_n$
and $\mathring{\gamma}_n$, and taking into account
the restrictions on $B_2$, we see that the above solutions
are given by infinite series beginning at $n=0$.

>From the convergence of the two-sided solutions for the DCHE, 
we find that 
\begin{itemize}
\itemsep-3pt
\item $\mathring{U}_{\ 1}^{(1)}$
and $\mathring{U}_{\ 2}^{(1)}$ converge for all $z$ but,
to decide if they are bounded or unbounded at the
singular points, we have to compute the limits 
when $z\to 0$ and $z\to\infty$.
\item $\mathring{U}_{\ 1}^{(j)}$
and $\mathring{U}_{\ 2}^{(j)}$ ($j=1,\ 2,\ 3$) 
converge for $|z|>0$ but their
limits when $z\to\infty$ may be unbounded
due to relations (\ref{comportamento-infinito}).
\end{itemize}
%

%
\section{Possible Applications}

In this section we consider two examples which use solutions for
the CHE. In the first example we discuss solutions for
the Klein-Gordon equation for a scalar test field $\Phi$ in the gravitational backgound
of a singular and a non-singular spacetimes. As the time dependence 
of the scalar field obeys Mathieu equations without any
arbitrary constant, we have to use two-side series solutions.
The parameter $\nu$ must be determined from the characteristic
equation. Then, the Raabe test assures that the wavefunction for the
non-sigular spacetime is bounded and convergent for
all values of the time variable.

The second example deals with the one-dimensional
Schr\"{o}dinger equation for a family of quasi-exactly solvable
potentials. In addition to the expected solutions given
by finite series, for a subfamily of the potentials 
we find infinite-series solutions which, due to the Raabe test, 
are bounded and convergent for all values of the independent variable.
If the equation reduces to a Whittaker-Hill equation (WHE),
we express such solutions by one-sided expansions in
series of Coulomb functions; otherwise, the solutions are
expressed by two-sided series. For the latter case the parameter $\nu$ 
may be chosen conveniently, while the characteristic equation
must be used for determining the energy levels.

\subsection{Klein-Gordon equation in curved spacetimes}

In its conformally static form, the
line element $ds^2=g_{\mu\nu}dx^{\mu}dx^{\nu}$ 
for nonflat Friedmann-Robertson-Walker spacetimes is written as \cite{birrel}
\antiletra
\begin{eqnarray}
\label{le}
ds^2=\left[A(\tau)\right]^{2}\left[d\tau^2-d\chi^{2}-\frac{\sin^2
(\sqrt\epsilon\chi)}{\epsilon}\left(d\theta^2+\sin^2\theta d\varphi^2\right)
\right],\quad x^{\mu}=(\tau,\chi,\theta,\varphi)
\end{eqnarray}
where $\epsilon=\pm 1$ is the spatial curvature, $\tau$ is the time
variable, whereas $\chi$, $\theta$ and $\varphi$ are 
the spatial coordinates. The Klein-Gordon equation for a 
field $\Phi$ with mass $M$ ($\hbar=c=1$) is
\begin{eqnarray*}
\partial_{\mu}\left(\sqrt{-g}g^{\mu\nu}\partial_{\nu}\Phi\right)
+\sqrt{-g}\left(M^2+\varrho R\right)\Phi=0,\qquad 
\partial_{\mu}={\partial}/{\partial x^{\mu}},
\end{eqnarray*} 
where $g$ is the determinant associated with $g_{\mu\nu}$,
$R$ is the Ricci scalar, $\varrho=1/6$ for conformal coupling, and 
$\varrho=0$ for minimal coupling.  By performing the separation
of variables
\begin{eqnarray}\label{full}
\Phi(\chi,\theta,\varphi,\tau)=[A(\tau)]^{-1}\ T(\tau)\ X(\chi)\ \Theta(\theta)
\ e^{im\varphi},\qquad m=0,\pm 1,\pm 2,\cdots,
\end{eqnarray}
one finds that $X$ and $\Theta$ are given by the same 
special functions for any scale factor $A(\tau)$ \cite{birrel},
while $T$ obeys the equation
\begin{eqnarray*}\label{Ttau}
\frac{d^2T}{d\tau^2}+\left[\kappa^2+M^2A^2 +(6\varrho-1)
\left(\frac{1}{A}\frac{d^2A}{d\tau^2}+\epsilon\right)\right]T=0,
\end{eqnarray*}
The constant of separation  $\kappa$, determined from the 
spatial dependence of $\Phi$, is given by
\[\kappa=1,2,3,\cdots \ \text{ if } \epsilon=1,\quad \text{ and } 0<\kappa<\infty
\text{ if } \epsilon=-1.\]
%
For a nonsingular model of universe 
and for (singular) radiation-dominated models, 
Eq. (\ref{Ttau}) reduces to Mathieu equations. 

\subsubsection{Nonsingular metric}

For the nonsingular case, the scale factor $A(\tau)$
is given by \cite{novello}
\[A(\tau)=\mathrm{a}_{0}\cosh{\tau},\qquad \epsilon=-1,\qquad -\infty
<\tau<\infty,\]
where $\mathrm{a}_0$ is a positive constant, 
leads to the modified Mathieu equation
\begin{eqnarray*}
\frac{d^2T}{d\tau^2}+\left[\kappa^2+\frac{1}{2}M^2\mathrm{a}_0^{2} 
+\frac{1}{2}M^2\mathrm{a}_0^{2}\cosh(2\tau)\right]T=0.
\end{eqnarray*}
So, in Eq. (\ref{mathieu}) we have
\[\sigma=i,\qquad \mathrm{a}=-\kappa^2-
\left(M^2\mathrm{a}_0^2/2\right),
\qquad k=M\mathrm{a}_0/2,\qquad u=\tau.\]
Solutions for this problem have already been given in 
Ref. \cite{nami} where the convergence at $\tau=0$ is not discussed.
Here this question is solved by using the Raabe test. 

From the Heine-type solutions, only $w_{\ 1}^{(j)}$ 
given in equation (\ref{Heine-1})
afford convergent and bounded wave functions for all 
$\tau\in(-\infty,\infty)$. We find the 
solutions 
\letra
\begin{eqnarray}\label{salim}
T^{(j)}(\tau) =
\displaystyle \sum_{n=-\infty}^{\infty} (-1)^nc_{n}
Z_{2n+2\nu+1}^{(j)}\left(M \mathrm{a}_0\cosh{\tau}\right),
\qquad \left[2\nu\notin \mathbb{Z}\right]
\end{eqnarray}
where the recurrence relations for $c_n$ are
\begin{eqnarray}
M^2\mathrm{a}_0^2\ c_{n+1}{+}\left[\left(4n+4\nu+2\right)^2+
4\kappa^2+2M^2\mathrm{a}_0^2
\right]c_{n}+M^2\mathrm{a}_0^2\ c_{n-1}=0.
\end{eqnarray}
The relations among the Bessel functions \cite{nist}
imply that only two of the four solutions (\ref{salim}) 
are linearly independent. Similar results are found
by treating the Mathieu equation as a CHE. In
effect, by using $w_{\ 1}^{(j)}$ given in (\ref{?}) 
we obtain 
\antiletra\letra
\begin{eqnarray}\label{salim-2}
\mathrm{T}^{(j)}(\tau) =
\displaystyle \sum_{n=-\infty}^{\infty} a_{n}
Z_{n+\nu+(1/2)}^{(j)}\left[2M \mathrm{a}_0\cosh^2\left({\tau}/{2}\right)\right],
\qquad \left[2\nu\notin \mathbb{Z}\right]
\end{eqnarray}
where the recurrence relations for $a_n$ are
\begin{eqnarray}\begin{array}{l}
M\mathrm{a}_0[n+\nu+1] a_{n+1}-\left[\left(n+\nu+\frac{1}{2}\right)^2+
\kappa^2+M^2\mathrm{a}_0^2
\right]a_{n}+M\mathrm{a}_0[n+\nu]a_{n-1}=0.
\end{array}
\end{eqnarray}
%
%
%
%
Thus, the Raabe test assures that $T^{(j)}$ and  $\mathrm{T}^{(j)}$, as well as
the corresponding wavefunctions (\ref{full}), 
are bounded and convergent for all $\tau\in(-\infty,\infty)$.
Notice that the solutions (\ref{mathieu-dche}), resulting
from the DCHE, do not satisfy the conditions 
(\ref{mathieu-curved}) for all $\tau\in(-\infty,\infty)$. 

\subsubsection{Singular metric}

For radiation-dominated spacetimes, 
$A(\tau)= \mathrm{a}_0\sin{(\epsilon\tau)}/ \sqrt{\epsilon}$ 
($\tau\geq 0$)
and, so,
\antiletra
\begin{eqnarray}
\frac{d^2T}{d\tau^2}+\left[\kappa^2+\frac{\epsilon}{2}M^2\mathrm{a}_0^{2} 
-\frac{\epsilon}{2}M^2\mathrm{a}_0^{2}\cos\left(2
\sqrt{\epsilon}\tau\right)\right]T=0.
\end{eqnarray}
We consider only the case  $\epsilon=-1$. We take
$\sigma=i$, $\mathrm{a}=\left(M^2\mathrm{a}_0^2/2\right)-\kappa^2$,
$k=M\mathrm{a}_0/2$ and $u=\tau$. Then the Heine
solutions $w_{\ 1}^{(j)}(u)$ 
given in equation (\ref{Heine-1}) lead 
\letra
\begin{eqnarray}
{T}^{(j)}(\tau) =
\displaystyle \sum_{n=-\infty}^{\infty} (-1)^n{c}_{n}
Z_{2n+2\nu+1}^{(j)}\left(M \mathrm{a}_0\cosh{\tau}\right),
\qquad \left[2\nu\notin \mathbb{Z}\right]
\end{eqnarray}
where the recurrence relations are
\begin{eqnarray}
M^2\mathrm{a}_0^2\ {c}_{n+1}{+}\left[\left(4n+4\nu+2\right)^2+
4\kappa^2-2M^2\mathrm{a}_0^2
\right]{c}_{n}+M^2\mathrm{a}_0^2\ {c}_{n-1}=0.
\end{eqnarray}

On the other side, from the solutions $w_{\ 1}^{(j)}(u)$ given 
in (\ref{?}) (CHE) we find 
\antiletra\letra
\begin{eqnarray}
\mathrm{T}^{(j)}(\tau) =
\displaystyle \sum_{n=-\infty}^{\infty} a_{n}
Z_{n+\nu+(1/2)}^{(j)}\left[2M \mathrm{a}_0\cosh^2\left({\tau}/{2}\right)\right],
\qquad \left[2\nu\notin \mathbb{Z}\right]
\end{eqnarray}
where the relations for $a_n$ are
\begin{eqnarray}\begin{array}{l}
M\mathrm{a}_0[n+\nu+1] a_{n+1}-\left[\left(n+\nu+\frac{1}{2}\right)^2+
\kappa^2\right]a_{n}+M\mathrm{a}_0[n+\nu]a_{n-1}=0.
\end{array}
\end{eqnarray}
Once more, ${T}^{(j)}$ and $\mathrm{T}^{(j)}$ 
are convergent and bounded for all 
$\tau\geq 0$ but now the wavefunctions (\ref{full})
become unbounded at $\tau=0$ due to the factor
$1/A(\tau)=1/(\mathrm{a}_0\sinh{\tau})$.
Again, the solutions (\ref{mathieu-dche}) resulting
from the DCHE are inappropriate for $\epsilon=-1$.

Therefore, if $\epsilon=-1$ the solutions for the
modified Mathieu obtained from the CHE and its Whittaker-Hill
limit are suitable for the singular and nonsingular
metrics, while the solutions obtained from
the DCHE are unsuitable. For the singular metric,
the unboundedness of the solutions (\ref{full}) 
at $\tau=0$ is expected since at this point
there is a physical singularity in the sense that
the pressure and density energy diverge.


\subsection{Schr\"{o}dinger equation for quasi-exactly solvable potentials}
 
 
 Now we consider problems involving solutions
 given by finite and infinite series for the CHE. 
 For this end, we write the one-dimensional 
 stationary Schr\"{o}dinger
 equation for a particle with mass $M$
 and energy $E$ as  
\antiletra
  \begin{eqnarray}
  \label{schr}
  \frac{d^2\psi}{du^2}+\big[{\cal E}-\mathcal{V}(u)\big]\psi=0, 
  \quad u={a} x, \
   \quad {\cal E}=\frac{2M }{\hbar^2 a^2} E,
   \qquad \mathcal{V}(u)=\frac{2M }{\hbar^2 a^2} \mathrm{V}(x),
  \end{eqnarray}
 where ${a}$ is a constant with inverse-length dimension, 
 $\hbar$ is the Plank constant
 divided by $2\pi$, $x$ is the spatial coordinate and 
 ${V}(x)$ is the potential. 
 For $\mathcal{V}(u)$ we choose the 
  Ushveridze quasi-exact solvable potential \cite{ushveridze1}
  \begin{eqnarray}
  \label{potencial}
  \mathcal{V}(u)&=&\begin{array}{l}4\beta^{2}{\sinh^{4}u}+ 4\beta\big[\beta-
  2(\gamma+\delta)-2\ell \big]{\sinh^{2}u}+4\left[ \delta-\frac{1}{4}\right]
   \left[\delta-\frac{3}{4}\right]\frac{1}{\sinh^{2}u}
   \end{array}
  \nonumber\vspace{2mm}\\
 &-& \begin{array}{l}
  4\left[ \gamma-\frac{1}{4}\right] \left[\gamma-\frac{3}{4}\right]\frac{1}
  {\cosh^{2}u}, 
  \qquad[\ell=0,1,2,\cdots]
  \end{array}
   \end{eqnarray}
where $\beta$, $\gamma$ and $\delta$ are real constants
with $\beta> 0$ and $\delta\geq 1/4$. 

When $\delta\geq1/4$ and $\ell$ is 
zero or a natural number, the above family 
of potentials is quasi-exactly
solvable  because it admits
bounded wavefunctions given by finite series which   
allow to determine only a finite
number of energy levels. {However, 
for $1/4\leq\delta<1/2$ and $1/2<\delta\leq 3/4$
we also find infinite-series solutions which are
convergent and bounded for all values of the independent
variable: this suggests the possibility of determining
the remaining part of
the energy spectra as solutions of a
characteristic equation. 
For $\delta>3/4$ we find no solutions like these.}

{Notice that Ushveridze supposed that $\ell=0,1,2,\cdots$, 
but we will find that
if 
\begin{eqnarray}\label{4casos}
\begin{array}{l}
\left(\gamma,\delta\right)=\left(\frac{1}{4},\frac{1}{4}\right), \ 
\left(\frac{1}{4},\frac{3}{4}\right), \ 
\left(\frac{3}{4},\frac{1}{4}\right), \ 
\left(\frac{3}{4},\frac{3}{4}\right),
\end{array}
\end{eqnarray}
then we also have quasisolvability when $\ell$ 
is a positive half-integer.  
In addition,} Ushveridze 
supposed that $u\in(-\infty,\infty)$. 
However, we get
\begin{eqnarray}\label{z=0}
\lim_{u\to\pm\infty}\mathcal{V}(u)=\infty,\qquad
\lim_{u\to 0} \mathcal{V}(u)=
\begin{cases} 
-4\left[\gamma-\frac{1}{4}\right]\left[\gamma-\frac{3}{4}\right], \text{ if } \delta=\frac{1}{4}
\text { or } \delta=\frac{3}{4};\vspace{2mm}\\ %
-\infty, \text{ if } \delta\in\left(\frac{1}{4},\frac{3}{4}\right);
\vspace{2mm}\\
+\infty, \text{ if } \delta\notin\left[\frac{1}{4},\frac{3}{4}\right].
\end{cases}
\end{eqnarray}
Hence, for $\delta\notin[1/4,3/4]$ there is an infinite 
barrier at $u=0$ and, so, we can suppose that $u\geq 0$
or $u\leq 0$.



\subsubsection{Wavefunctions for the Whittaker-Hill equation}
If $\gamma$ and $\delta$ take the values (\ref{4casos}),
the potential reads
\letra
\begin{eqnarray} 
\label{potencial4}
\mathcal{V}(u)=4\beta^{2}{\sinh^{4}u}+ 4\beta\big[\beta-
2(\gamma+\delta)-2\ell \big]{\sinh^{2}u}, 
\qquad \left[\ell=0,\ \frac{1}{2},\ 1,\ \frac{3}{2}, \cdots\right]
\end{eqnarray} 
where $u\in(-\infty,\infty)$.  
Thence, by using $\sinh^2{u}=[\cosh(2u)-1]/2$
and $\sinh^4{u}=[\cosh(4u)-4\cosh(2u)+3]/8$,
Eq. (\ref{schr}) becomes a  
modified WHE (\ref{whe}) 
with the parameters
\begin{equation}
\sigma=i,\qquad 
\vartheta=-\mathcal{E}+ 4 \beta (\ell+\gamma+\delta),\qquad
p+1=2(\ell+\gamma+\delta),\qquad 
\xi=2 \beta.
\end{equation}
{In fact,  the WHE 
also occurs in the cases of the Razavy potential  \cite{razavy}
and the symmetric double-Morse potential 
considered by Zaslavskii and Ulyanov \cite{zaslavskii}.}

{On the other side}, the substitutions
\antiletra
 \begin{eqnarray}
 z=\cosh^2u,\qquad \psi(u)={\psi}[u(z)]=U(z),
 \qquad[z\geq 1]
 \end{eqnarray}
 transform the Schr\"odinger equation for the preceding potential into
 the CHE (\ref{gswe}) with 
 \begin{eqnarray}\label{whe-parametros}
\begin{array}{l}
 z_0=1,\quad B_{1}=-\frac{1}{2}, \qquad B_{2}=1, \qquad 
 B_{3}=\frac{{\cal E}}{4},\qquad
 i\omega=\pm\beta,\qquad i\eta= \pm\left(\ell+\gamma+\delta\right),
 \end{array}
 \end{eqnarray}
where the plus or minus sign must be used throughout. 
{Thus, we can attempt to solve
the problem by
using known solutions for the CHE.
For example, from the
Baber-Hass\'{e} expansions in power series,
the Hylleraas solutions 
or from the Jaff\'{e} solutions \cite{leaver,baber,hylleraas,jaffe} 
we obtain even and odd
finite-series solutions bounded for $z\geq 1$: such
solutions allow to find only a finite number of
energy levels. There are also infinite-series solutions
which, however, must be discarded because they are not bounded 
for any admissible value of $z$. }

{On the other side, if we use  
one-sided series solutions 
$\mathring{\mathbb{U}}_i(z)$ in
terms of Coulomb wavefunctions, we shall find that 
(for $\ell$ equal to a  
non-negative integer or a positive 
half-integer)
 \begin{itemize}
 \itemsep-3pt
 \item Even and odd finite-series solutions which are 
 convergent and bounded for all $z\geq1$;
 these finite-series can be written in terms of 
 generalized Laguerre polynomials by means of (\ref{laguerre-1}).
 \item  Even infinite-series solutions which, 
 due to the Raabe test, are 
 convergent and bounded for
 all $z\geq1$.
 \item  Odd infinite-series
 solutions which 
 converge and are bounded only for
 $z>1$, and odd solutions which converge and are bounded
 only for finite values of $z$. To cover the
 entire interval $z\geq1$ it is necessary to consider
 two of such solutions. 
\end{itemize}
Therefore, in this case we could find additional
energy levels by solving a transcendental
characteristic equation.} 

{In effect, since} $B_1=-1/2$ and $B_2=1$, we find that
all the confluent hypergeometric functions which appear
in $\mathring{\mathbb{U}}_i(z)$ are linearly independent.
It is sufficient to consider the solutions
$\mathring{\mathbb{U}}_i=
(\mathring{U}_i,\mathring{U}_i^{\pm})$ with $i=1,2,3,4$.
We take $\gamma=\delta=1/4$ and consider the other cases
in the paragraph containing Eqs. (\ref{outros-casos}). As 
the expansions $\mathring{\mathbb{U}}_i(z)$
are not affected when we change ($\omega,\eta$) by
 ($-\omega,-\eta$), in Eqs. (\ref{whe-parametros}) we select
\begin{eqnarray}\label{gamma=delta}
i\omega=-\beta,\qquad i\eta=-\ell-(1/2), \qquad [\gamma=\delta=1/4].
\end{eqnarray}
Then, by using $\mathring{\mathbb{U}}_1=(\mathring{U}_1,
\mathring{U}_1^{\pm})$ given in
(\ref{par1}), we find 
\begin{eqnarray}
\begin{array}{l}
\mathring{\psi}_1(u)=  e^{-\beta \cosh^2u}\displaystyle
\sum_{n=0}^{\infty}\begin{array}{l}
\frac{\mathring{b}_{n}^{1}\left[-2\beta \cosh^2 {u}\right]^{n}}{\Gamma[2n+1]}
\Phi\left[n-\ell,2n+1;2\beta \cosh^2u\right]\end{array},\vspace{2mm}\\
\mathring{\psi}_1^{\pm}(u)=  e^{\mp\beta \cosh^2u}\displaystyle
\sum_{n=0}^{\infty}\begin{array}{l}
\frac{\mathring{b}_{n}^{1} \left[2\beta \cosh^2 {u}\right]^{n}}
{\Gamma[n+\frac{1}{2}\pm(\ell+\frac{1}{2})]}
\Psi\left[n+\frac{1}{2}\mp\left(\ell+\frac{1}{2}\right),2n+1;\pm 2\beta \cosh^2u\right]\end{array}
\end{array}
\end{eqnarray}
which are even solutions with respect to 
$u\to-u$. According to Eqs. (\ref{coef-3'}) and (\ref{s-che-1}),
if the series begin at $n=0$
the coefficients $\mathring{b}_{n}^{1}$ satisfy 
the recurrence relations (\ref{r2a}) with 
 \begin{eqnarray*}
 \begin{array}{l}
 \mathring{\alpha}_{n}^{1}  =  -\frac{\beta}{2},\qquad
 \mathring{\beta}_{n}^{1}  =- n^2 
 -\frac{\mathcal{E}}{4}
 +\beta\left(\ell+\frac{1}{2}\right),\qquad
 \mathring{\gamma}_{n}^{1}  = \frac{\beta}{2}
 \left(n+\ell\right)
 \left(n-\ell-1 \right)
 \end{array}
 \end{eqnarray*}
If $\ell$ is a non-negative integer  $\mathring{\psi}_1\propto\mathring{\psi}_1^{+}$ 
represents bounded finite-series solutions with $0\leq n\leq\ell$, 
while $\mathring{\psi}_1^{-}$ represents infinite-series solutions
with $n\geq\ell+1$ which become unbounded when 
$z=\cosh^2u\to\infty$ due to the factor $\exp{(\beta z)}$ 
($\beta>0$). On the other hand, if $\ell$ is a positive
half-integer $\mathring{\psi}_1^{+}$ represents 
infinite-series solutions convergent for $z\geq 1$ because
$\text{Re}[B_2+(B_1/z_0)]=1/2<1$ as required by the Raabe test;
these solutions are bounded also at $z=\infty$ since
relation (\ref{asymptotic-confluent1}) leads to
\begin{eqnarray*}
\mathring{\psi}_1^{+}(u)\sim
e^{-\beta \cosh^2u}(\cosh{u})^{2\ell}
\sum_{n=0}^{\infty}\frac{b_n^1}{\Gamma[n+\ell+1]}\to 0,\qquad
\ell=\frac{1}{2},\frac{3}{2},\cdots\qquad[u\to\pm \infty].
\end{eqnarray*}
By using $\mathring{\mathbb{U}}_2(z)$ 
given in (\ref{par2}), we again get even solutions
but now the finite series hold if $\ell$ is half-integer
whereas the infinite series hold if $\ell$
is integer. We find
\begin{eqnarray}
\begin{array}{l}
\mathring{\psi}_2= e^{-\beta \cosh^2u}\cosh {u}
\displaystyle\sum_{n=0}^{\infty}\begin{array}{l}
\frac{\mathring{b}_{n}^{2} \left[-2\beta \cosh^2u\right]^{n}}
{\Gamma\left[2n+2\right]} 
\Phi\left[n-\ell+\frac{1}{2},2n+2;2\beta \cosh^2 u\right],
\end{array}\vspace{2mm}\\
\mathring{\psi}_2^{\pm}= e^{\mp \beta \cosh^2u}\cosh {u}
\displaystyle\sum_{n=0}^{\infty}\begin{array}{l}
\frac{\mathring{b}_{n}^{2} \left[2\beta \cosh^2u\right]^{n}}{\Gamma\left[n+1\pm\left(\ell+\frac{1}{2}\right)\right]} 
\Psi\left[n+1\mp\left(\ell+\frac{1}{2}\right),2n+2;\pm2\beta \cosh^2 u\right],
\end{array}
\end{array}
\end{eqnarray}
where, for series beginning at $n=0$, the coefficients $\mathring{b}_{n}^{2}$ satisfy 
the relations (\ref{r3a}) with
\begin{eqnarray*}
\begin{array}{l}
\mathring{\alpha}_{n}^{2}  =  -\frac{\beta}{2},\quad
\mathring{\beta}_{n}^{2}  =- \left[n+\frac{1}{2}\right]^2 
-\frac{\mathcal{E}}{4}
+\beta\left[\ell+\frac{1}{2}\right],\quad
\mathring{\gamma}_{n}^{2}  = \frac{\beta}{2}
\left[n+\ell+\frac{1}{2}\right]
\left[n-\ell-\frac{1}{2} \right].
\end{array}
\end{eqnarray*}
Thus, if $\ell$ is a half-integer then  $\mathring{\psi}_2\propto\mathring{\psi}_2^{+}$ 
gives bounded finite-series solutions with $0\leq n\leq\ell-(1/2)$. 
If $\ell$ is integer $\mathring{\psi}_1^{+}$ gives  
infinite-series solutions convergent for $z\geq 1$ because
$\text{Re}[B_2+(B_1/z_0)]<1$ as required by the Raabe test;
these solutions are bounded also at $z=\infty$ due to 
relation (\ref{asymptotic-confluent1}). 

Odd solutions are generated from the expansions
$\mathring{\mathbb{U}}_3$ and $\mathring{\mathbb{U}}_4$,
but in these cases we find no single infinite-series
solution converging for all $z\geq 1$
because the Raabe test is inapplicable
for $U_3^{\pm}$ and $U_4^{\pm}$.  
Thus, from $\mathring{\mathbb{U}}_3$
given in (\ref{par3}), we find the odd solutions
\begin{eqnarray}\begin{array}{l}
\mathring{\psi}_3= 
e^{-\beta \cosh^2u} \sinh{(2u)}
\displaystyle\sum_{n=0}^{\infty}\begin{array}{l}
\frac{\mathring{b}_{n}^{3}\left[-2\beta \cosh^2u\right]^{n}}{\Gamma[2n+3]}
\Phi\left[n+1-\ell,
2n+3;2\beta \cosh^2u\right],\end{array}\vspace{2mm}\\
\mathring{\psi}_3^{\pm}= 
e^{\mp \beta \cosh^2u} \sinh{(2u)}
\displaystyle\sum_{n=0}^{\infty}
\begin{array}{l}
\frac{\mathring{b}_{n}^{3}\left[2\beta \cosh^2u\right]^{n}}
{\Gamma\left[n+\frac{3}{2}\pm\left(\ell+\frac{1}{2}\right)\right]}
\times\end{array}\vspace{2mm}\\
\hspace{5.2cm}\begin{array}{l}
\Psi\left[n+\frac{3}{2}\mp\left(\ell+\frac{1}{2}\right),2n+3;\pm 2\beta \cosh^2u\right],\end{array}
\end{array}
\end{eqnarray}
where, for series beginning at $n=0$, 
the $\mathring{b}_{n}^{3}$ satisfy 
the relations (\ref{r1a}) with 
\begin{eqnarray*}
\begin{array}{l}
\mathring{\alpha}_{n}^{3}  =  -\beta\ \frac{(n+1)}{(2n+4)},\qquad
\mathring{\beta}_{n}^{3}  =- (n+1)^2 
-\frac{\mathcal{E}}{4}
+\beta\left(\ell+1\right),\qquad
\mathring{\gamma}_{n}^{3}  = \beta\ \frac{(n+1)
\left(n+\ell+1\right)
\left(n-\ell \right)}{2n}.
\end{array}
\end{eqnarray*}
So, if $\ell$ is integer then $\psi_3\propto \psi_3^{+}$
is given finite-series with $0\leq n\leq\ell-1$
($\ell\geq 1$), whereas  
$\psi_3^{-}$ is given by infinite 
series with $n\geq \ell$ (unbounded at 
$z=\infty$). If $\ell$ is half-integer
the three solutions are given by infinite series
and $\psi_3$ stands for $z\neq\infty$, 
$\psi_3^{+}$ stands for $z\neq 1$
whereas $\psi_3^{-}(u)$ (linear combination
of $\psi_3$ and $\psi_3^{+}$) 
stands for $z\neq 1$ and $z\neq \infty$.

By using $\mathring{\mathbb{U}}_4(z)$ 
given in (\ref{par4}), we again get odd solutions
but now the finite series hold if $\ell$ is half-integer
whereas the infinite series hold if $\ell$
is integer. We find
\begin{eqnarray}
\begin{array}{l}
\mathring{\psi}_4=e^{-\beta \cosh^2u}\sinh{u}
\displaystyle\sum_{n=0}^{\infty}\begin{array}{l}
\frac{\mathring{b}_{n}^{4}\ [-2\beta \cosh^2u]^{n}}{\Gamma[2n+2]}
\Phi\left[n-\ell+\frac{1}{2},2n+2;2\beta \cosh^2u\right],
\end{array}
\vspace{2mm}\\
\mathring{\psi}_4^{\pm}=e^{\mp \beta \cosh^2u}\sinh{u}
\displaystyle
\sum_{n=0}^{\infty}\begin{array}{l}
\frac{\mathring{b}_{n}^{4}\ [2\beta \cosh^2u]^{n}}
{\Gamma\left[n+1\pm \left(\ell+\frac{1}{2}\right)\right]}
\Psi\left[n+1\mp\left(\ell+\frac{1}{2}\right),2n+2;\pm2\beta \cosh^2u\right],
\end{array}
\end{array}
\end{eqnarray}
where the coefficients $\mathring{b}_{n}^{4}$ satisfy 
the recurrence relations (\ref{r3a}) with 
\begin{eqnarray*}
\begin{array}{l}
\mathring{\alpha}_{n}^{4}  =  -\frac{\beta}{2}\left[ \frac{2n+1}{2n+3}\right],
\qquad
\mathring{\beta}_{n}^{4}  = \mathring{\beta}_{n}^{2},\qquad
\mathring{\gamma}_{n}^{4}  = \frac{\beta}{2} 
\left[\frac{2n+1}{2n-1}\right]
\left[n+\ell+\frac{1}{2}\right]
\left[n-\ell -\frac{1}{2}\right].
\end{array}
\end{eqnarray*}
Thus, if $\ell$ is integer it is necessary to take two
infinite-series solutions to cover the interval $z=\cosh^2u\geq 1$ since 
$\psi_4(u)$  holds for $z\neq\infty$,
$\psi_4^{+}(u)$ for $z\neq 1$, while $\psi_4^{-}(u)$
holds for $z\neq 1$ and $z\neq \infty$. 

For the other values (\ref{4casos}) for the pair ($\delta,\gamma$), the solutions
can be obtained from the above ones ($\gamma=\delta=1/4$) by a change in the 
parameter $\ell$. In effect, we find that
\begin{eqnarray}\label{outros-casos}
\begin{array}{l}
 \left(\gamma,\delta\right)=\left(\frac{1}{4},\frac{3}{4}\right) \text{ or } 
\left(\frac{3}{4},\frac{1}{4}\right) \Rightarrow i\eta=-\ell-1;\qquad
 \left(\gamma,\delta\right)=\left(\frac{3}{4},\frac{3}{4}\right) \Rightarrow i\eta=-\ell-\frac{3}{2},
\end{array}
\end{eqnarray}
without any other change in the parameters (\ref{whe-parametros}). Hence, by comparing
the previous expressions for $i\eta$ with the one given in (\ref{gamma=delta}), 
we see that in the above solutions it is sufficient to replace: (i) $\ell$ by $\ell+(1/2)$
in the first two cases, (ii) $\ell$ by $\ell+1$ in the last case. Anyway, the important
conclusion is the same: there are even infinite-series solutions
that are convergent and bounded for all $z\geq 1$.


\subsubsection{Wavefunctions for the cases $1/4\leq\delta<1/2$
and $1/2<\delta\leq3/4$}
For the Ushveridze potential (\ref{potencial}), the substitutions
\begin{eqnarray} \label{Ush}
z=\cosh^2u,\qquad \psi(u)={\psi}[u(z)]=
z^{\gamma-\frac{1}{4}}(z-1)^{\delta-\frac{1}{4}}U(z),\qquad[z\geq 1]
\end{eqnarray}
transform the Schr\"odinger equation (\ref{schr}) 
into a confluent Heun equation with 
\begin{eqnarray}
&\begin{array}{l}
z_0=1,\qquad B_{1}=-2\gamma, \qquad  B_{2}=2\gamma+2\delta, \qquad 
B_{3}=\frac{{\cal E}}{4}+\left(\gamma+\delta-\frac{1}{2}\right)^{2},\end{array} 
\vspace{2mm}\nonumber\\
&i\omega=\pm\beta,\qquad i\eta= \pm(\ell+\delta+\gamma).
\end{eqnarray}
Now we exclude the cases (\ref{4casos}) and suppose that $\ell$ is a non negative
integer. We select
\[i\omega=-\beta,\qquad i\eta=-\ell-\gamma-\delta.\]
Then, by using for $U(z)$ the power series written in
Eqs. (29a-b) of Ref. \cite{lea-1}, we find 
\letra
\begin{eqnarray} \label{baber}
\psi_1^{\text{baber}}[u(z)]=
e^{-\beta z}z^{\gamma-\frac{1}{4}}(z-1)^{\delta-\frac{1}{4}}
\sum_{n=0}^{\ell}\mathrm{a}_n(z-1)^n,\qquad [\ell=0,1,2,\cdots]
\end{eqnarray}
where the series coefficients satisfy ($\mathrm{a}_{-1}=0$)
\begin{eqnarray}
&\begin{array}{l}
(n+1)(n+2\delta)\mathrm{a}_{n+1}+ 
\left[n(n+2\gamma+2\delta-1-2\beta)+\frac{{\cal E}}{4}+
\left(\gamma+\delta-\frac{1}{2}\right)^2 -2\beta\delta\right]\mathrm{a}_n+
\end{array}\nonumber\vspace{2mm}\\
&-2\beta(n-\ell-1)\mathrm{a}_{n-1}=0.
\end{eqnarray}
Since $\beta>0$, the previous finite-series solutions are bounbed 
for all $z\geq 1$ provided that $\delta\geq 1/4$. According
to theory of three-term recurrence relations \cite{arscott},  
$\psi_1^{\text{baber}}$ represents $\ell+1$ distinct solutions, 
each one with a different energy.

By the other side, we find cases for which there are 
infinite-series solutions appropriate for any $z\geq 1$. 
For this we insert into (\ref{Ush}) the two-sided solutions ${U}_1^+$
and ${U}_2^+$ given in (\ref{par1-nu}) and (\ref{par2-nu}), respectively,
and use the Raabe test along with the limit (\ref{asymptotic-confluent1}).
Thence by scribing convenient values to the parameter $\nu$, we
find the solutions $\psi_{ 1}^{+}$ and $\psi_{ 2}^{+}$ having the following 
properties
\begin{itemize}
\itemsep-3pt
\item The solutions $\psi_{ 1}^{+}$ are 
convergent and bounded for all $z\geq1$ if $1/4\leq\delta<1/2$.
\item  The solutions $\psi_{ 2}^{+}$ are 
convergent and bounded for all $z\geq1$ if $1/2<\delta\leq 3/4$.
\item  If $\delta=1/2$, then $\psi_{ 1}^{+}=\psi_{ 2}^{+}$.
For this case the Raabe test is inconclusive as to the
convergence at $z=1$.
\end{itemize}
In effect, if $U(z)=U_1^{+}(z)$ we obtain
\antiletra\letra
\begin{eqnarray}\label{bottom1}
%
%
\psi_{ 1}^{+}(z)& =&e^{-\beta z}z^{\nu-\delta+\frac{3}{4}}
(z-	1)^{\delta-\frac{1}{4}} \displaystyle
\sum _{n=-\infty}^{\infty} 
\frac{b_n^{1}\ 
[2\beta z]^{n}}{\Gamma[n+\nu+1+\ell+\gamma+\delta]}
\vspace{2mm}\nonumber\\
&\times&\Psi\left[n+\nu+1-\ell-\gamma -\delta,2n+2\nu+2; 
2\beta z\right],\qquad {1}/{4}\leq\delta<{1}/{2}
\end{eqnarray}
where, in the recurrence relations (\ref{recursion-1}) for
$b_{n}^{1}$ we have  
\begin{eqnarray}
\begin{array}{l}
\alpha_{n}^{1}  =  -\frac{2\beta\left[n+\nu+2-\gamma-\delta\right]
\left[n+\nu+1+\gamma-\delta\right]}
{(2n+2\nu+2)(2n+2\nu+3)},
\vspace{.1cm} \\
\beta_{n}^{1} =  -\frac{\cal E}{4}+\beta(\ell+\gamma+\delta)-
\left[n+\nu+\frac{1}{2}\right]^2
-\frac{\beta\left[\ell+\gamma+\delta\right][\gamma+\delta-1]
\left[\gamma-\delta\right]}
{[n+\nu][n+\nu+1]},
\vspace{.2cm} \\
\gamma_{n}^{1}  =
\frac{2\beta\left[n+\nu-1+\gamma+\delta\right]
\left[n+\nu-\gamma+\delta\right]
[n+\nu+\ell+\gamma+\delta][n+\nu-\ell-\gamma-\delta]}
{[2n+2\nu-1][2n+2\nu]}.
\end{array}
\end{eqnarray}
By the Raabe test the condition $\delta<1/2$ assures that
the series converge at $z=1$, while the condition $\delta\geq 1/4$
assures that the factor $(z-1)^{\delta-(1/4)}$ is bounded at
$z=1$. Similarly, if $U(z)=U_2^{+}(z)$ we obtain
\antiletra\letra
\begin{eqnarray}\label{bottom2}
%
%
\psi_{ 2}^{+}(z)&=&e^{-\beta z}z^{\nu+\delta-\frac{1}{4}}
(z-	1)^{-\delta+\frac{3}{4}} \displaystyle
\sum _{n=-\infty}^{\infty} 
\frac{b_n^{2}\ 
[2\beta z]^{n}}{\Gamma[n+\nu+1+\ell+\gamma+\delta]}\nonumber
\vspace{2mm}\\
&\times& \Psi\left[n+\nu+1-\ell-\gamma -\delta,2n+2\nu+2; 
2\beta z\right], \qquad 1/2< \delta\leq 4/3 
\end{eqnarray}
where, in the recurrence relations (\ref{recursion-1}) for
$b_{n}^{2}$  
\begin{eqnarray}
\begin{array}{l}
\alpha_{n}^{2}  =  -\frac{2\beta\left[n+\nu+\gamma+\delta\right]
\left[n+\nu+1-\gamma+\delta\right]}
{(2n+2\nu+2)(2n+2\nu+3)},\qquad
\beta_{n}^{2} =  \beta_{n}^{1} 
\vspace{.2cm} \\
\gamma_{n}^{2}  =
\frac{2\beta\left[n+\nu+1-\gamma-\delta\right]
\left[n+\nu+\gamma-\delta\right]
[n+\nu+\ell+\gamma+\delta][n+\nu-\ell-\gamma-\delta]}
{[2n+2\nu-1][2n+2\nu]},
\end{array}
\end{eqnarray}
where $\beta_{n}^{2} =  \beta_{n}^{1}$ is a functional
identity; in fact, $\beta_{n}^{1}$ and $\beta_{n}^{2}$
are different of each other because they hold for distinct intevals
of $\delta$. 

To assure that all the terms of the series are linearly independent 
and that the summation extends from minus to plus infinity,
the parameter $\nu$ must be chosen such that
\antiletra
\begin{eqnarray}
2\nu,\quad \nu\pm(\gamma+\delta)\quad \text{and}\quad \nu\pm(\gamma-\delta)
\quad \text{are not integers},
\end{eqnarray}
where the values for $\delta$ are different for solutions (\ref{bottom1}) 
and (\ref{bottom2}). The linear independence is
assured by requiring that $2\nu$ is not integer, 
without any restrictions 
on the parameters of the potential.
Thus, for fixed values of $\gamma$ and $\delta$, 
we can choose for $\nu$ any value in the open 
interval $(0,1/2)$ convenient to satisfy 
the above conditions. The use of one-sided series 
would lead to restrictions on $\gamma$ and $\delta$.

\section{Conclusion}

We have dealt with the convergence
of Leaver's expansions in series of Coulomb wave functions
for solutions of the CHE and DCHE. By redefining the Coulomb 
functions, we have completed the 
proof of convergence delineated by Leaver and, in addition, 
have found that the Raabe test improves the regions of
convergence for solutions of the CHE and its
Whittaker-Ince limit (\ref{incegswe}) if certain conditions 
are fulfilled. It is worth noticing that 
in using the convergence tests we suppose that the independent
variable $z$ is finite. So, when $z$ tends to infinite 
the behavior of each solution 
must be analyzed carefully.

Other points that have also extended the meaning of 
Leaver solutions are: (i) the transformations
of variables which generate new solutions 
(Leaver has used only transformation $t_2$ for 
the DCHE); (ii) the Whittaker-Ince
limit which gives solutions for Eqs. (\ref{incegswe}) and (\ref{incedche});
(iii) the construction of one-sided series solutions
which, in most cases, lead to finite series solutions. 
These issues have been 
partially discussed elsewhere, but now
they incorporate the consequences of the Raabe test
for the solutions of the CHEs.

The transformations of variables 
lead to solutions with different
domains of convergence and/or different behaviors
at the singular points. By this procedure, from the solutions
for the Whittaker-Hill limit of the CHE we 
can recover all the Heine solutions in series of Bessel
functions for the Mathieu equation \cite{nist} but  
the Raabe test improves the regions of convergence
for some of these solutions. For example, in section 4,
we have found that
the Klein-Gordon equation in a non-singular model of universe 
has solutions bounded and convergent 
for all values of the time variable.

In section 4  
we have also considered the Schr\"{o}dinger equation for 
the QES potentials given in 
Eq. (\ref{potencial}). For 
certain intervals of the parameter $\delta$ we have found 
some eigenfunctions given by one-sided and two-sided infinite
series which, by the Raabe test, are convergent and bounded for
all values of the independent variable. 
This type of solutions are not obtained by using
other known solutions for the CHE as, for example, 
the Hylleraas and Jaff\'e solutions \cite{leaver,hylleraas, jaffe}.

In fact, there is another 
QES problem \cite{lea-2} admitting well-behaved 
infinite-series eigenfunctions, 
and these result from solutions for the Whittaker-Ince 
limit of the CHE.
However, for the problem of
section 4, the expansions in series of Coulomb functions
do not give appropriate infinite-series wavefunctions
if $\delta>3/4$. We do not know whether it is possible
to find new solutions for the CHE suitable for
this case.

On the other side, as particular cases of the solutions for
the CHE we have obtained 
the Meixner two-sided  solutions
in series of Bessel functions  
for the spheroidal equation \cite{meixner}. These solutions give 
one-sided infinite-series solutions which, in turn, 
lead to finite-series solutions as
discussed after Eqs. (\ref{esferoidal-finitas}).
As far as we are aware, up to now these are
the only solutions for
the spheroidal equation which admit finite series
despite the fact that we have found no application
for them. 

We have also seen that the Raabe test is useless 
for solutions of the DCHE. On the other hand, 
we have found two subgroups of solutions for the DCHE -- see Eqs. 
(\ref{hipergeometricas}) and (\ref{hipergeometricas-DCHE}).
One subgroup is obtained from solutions of 
the CHE when $z_0\to 0$; the other follows from 
this subgroup by the transformation $t_2$
given in Eqs. (\ref{transformacao-dche}) but cannot 
be derived as limit of expansions 
in series of Coulomb functions. 
Thus, we can ask if there are solutions for
the CHE which yield such subgroup for the 
DCHE when $z_0\to 0$. 

It seems that is possible to solve the preceding problem
by further investigation on known solutions in series of 
Gauss hypergeometric functions for the CHE \cite{eu2,eu-novello,otchik}.
For instance, we have the two-sided series expansion \cite{eu2}
 \begin{eqnarray}
 \mathcal{U}_{1}^{+}(z)=
 e^{i\omega z}\sum_{n}
 \begin{array}{l}
 d_{n}^{1}\  F\left(\frac{B_{2}}{2}-n-\nu-1,
 n+\nu+\frac{B_{2}}{2};B_{2}+\frac{B_{1}}{z_{0}};1-
 \frac{z}{z_{0}}\right),
 \end{array}
 \end{eqnarray}
 where $F(a,b;c;\zeta)$ denotes the Gauss hypergeometic functions, and 
 the coefficients $d_n^1$ are multiples of
 the coefficients $b_n^1$ which appear in 
 solutions (\ref{par1-nu}). This solution
 is convergent for finite values of $z$
 and, so, cover a region of the complex plane
 left off by the expansions in series of Coulomb wave functions.
 On the other side, from the limit \cite{erdelyi1}
 \begin{eqnarray*} 
 \lim_{c\rightarrow \infty}
 F\left[a,b;c;1-({c}/{y})\right]=y^b\Psi(b,1+b-a;y), 
 \end{eqnarray*}
 we find $\lim_{z_0\to 0}\mathcal{U}_{1}^{+}(z) \propto U_{3}^{+}(z)$, where 
 $U_{3}^{+}(z)$ is just the solution given in
 Eqs. (\ref{dche-3a}) for the DCHE, which does not
 result of any expansion in series of 
 Coulomb functions. To get the solutions
 $U_{3}(z)$ and $U_{3}^{-}(z)$ for the DCHE, we would need to 
 find two additional expansions in series of hypergeometric
 functions for the CHE. We could also examine if such solutions 
supply the Meixner expansions in series of 
Legendre functions for the spheroidal equation \cite{meixner}, having the same coefficients as
his expansions in series of Bessel functions.

Finally, notice that we have not considered the second type 
of Leaver solutions in series of confluent hypergeometric functions 
for the CHE. In fact, these solutions have been discussed in a previous paper \cite{lea-1} where, however, 
the Raabe test has not taken into
account. First, we can show that the test is applicable to
the expansions in series of irregular confluent
hypergeometric functions. Second, for the DCHE  
the transformation $t_2$ gives solutions which 
cannot been derived from
Leaver's solutions of the CHE through the limit
$z_0\to 0$. Thence, we have another unsolved problem connected
with Leaver's solutions. 
 

   %
%
\appendix
%
\section{Confluent-hypergeometric and Coulomb Functions}
\protect\label{A}
\setcounter{equation}{0}
\renewcommand{\theequation}{A.\arabic{equation}}
Here we write some useful formulas concerning the
confluent hypergeometric functions and,  in Eqs. (\ref{U-n}) 
and (\ref{fi}), redefine the Coulomb 
wave functions. At the end we obtain the relations 
(\ref{bessel-teste}) which are important to 
apply the convergence tests for
infinite-series solutions of the CHE and DCHE.

The regular and irregular confluent
hypergeometric functions, $\Phi({a,c};u)$
and $\Psi({a},{c};u)$, are solutions of
the confluent hypergeometric equation \cite{erdelyi1}
\begin{eqnarray}\label{confluent0}
\begin{array}{l}
y\frac{d^2\varphi}{dy^2}+(c-y)\frac{d\varphi}{dy}-
a\ \varphi=0.
\end{array}
\end{eqnarray}
The functions $\Phi({a,c};y)$ and $\Psi({a},{c};y)$ are also
denoted by $M({a,c},y)$ and $U({a},{c},y)$, respectively \cite{abramowitz}.
In fact, the following four solutions for Eq. (\ref{confluent0})
\begin{eqnarray}\label{confluent1}
\begin{array}{ll}
\varphi^{(1)}(y)=\Phi(a,c;y),\qquad&\varphi^{(2)}(y)=\Psi(a,c;y),
\vspace{3mm}\\
\varphi^{(3)}(y)=e^{y} \Psi(c-a,c;-y),&
\varphi^{(4)}(y)= y^{1-c}\Phi(1+a-c,2-c;y),
\end{array}
\end{eqnarray}
are all of them defined and distinct only if
$c$ is not an integer \cite{erdelyi1}. 
Different forms for $\varphi^{(i)}$ follow from the
Kummer transformations
\begin{eqnarray}\label{kummer}
\begin{array}{l}
\Phi(a,c;y)=e^{y}\Phi(c-a,c;-y),
\qquad \Psi(a,c;y)=y^{1-c}\Psi(1+a-c,2-c;y).
\end{array}
\end{eqnarray}
In this article, we use only $\varphi^{(1)}$, $\varphi^{(2)}$ and 
$\varphi^{(3)}$. Their Wronskians are \cite{erdelyi1}
\begin{eqnarray}\label{wronskians}
\begin{array}{l}
\mathscr{W}\left[\varphi^{(1)},\varphi^{(2)}\right]=
\mathscr{W}\left[\Phi(a,c;y),\Psi(a,c;y)\right]=
-\frac{\Gamma(c)}{\Gamma(a)}\ y^{-c}\ e^{y},
\vspace{2mm}\\
\mathscr{W}\left[\varphi^{(1)},\varphi^{(3)}\right]=
\mathscr{W}\left[\Phi(a,c;y),e^y\Psi(c-a,c;e^{\pm i\pi}y)\right]=
\frac{\Gamma(c)}{\Gamma(c-a)}\ e^{\mp i\pi c}\ y^{-c}\ e^{y},
\vspace{2mm}\\
\mathscr{W}\left[\varphi^{(2)},\varphi^{(3)}\right]=
\mathscr{W}\left[\Psi(a,c;y),e^y\Psi(c-a,c;e^{\pm i\pi}y)\right]=
e^{\pm i\pi (a-c)}\ y^{-c}\ e^{y}.
\end{array}
\end{eqnarray}
Therefore, if
$a$, $c$ and $c-a $ are not zero or negative integers, 
the three solutions are well defined and
any two of them 
form a fundamental system of solutions for
confluent hypergeometric equation \cite{erdelyi1}.
If $c$ is not zero or a negative integer, 
the solutions are connected by 
\cite{erdelyi1},
\begin{equation}\label{continuation2}
\begin{array}{l}
\frac{\Phi(a,c;y)}{\Gamma(c)}=  \frac{e^{\mp i\pi a}}
{\Gamma(c-a)}\  \Psi(a,c;y)+ \frac{e^{\pm i\pi (c-a)}}
{\Gamma(a)}
e^{y}\ \Psi\left(c-a,c;e^{\pm i\pi}y\right), 
%
\end{array}
\end{equation}

When $y\to \infty$, the behavior of  of $\Psi(a,c;y)$ is given by 
\cite{abramowitz}
\begin{eqnarray}\label{asymptotic-confluent1}
\begin{array}{l}\displaystyle
\Psi(a,c;y)\sim y^{-a}\sum_{m=0}^{\infty}\frac{(a)_m(a-c+1)_m}{m!}(-y)^{-m},
\qquad  -\frac{3\pi}{2}< \arg{y}<\frac{3\pi}{2};
\end{array}
\end{eqnarray}
while the behaviour of $\Phi(a,c;y)$ is given by
\begin{eqnarray} \label{asymptotic-confluent}
&\begin{array}{l} \displaystyle
\frac{\Phi(a,c;y)}{\Gamma(c)}\sim
\frac{e^{y}y^{a-c}}{\Gamma(a)}\sum_{m=0}^{\infty}\frac{(1-a)_m(c-a)_m}{m!}y^{-m}+
\end{array}\nonumber\\
&\frac{e^{\pm i\pi a}y^{-a}}{\Gamma(c-a)}\sum_{m=0}^{\infty}\frac{(a)_m(a-c+1)_m}{m!}(-y)^{-m},
\quad a\neq 0,-1,\cdots, \ c-a\neq 0,-1,\cdots,\quad
\end{eqnarray}
where the upper sign holds for $-\pi/2<\arg{y}<3\pi/2$
and the lower sign, for $-3\pi/2<\arg{y}\leq-\pi/2$.
In these limits $(\text{x})_m$ denotes the Pochhammer symbol whose definition is
\[(\text{x})_0=1,\qquad (\text{x})_1=\text{x},\qquad
(\text{x})_m=\text{x}(\text{x}-1)(\text{x}-2)\cdots(\text{x}+m-1)=
{\Gamma(\text{x}+m)}/{\Gamma(\text{x})}.\]

On the other hand the Coulomb wave functions are solutions of
the equation
\begin{eqnarray}\label{coulomb1}
\begin{array}{l}
\frac{d^2\mathscr{U}_{n+\nu}}{dy^2}+
\left[1-\frac{2\eta}{y}-\frac{(n+\nu)(n+\nu+1)}{y^2}\right]
\mathscr{U}_{n+\nu}=0.
\end{array}
\end{eqnarray}
If $\eta=0$, this can be written in the usual form of the
Bessel equation by a substitution of variable.
If $\eta\neq 0$ the solutions $\mathscr{U}_{n+\nu}(y)=\mathscr{U}_{n+\nu}(\eta,y)$
are written in terms of one regular confluent hypergeometric
function $\Phi$ and two irregular  functions
$\Psi$, that is,
\begin{eqnarray}\label{U-n}
\mathscr{U}_{n+\nu}(\eta,y)=
\left[\phi_{n+\nu}\big(\eta,y\big),\ \psi_{n+\nu}^{\ +}\big(\eta,y\big),\
\psi_{n+\nu)}^{\ -}\big(\eta,y\big)
\right]
\end{eqnarray}
where, by definition, we take
\begin{eqnarray}\label{fi}
\begin{array}{l}
\phi_{n+\nu}(\eta,y)
%
=\frac{e^{iy}}{\Gamma[2n+2\nu+2]}\ 
[2iy]^{n+\nu+1}{\Phi}[n+\nu+1+ i\eta,2n+2\nu+2;- 2iy],\vspace{2mm}\\
%
\displaystyle
\psi_{n+\nu}^{\ \pm}\big(\eta,y\big)=
\frac{\pm 2ie^{\eta\pi}\ e^{\pm iy}}{\Gamma[n+\nu+1\mp i\eta]}
[-2iy]^{n+\nu+1}\Psi[n+\nu+1\pm i\eta,2n+2\nu+2;\mp 2iy].
\end{array}
\end{eqnarray}
In $\psi_{n+\nu}^{\ \pm}$ the irrelevant factors $\pm 2i\exp(\eta\pi)$
are maintained just to connect the above definitions with the
ones used by Leaver. In fact, for $\mathscr{U}_{n+\nu}$  Leaver 
used the functions $F_{n+\nu}(\eta,y)$ and $G_{n+\nu}(\eta,y)$, defined as
\begin{eqnarray}\label{L-1}
\begin{array}{l}
F_{n+\nu}(\eta,y)=
\frac{[\Gamma(n+\nu+1+i\eta)\Gamma(n+\nu+1-i\eta)]^{1/2}}
{2e^{\pi\eta/2}\Gamma(2n+2\nu+2)}\times\vspace{2mm}\\
\hspace{2cm}
e^{iy}
(2y)^{n+\nu+1}{\Phi}(n+\nu+1+ i\eta,2n+2\nu+2;- 2iy),
\end{array}
\end{eqnarray}
\begin{eqnarray}\label{L-2}
\begin{array}{l}
G_{n+\nu}(\eta,y)\pm iF_{n+\nu}(\eta,y)=
e^{\pi\eta/2}e^{\mp i\pi(n+\nu+1/2)}
\left[\frac{\Gamma(n+\nu+1\pm i\eta)}{\Gamma(n+\nu+1\mp i\eta)}\right]^{1/2}\times
\vspace{2mm}\\
\hspace{2.5cm}
e^{\pm iy}
(2y)^{n+\nu+1}{\Psi}\left[n+\nu+1\pm i\eta,2n+2\nu+2;\mp 2iy\right].
\end{array}
\end{eqnarray}
Thus, $\phi_{n+\nu}$ and $\psi_{n+\nu}^{\pm}$ 
are obtained by dividing the above expressions by $\Gamma_n$, defined as
\begin{equation}\label{fator}
\Gamma_n=({1}/{2})e^{-\eta\pi/2}(-i)^{n+\nu+1}
[\Gamma(n+\nu+1+i\eta)\Gamma(n+\nu+1-i\eta)]^{1/2}.
\end{equation}
Inversely, when 
$\phi_{n+\nu}$ and $\psi_{n+\nu}^{\pm}$
are multiplied by $\Gamma_n$, we
recover the Leaver normalization.

>From the properties of the functions
$F_{n+\nu}(\eta,y)$ and $G_{n+\nu}(\eta,y)$ given in Eqs. (126) and (125)
of Leaver's paper,
we find that the functions (\ref{U-n}) satisfy the equations
\begin{eqnarray}\label{coulomb2}
\frac{d\mathscr{U}_{n+\nu}}{dy}
&=&
\frac{i(n+\nu)(n+\nu+1+i\eta)(n+\nu+1-i\eta)}{(n+\nu+1)(2n+2\nu+1)}\
\mathscr{U}_{n+\nu+1}
\nonumber\\
&-&
\frac{\eta}{(n+\nu)(n+\nu+1)}\ \mathscr{U}_{n+\nu}+
\frac{i(n+\nu+1)}{(n+\nu)(2n+2\nu+1)}\ \mathscr{U}_{n+\nu-1}
\end{eqnarray}
and
\begin{eqnarray}\label{coulomb3}
&\frac{(n+\nu)(n+\nu+1+i\eta)(n+\nu+1-i\eta)}
{(2n+2\nu+1)}\ \mathscr{U}_{n+\nu+1}-
\nonumber\\
%
&i\left[\frac{(n+\nu)(n+\nu+1)}{y}+
\eta\right]\mathscr{U}_{n+\nu}
%
-\frac{(n+\nu+1)}{(2n+2\nu+1)}\ \mathscr{U}_{n+\nu-1}=0.
\end{eqnarray}
For $\nu$ and $\eta$ fixed, by dividing all
terms of (\ref{coulomb3}) by $(n^2/2)\mathscr{U}_{n+\nu}$ and letting $n\to\pm\infty$
we find
\begin{eqnarray*}\begin{array}{l}
\left[1+\frac{1}{n}\left(2\nu+\frac{3}{2}\right)\right]
\frac{\mathscr{U}_{n+\nu+1}}{\mathscr{U}_{n+\nu}}-
\frac{2i}{y}\left[1+\frac{1}{n}\left(2\nu+1\right)\right]-
\frac{1}{n^2}
{\left[1+\frac{1}{2n}\right]}
\frac{\mathscr{U}_{n+\nu-1}}{\mathscr{U}_{n+\nu}}=0,
\end{array}
\end{eqnarray*}
whose solutions are
\begin{eqnarray}\label{bessel}
\begin{array}{l}\frac{\mathscr{U}_{n+\nu+1}}{\mathscr{U}_{n+\nu}}
\sim \frac{iy}{2n^2}\left[1-\frac{1}{n}\left(2\nu+\frac{5}{2}\right)\right]
 \
\Leftrightarrow\
\frac{\mathscr{U}_{n+\nu-1}}{\mathscr{U}_{n+\nu}}\sim
-\frac{2in^2}{y}\left[1+\frac{1}{n}\left(2\nu+\frac{1}{2}\right)\right]
\end{array}
\end{eqnarray}
and
\begin{eqnarray}\label{hankel}
\begin{array}{l}
\frac{\mathscr{U}_{n+\nu+1}}{\mathscr{U}_{n+\nu}}\sim \frac{2i}{y}
\left[1-\frac{1}{2n}\right] \
\Leftrightarrow\
\frac{\mathscr{U}_{n+\nu-1}}{\mathscr{U}_{n+\nu}}\sim \frac{y}{2i}
\left[1+\frac{1}{2n}\right],\end{array}
\end{eqnarray}
provided that $y/n^2= 0$ when $n\to\pm\infty$ (this condition is satisfied if $y$ is finite). 
Thus, there are two
possibilities for the ratios between successive Coulomb
functions. By demanding that these relations are valid also
for $\eta=0$, we find only one ratio: (i)  
the first expressions
in (\ref{bessel}) and (\ref{hankel}) hold,
respectively, for
$\phi_{n+\nu}$ and $\psi_{n+\nu}^{\pm}$ when $n\to\infty$,
(ii) the second expression in (\ref{bessel}) is valid for 
the three functions when $n\to-\infty$. In other words,
\begin{eqnarray}\label{bessel-teste}
\begin{array}{lll}
\frac{\phi_{n+\nu+1}}{\phi_{n+\nu}}
\sim \frac{iy}{2n^2}\left[1-\frac{1}{n}\left(2\nu+\frac{5}{2}\right)\right],&
\frac{\psi_{n+\nu+1}^{\pm}}{\psi_{n+\nu}^{\pm}}\sim \frac{2i}{y}\left[1-\frac{1}{2n}\right],&
\quad [n\to \infty],
\vspace{2mm}\\
\frac{\mathscr{U}_{n+\nu-1}}{\mathscr{U}_{n+\nu}}\sim
-\frac{2in^2}{y}\left[1+\frac{1}{n}\left(2\nu+\frac{1}{2}\right)\right],\quad&
\mathscr{U}_{n+\nu}=\left(\phi_{n+\nu}, \psi_{n+\nu}^{\pm}\right),&\quad[n\to -\infty].
\end{array}
\end{eqnarray}

The above conclusions are obtained as follows. 
In the first place, if $\eta=0$
the functions $\phi_{n+\nu}$ and $\psi_{n+\nu}^{\pm}$
can be rewritten in terms of
Bessel functions since 
\cite{erdelyi1} 
\begin{eqnarray}\label{bessel-1}
\begin{array}{l}
\Phi(n+\nu+1,2n+2\nu+2;-2iy)=\Gamma\left[n+\nu+({3}/{2})\right]
\left[{y}/{2}\right]^{-n-\nu-\frac{1}{2}}
e^{-iy}J_{n+\nu+\frac{1}{2}}(y),\vspace{2mm}\\
\Psi(n+\nu+1,2n+2\nu+2;-2iy)=\frac{i\sqrt{\pi}}{2}
e^{-iy+i\pi(n+\nu+\frac{1}{2})}
(2y)^{-n -\nu-\frac{1}{2}}H_{n+\nu+\frac{1}{2}}^{(1)}(y),
\vspace{2mm}\\
\Psi(n+\nu+1,2n+2\nu+2;+2iy)=-\frac{i\sqrt{\pi}}{2}
e^{iy-i\pi(n+\nu+\frac{1}{2})}
(2y)^{-n -\nu-\frac{1}{2}}H_{n+\nu+\frac{1}{2}}^{(2)}(y),
\end{array}
\end{eqnarray}
where $J_{\kappa}$ is the Bessel function of the first kind,
and $H_{\kappa}^{(1)}$ and $H_{\kappa}^{(2)}$ are the first
and the second Hankel functions. Thence
\begin{eqnarray}\label{zero}
\begin{array}{l}
\phi_{n+\nu}(0,y)=
\frac{i^n \ C}{\Gamma[n+\nu+1]}\sqrt{y} J_{n+\nu+\frac{1}{2}}(y),
\quad \psi_{n+\nu}^{\pm}(0,y)=\frac{i^n\ C^{\pm}}{\Gamma[n+\nu+1]}\sqrt{y}
H_{n+\nu+\frac{1}{2}}^{(1,2)}(y)
\end{array}
\end{eqnarray}
where the constants $C$ and $C^{\pm}$ do not depend on $n$.
In the second place, if $y$ is bounded and 
$\kappa\rightarrow \infty$ \cite{arscott}
\begin{eqnarray}\label{Z}
\begin{array}{l}
J_{\kappa}(y)\sim\frac{1}{\Gamma(\kappa+1)}
\left( \frac{y}{2}\right)^{\kappa}, \qquad
 H_{\kappa}^{(1)}(y)
\sim -H_{\kappa}^{(2)}(y)\sim-\frac{i}{\pi}\Gamma(\kappa)
\left( \frac{2}{y}\right)^{\kappa}.\end{array}
\end{eqnarray}
Combining (\ref{zero}) with (\ref{Z}), we 
establish (\ref{bessel-teste})
for $\kappa=n+\nu+(1/2)$ when $n\to\infty$ ($\eta=0$). 
On the other side, if $\kappa\to -\infty$,
we use the previous
relations for $H_{\kappa}^{(1,2)}(y)$ in conjunction with
\cite{nist}
\begin{equation}\label{Z-2}
\begin{array}{l}
H_{-\kappa}^{(1)}(y)=e^{i\pi\kappa}H_{\kappa}^{(1)}(y),\qquad
H_{-\kappa}^{(2)}(y)=e^{-i\pi\kappa}H_{\kappa}^{(2)}(y),\quad
%
\end{array}
\end{equation}
Thus, we find (\ref{bessel-teste}) for $\mathscr{U}_{n+\nu}=\psi_{n+\nu}^{\pm}$
when $\kappa=n+\nu+(1/2)$ with $n\to -\infty$ ($\eta=0$).
For $\mathscr{U}_{n+\nu}=\phi_{n+\nu}$, if $y$ is bounded
and $\kappa\to -\infty$, once more
we use the relation given in (\ref{Z}) for $\displaystyle J_{\kappa}(y)$
since
\begin{eqnarray}
J_{\kappa}(y)&=&\left(\frac{y}{2}\right)^{\kappa}\sum_{m=0}^{\infty}
\frac{(-1)^m}{m \ ! \ \Gamma(\kappa+m+1)}\left(\frac{y}{2}\right)^{2m}\nonumber\\
&=&\left(\frac{y}{2}\right)^{\kappa}\left[\frac{1}{\Gamma(\kappa+1)}+\sum_{m=1}^{\infty}
\frac{(-1)^m}{m \ ! \ \Gamma(\kappa+m+1)}\left(\frac{y}{2}\right)^{2m}\right].
\end{eqnarray}
In this manner, we establish the ratio (\ref{bessel-teste}) for
the three Coulomb functions when $n\to-\infty$.
%

%
%
%
%
%
%
%
%
%
%
%
%
\section{Recurrence Relations for the Series Coefficients}
\protect\label{B}
\setcounter{equation}{0}
\renewcommand{\theequation}{B.\arabic{equation}}
Now we present the derivation 
of the recurrence relations for the series coefficients
of the two-sided solutions 
$\mathbb{U}_1(z)$ of the CHE, as well as of 
the one-sided solutions  $\mathbb{\mathring{U}}_1(z)$.  
The relations for the other sets of solutions may be obtained 
from these by transformations of variables. 
Notice that: (i) the derivation is formal   
in the sense that, in each series, we suppose linear independence 
of all Coulomb wave functions; (ii) the importance
of the three types of relations for one-sided solutions
becomes apparent in the cases of 
the Whittaker-Hill and Mathieu equations. Details are afforded 
throughout the paper.

The Leaver substitutions \cite{leaver}
\begin{eqnarray}
U(z)=z^{-B_2/2}H(y),\qquad y=\omega z
\end{eqnarray}
transform the CHE (\ref{gswe}) into
\begin{eqnarray}
&&\begin{array}{l}
y(y-\omega z_0)\left[\frac{d^2H}{dy^2}+\left(1-\frac{2\eta}{y}\right)H\right]+
\mathsf{C}_1\omega\frac{dH}{dy}+\left[\mathsf{C}_2+\frac{\mathsf{C}_3\omega}{y}\right]H=0
\quad \mbox{ where}\end{array}\\
&&\begin{array}{l}
\mathsf{C}_1=B_1+B_2z_0,\qquad \mathsf{C}_2=B_3-\frac{B_2}{2}\left[\frac{B_2}{2}-1\right],
\qquad
\mathsf{C}_3=-\frac{B_2z_0}{2}\left[1+\frac{B_2}{2}+\frac{B_1}{z_0}\right].
\end{array}\nonumber
\end{eqnarray}
Expanding $H(y)$ as
\begin{eqnarray}\label{H}
H(y) =\displaystyle\sum_{n=-\infty}^{\infty} b_{n}^{1}
\mathscr{U}_{n+\nu}\left(\eta,y\right)
\quad\Leftrightarrow\quad 
\mathbb{U}_1(z)=z^{-\frac{B_2}{2}}\sum_{n=-\infty}^{\infty} b_{n}^{1}
\mathscr{U}_{n+\nu}\left(\eta,y\right)
\end{eqnarray}
and using Eqs. (\ref{coulomb1}), (\ref{coulomb2}) and (\ref{coulomb3}) we find
\begin{eqnarray}\label{leaver-1}
\displaystyle\sum_{n=-\infty}^{\infty} b_{n}^{1}
\big[ \alpha_{n-1}^{1}\mathscr{U}_{n+\nu-1}(\eta,y)
+\beta_{n} ^{1}\mathscr{U}_{n+\nu}(\eta,y)+
\gamma_{n+1}^{1} \mathscr{U}_{n+\nu+1}(\eta,y)\big]
=0,
\end{eqnarray}
where $\alpha_n^{(1)}$, $\beta_n^{(1)}$ and $\gamma_n^{(1)}$
are defined in Eqs. (\ref{nu10}).

If $\nu$ is such that the summation runs from minus to 
plus infinity, the
preceding equation takes the form
\begin{eqnarray}\label{leaver-2}
\displaystyle\sum_{n=-\infty}^{\infty}
\left[ \alpha_{n}^{1}\ b_{n+1}^{1}+\beta_{n}^{1}\ b_{n}^{1}+
\gamma_{n}^{1}\ b_{n-1}^{1} \right]\mathscr{U}_{n+\nu}(\eta,y)
=0,
\end{eqnarray}
which is satisfied by the three-term recurrence relations (\ref{recursion-1}) 
provided that all the functions $\mathscr{U}_{n+\nu}(\eta,y) $
are linearly independent. 

On the other hand, the series summation 
can begin at $n=0$ if $\nu$ is chosen 
conveniently. In this case Eq. (\ref{leaver-1})
can be rewritten as
\begin{eqnarray*}\label{r00}
\displaystyle\sum_{m=-1}^{\infty}
 \alpha_{m}^{1}b_{m+1}^{1}\mathscr{U}_{m+\nu}(\eta,y)
+\sum_{m=0}^{\infty}\beta_{m}^{1}b_{m}^{1} \mathscr{U}_{m+\nu}(\eta,y)+
\sum_{m=1}^{\infty}\gamma_{m}^{1} b_{m-1}^{1}\mathscr{U}_{m+\nu}(\eta,y)
=0
\end{eqnarray*}
or, equivalently, as
\begin{eqnarray}\label{r0}
&&\alpha_{-1}^{1} b_{0}^{1} \mathscr{U}_{\nu-1}(\eta,y)+
\left[\alpha_{0}^{1} b_{1}^{1}+\beta_{0}^{1} b_{0}^{1}\right]\mathscr{U}_{\nu}(\eta,y) 
%
+\left[\alpha_{1}^{1} b_{2}^{1}+\beta_{1}^{1} b_{1}^{1}+
\gamma_{1}^{1} b_{0}^{1}\right]\mathscr{U}_{\nu+1}(\eta,y)\nonumber\\
&&+\displaystyle\sum_{n=2}^{\infty}
\left[\alpha_{n}^{1} b_{n+1}^{1}+\beta_{n}^{1} b_{n}^{1}+
\gamma_{n}^{1} b_{n-1}^{1}\right]
\mathscr{U}_{n+\nu}(\eta,y)=0.
\end{eqnarray}
If $\nu$ is such that $\alpha_{-1}^1\mathscr{U}_{\nu-1}=0$,
Eq. (\ref{r0}) is obviously satisfied by the recurrence relations
(\ref{r1a}). If $\alpha_{-1}^1\mathscr{U}_{\nu-1}\neq 0$, the 
above equation is also fulfiled when
$\mathscr{U}_{\nu-1}$ and $\mathscr{U}_{\nu+1}$, or 
$\mathscr{U}_{\nu-1}$ and $\mathscr{U}_{\nu}$, are linearly dependent:
these cases give the recurrence relations (\ref{r2a}) and 
(\ref{r3a}), respectively,
as we will see. 

In effect, from (\ref{nu10}), 
in  $\alpha_{-1}^{1}\mathscr{U}_{\nu-1}$ we have
\begin{eqnarray}\label{Alfa-1}
\begin{array}{l}
\alpha_{-1}^{1}=i\omega z_{0} \left[\nu+1-\frac{B_{2}}{2}\right]
\left[\nu-\frac{B_{1}}{z_{0}}-\frac{B_{2}}{2}\right]
\Big/ \left[2\nu\left(\nu+\frac{1}{2}\right)\right].
\end{array}
\end{eqnarray}
If there is no cancellation between numerators and denominators,
we find  $\alpha_{-1}^1=0$ ($\alpha_{-1}^{1}\mathscr{U}_{\nu-1}=0$)
by choosing 
\begin{eqnarray}
\nu=\nu_1=({B_{2}}/{2})-1\quad \text{or}\quad
\nu=\nu_2=({B_{1}}/z_{0})+({B_{2}}/{2}).
\end{eqnarray}
Let us consider only $\nu=\nu_1$ corresponding to
the solutions $
\mathring{\mathbb{U}}_1$ given in Eq. (\ref{forma-de-leaver-truncada})
(for $\nu=\nu_2$, we get  
$\mathring{\mathbb{U}}_2= T_1\mathring{\mathbb{U}}_1$). 
Then, we find that 
\begin{eqnarray}
\nu=\nu_1=\frac{B_{2}}{2}-1\quad \Rightarrow\quad
 \alpha_{-1}^{1}=
\begin{cases}
0, \mbox{ if }\ B_{2}\neq 1,\ 2;\vspace{2mm}\\
i\omega z_0\left[1+({B_1}/{z_0})\right],\text{ if } B_2=1;
\vspace{2mm}\\
-i\omega z_0\left[1+({B_1}/{z_0})\right], \text{ if } B_2=2,
\end{cases}
\end{eqnarray}
whereof we see that, in general, 
$\alpha_{-1}^{1}\mathscr{U}_{\nu-1}\neq 0$ if
$B_2=1$ or $B_2=2$ -- for these cases there are
cancellations between numerators and denominators
in (\ref{Alfa-1}). Thus, we find that Eq. (\ref{r0})
is satisfied by the recurrence relations 
(\ref{r1a}) for $b_n^{1}$ if $B_2\neq 1,2$  
(${b}_n^{1}=\mathring{b}_n^{1}$ for one-sided solutions). 
Next we will find  the relations
(\ref{r2a}) if $B_2=1 $, and (\ref{r3a}) if $B_2=2$.
In the first place, for the Coulomb 
functions (\ref{fi}) we have
\begin{eqnarray*}\label{coulomb-fim}
\begin{array}{l}
\phi_{n+\nu}(\eta,y)\Big|_{\nu=(B_2/2)-1}
=\frac{e^{iy}}{\Gamma[2n+B_2]} \
[2iy]^{n+\frac{B_2}{2}}{\Phi}\left[n+\frac{B_2}{2}+ i\eta,2n+B_2;- 2iy\right],
\vspace{2mm}\\
%
\psi_{n+\nu}^{\pm}\big(\eta,y\big)\Big|_{\nu=(B_2/2)-1}=
\frac{\pm 2ie^{\eta\pi} e^{\pm iy}}{\Gamma[n+(B_2/2)\mp i{\eta}]}\
[-2iy]^{n+\frac{B_2}{2}}\Psi\left[n+\frac{B_2}{2}\pm i{\eta},2n+B_2;\mp2iy\right],
\end{array}
\end{eqnarray*}
where $y=\omega z$. Using the relations ($m=1,2,3,\cdots$)
\begin{eqnarray}\label{App-B}
\begin{array}{l}\displaystyle 
\lim_{c \rightarrow 1-m}\frac{\Phi(a,c;y)}{\Gamma(c)}=
\frac{\Gamma(a+m)}{\Gamma(a)\Gamma(m+1)}\  y^{m}\Phi(a+m,1+m;y)  \text{ and}
\vspace{2mm}\\
\Psi(a,1-m;y)=y^{m}\Psi(a+m,1+m,y)
\end{array}
\end{eqnarray}
we find that if $ B_2=1$ ($\nu=-1/2$)
\begin{equation*}\displaystyle
\lim_{m\to -1}\phi_{m-\frac{1}{2}}=-\left[\eta^2+\frac{1}{4}\right]\phi_{\frac{1}{2}}, \quad
\psi_{-\frac{3}{2}}^{\pm}=-\left[\eta^2+\frac{1}{4}\right]\psi_{\frac{1}{2}}^{\pm}
\ \Rightarrow \ \mathscr{U}_{\nu-1}=-\left[\eta^2+\frac{1}{4}\right]\mathscr{U}_{\nu+1}.
\end{equation*}
Therefore, if $B_2=1$, the first and the third terms in Eq. (\ref{r0})
are linearly dependent leading to the recurrence relations (\ref{r2a}). 
In the second place, if $B_2=2$ ($\nu=0$) we find
\begin{equation*}\displaystyle
\lim_{m\to -1}\phi_{m}=-i\eta\phi_{0}, \quad
\psi_{-1}^{\pm}=-i\eta\psi_{0}^{\pm}
\ \Rightarrow \ \mathscr{U}_{\nu-1}=-i\eta\mathscr{U}_{\nu},
\quad[\ B_2=2,\ \nu=0]
\end{equation*}
that is, if $B_2=2$ the first and the second terms in Eq. (\ref{r0})
are linearly dependent giving the recurrence relations (\ref{r3a}).
The previous limits suggest that $\mathscr{U}_{-1}=0$ 
if $\eta=0$ and, consequently, relations (\ref{r3a}) 
reduce to (\ref{r1a}). In fact, by putting $\nu=0$ and 
$B_2=2$ in Eqs. (\ref{zero}) we find $\mathscr{U}_{-1}=0$.

\section{Other One-Sided Solutions for the CHE}
\protect\label{C}
\setcounter{equation}{0}
\renewcommand{\theequation}{C.\arabic{equation}}
Now we complete the list of the one-sided solutions
for the CHE given in Sec. 3. First we write the solutions for $\eta\neq 0$ and,
by supposing that the series summation begins at $n=0$ we give
the conditions which assures the independence of the 
Coulomb wave functions as well as the recurrence relations for
the series coefficients. For series beginning at $n> 0$, 
see (\ref{N>1}) and (\ref{N2>1}). After this, the 
solutions for the case $\eta=0$ are expressed by series of Bessel functions.

%
\subsection{Solutions for $\eta\neq 0$}
{\it Second and sixth sets}: 
\begin{eqnarray*}\begin{array}{l}
\text{Conditions: }B_2+\frac{2B_1}{z_0}\neq -2,-{3},-4\cdots ;\vspace{2mm}\\
\text{Eqs. (\ref{r1a})} \text{ if } \frac{B_2}{2}+\frac{B_1}{z_0}\neq -\frac{1}{2},0; \quad  
\text{(\ref{r2a})} \text{ if } \frac{B_2}{2}+\frac{B_1}{z_0}=-\frac{1}{2}; 
\quad
\text{(\ref{r3a})} \text{ if } \frac{B_2}{2}+\frac{B_1}{z_0}= 0;
\end{array}
\end{eqnarray*}
%
%
%
\begin{eqnarray}
\label{par2}
\begin{array}{l}
\mathring{U}_{ 2}(z) =e^{i\omega z}z^{1+\frac{B_1}{z_0}}\displaystyle\sum_{n=0}^{\infty}
\begin{array}{l}
\frac{\mathring{b}_{n}^{2}\ [2i\omega z]^{n}}{\Gamma\left[2n+2+B_2+\frac{2B_1}{z_0}\right]}\end{array}
\times \\
\hspace{3cm}
\begin{array}{l}
\Phi\left[n+i\eta+1+\frac{B_{2}}{2}+\frac{B_1}{z_0},2n+2+B_2+\frac{2B_1}{z_0};-2i\omega z\right],\end{array}\vspace{2mm}\\
\mathring{U}_{2}^{\pm}(z) =e^{\pm i\omega z}z^{1+\frac{B_1}{z_0}}\displaystyle\sum_{n=0}^{\infty}
\begin{array}{l}
\frac{\mathring{b}_{n}^{2}\ [-2i\omega z]^{n}}{\Gamma\left[n\mp i\eta+1+\frac{B_2}{2}+\frac{B_1}{z_0}\right]}\end{array}
\times \\
\hspace{3cm}\begin{array}{l}
\Psi\left[n\pm i\eta+1+\frac{B_{2}}{2}+\frac{B_1}{z_0},2n+2+B_{2}+\frac{2B_1}{z_0};\mp 2i\omega z\right], \end{array}
\end{array} 
%
\end{eqnarray}
\begin{eqnarray*}
\begin{array}{l}
\mathring{\alpha}_{n}^{2}  =  \frac{2i\omega z_{0}[n+1]
\left[n+2+\frac{B_{1}}{z_{0}}\right]}
{\left[2n+2+{B_{2}}+\frac{2B_{1}}{z_{0}}\right]
\left[2n+3+{B_{2}}+\frac{2B_{1}}{z_{0}}
\right]},
\vspace{.2cm} \\
%
\mathring{\beta}_{n}^{2}  =-  
\left[n+1+\frac{B_{1}}{z_{0}}\right]
\left[n+B_{2}+\frac{B_{1}}{z_{0}}\right]-B_{3}-\eta \omega z_{0}-
\frac{\eta \omega z_{0}\left[B_{2}-2\right]
\left[B_{2}+\frac{2B_{1}}{z_{0}}\right]}
{\left[2n+{B_{2}}+\frac{2B_{1}}{z_{0}}\right]
\left[2n+2+B_{2}+\frac{2B_{1}}{z_{0}}\right]},
\vspace{.2cm} \\
%
\mathring{\gamma}_{n}^{2}  = -\frac{2i\omega z_{0}\left[n+B_{2}+
\frac{B_{1}}{z_{0}}-1\right]
\left[n+B_{2}+\frac{2B_{1}}{z_{0}}\right]
\left[n+\frac{B_{2}}{2}+\frac{B_{1}}{z_{0}}+i\eta\right]
\left[n+\frac{B_{2}}{2}+\frac{B_{1}}{z_{0}}-i\eta \right]}
{\left[2n-1+B_{2}+\frac{2B_{1}}{z_{0}}\right]\left[2n+B_{2}+\frac{2B_{1}}{z_{0}}\right]}.
\end{array}
\end{eqnarray*}
%

%
%
\begin{eqnarray}
\begin{array}{l}\displaystyle
\mathring{U}_{6}(z) =e^{i\omega z}z^{1+\frac{B_1}{z_0}}\displaystyle\sum_{n=0}^{\infty}
\begin{array}{l}
\frac{\mathring{b}_{n}^{6}\ \left[2i\omega (z-z_0)\right]^{n}}{\Gamma\left[2n+2+B_2+\frac{2B_1}{z_0}\right]}\end{array}
\times \\
\hspace{2,6cm}
\begin{array}{l}
\Phi\left[n+i\eta+1+\frac{B_{2}}{2}+\frac{B_1}{z_0},2n+2+B_2+\frac{2B_1}{z_0};-2i\omega (z-z_0)\right]
,\end{array}\vspace{2mm} \\
\displaystyle
\mathring{U}_{6}^{\pm}(z) =e^{\pm i\omega z}z^{1+\frac{B_1}{z_0}}\displaystyle\sum_{n=0}^{\infty}
\begin{array}{l}
\frac{\mathring{b}_{n}^{6}\ \left[-2i\omega (z-z_0)\right]^{n}}
{\Gamma\left[n\mp i\eta+1+\frac{B_2}{2}+\frac{B_1}{z_0}\right]}\end{array}
\times \\
\hspace{2,6cm}
\begin{array}{l}
\Psi\left[n\pm i\eta+1+\frac{B_{2}}{2}+\frac{B_1}{z_0},2n+2+B_2+\frac{2B_1}{z_0};\mp 2i\omega (z-z_0)\right]\end{array},
\end{array}
\end{eqnarray}
\begin{eqnarray*}
\begin{array}{l}
\mathring{\alpha}_{n}^{6}=-
\frac{2i\omega z_0\left[ n+1\right] 
\left[n+B_2+\frac{B_1}{z_0}\right]}
{\left[2n+2+B_2+\frac{2B_1}{z_0}\right]\left[2n+3+B_2+\frac{2B_1}{z_0}\right]},\qquad 
\mathring{\beta}_{n}^{6}=\mathring{\beta}_{ n}^{2},\vspace{2mm}\\
%
\mathring{\gamma}_{n}^{6}=\frac{2i\omega z_0
\left[n+B_2+\frac{2B_1}{z_0}\right]
\left[ n+1+\frac{B_1}{z_0}\right]\left[ n+i\eta+\frac{B_1}{z_0}+\frac{B_2}{2}\right]
\left[ n-i\eta +\frac{B_1}{z_0}+\frac{B_2}{2}\right]}
{\left[2n-1+B_2+\frac{2B_1}{z_0}\right]\left[2n+B_2+\frac{2B_1}{z_0}\right]}.
\end{array}
\end{eqnarray*}
%

%
%
%
{\it Third and seventh sets}: 
\begin{eqnarray*}\begin{array}{l}
\text{Conditions: }B_2\neq 4,5,6,\cdots.\vspace{2mm}\\
\text{Eqs. (\ref{r1a})} \text{ if } B_2 \neq  2,3; \qquad
 \text{(\ref{r2a})} \text{ if } B_2=3; \qquad
 \text{(\ref{r3a})} \text{ if } B_2=2.
\end{array}
\end{eqnarray*}
%
%
%
\begin{eqnarray}
\label{par3}
\begin{array}{l}
\mathring{U}_{ 3}(z) =e^{i\omega z}z^{1+\frac{B_1}{z_0}}(z-z_0)^{1-B_2-\frac{B_1}{z_0}}
\displaystyle\sum_{n=0}^{\infty}
\begin{array}{l}
\frac{\mathring{b}_{n}^{3}\ [2i\omega z]^{n}}{\Gamma[2n+4-B_2]} \times
\end{array} \\
\hspace{5.5cm}
\begin{array}{l}
\Phi\left[n+i\eta+2-\frac{B_{2}}{2},2n+4-B_2;-2i\omega z\right],
\end{array}\vspace{2mm}\\
\mathring{U}_{3}^{\pm}(z) =e^{\pm i\omega z}z^{1+\frac{B_1}{z_0}}(z-z_0)^{1-B_2-\frac{B_1}{z_0}}\displaystyle\sum_{n=0}^{\infty}
\begin{array}{l}
\frac{\mathring{b}_{n}^{3}\ [-2i\omega z]^{n}}
{\Gamma\left[n\mp i\eta+2-\frac{B_2}{2}\right]}
\times\end{array} \\
\hspace{5.5cm}\begin{array}{l}
\Psi\left[n\pm i\eta+2-\frac{B_2}{2},2n+4-B_2;\mp 2i\omega z\right], 
\end{array}\end{array} 
\end{eqnarray}
\begin{eqnarray*}
\begin{array}{l}
\mathring{\alpha}_{n}^{3} =  \frac{2i\omega z_{0}[n+1]
\left[n+2+\frac{B_{1}}{z_{0}}\right]}
{\left(2n+4-B_{2}\right)\left(2n+5-
B_{2}\right)},
\vspace{.2cm} \\
%
\mathring{\beta}_{ n}^{3}  =  -(n+1)(n+2-B_{2})-B_{3}-\eta \omega z_{0}
-\frac{\eta \omega z_{0}\left[B_{2}-2\right]
\left[B_{2}+\frac{2B_{1}}{z_{0}}\right]}
{\left(2n+2-B_{2}\right)\left(2n+4-
B_{2}\right)},
\vspace{.3cm} \\
%
\mathring{\gamma}_{n}^{3}  = -\frac{2i\omega z_{0}\left[n+2-B_{2}\right]
\left[n+1-B_{2}-\frac{B_{1}}{z_{0}}\right]
\left[n+1-\frac{B_{2}}{2}+i\eta\right]\left[n+1-\frac{B_{2}}{2}-i\eta \right]}
{\left(2n+1-B_{2}\right)\left(2n+2-B_{2}\right)}.
\end{array}
\end{eqnarray*}
%
%
%
%
\begin{eqnarray}
\label{par7}
\begin{array}{l}
\mathring{U}_{7}(z) =e^{i\omega z}z^{1+\frac{B_1}{z_0}}(z-z_0)^{1-B_2-\frac{B_1}{z_0}}\displaystyle
\sum_{n=0}^{\infty}
\begin{array}{l}
\frac{\mathring{b}_{n}^{7}\ \left[2i\omega (z-z_0)\right]^{n}}{\Gamma[2n+4-B_2]}\end{array}
\times \\
\hspace{4,5cm}\begin{array}{l}
\Phi\left[n+i\eta+2-\frac{B_{2}}{2},2n+4-B_2;-2i\omega (z-z_0)\right],
\end{array}
\vspace{2mm}\\
\mathring{U}_{7}^{\pm}(z) =e^{\pm i\omega z}z^{1+\frac{B_1}{z_0}}
(z-z_0)^{1-B_2-\frac{B_1}{z_0}} \displaystyle\sum_{n=0}^{\infty}
\begin{array}{l}
\frac{\mathring{b}_{n}^{6}\ \left[-2i\omega (z-z_0)\right]^{n}}
{\Gamma\left[n\mp i\eta+2-\frac{B_2}{2}\right]}
\times \end{array}\\
\hspace{4,5cm}
\begin{array}{l}
\Psi\left[n\pm i\eta+2-\frac{B_{2}}{2},2n+4-B_2;\mp 2i\omega (z-z_0)\right],
\end{array}
\end{array}
\end{eqnarray}
\begin{eqnarray*}
\begin{array}{l}
\mathring{\alpha}_{n}^{7} = -\frac{2i\omega z_{0}[n+1]
\left[n+2-B_2-\frac{B_{1}}{z_{0}}\right]}
{\left(2n+4-B_{2}\right)\left(2n+5-
B_{2}\right)},
\qquad
\mathring{\beta}_{ n}^{7}  =  \mathring{\beta}_{n}^{3},
\vspace{.2cm} \\
%
\mathring{\gamma}_{n}^{7}  = \frac{2i\omega z_{0}\left[n+2-B_{2}\right]
\left[n+1+\frac{B_{1}}{z_{0}}\right]
\left[n+1-\frac{B_{2}}{2}+i\eta\right]\left[n+1-\frac{B_{2}}{2}-i\eta \right]}
{\left(2n+1-B_{2}\right)\left(2n+2-B_{2}\right)}.
\end{array}
\end{eqnarray*}
%
%
%
%
{\it Fourth and eighth sets}:
\begin{eqnarray*}\begin{array}{l}
\text{Conditions: }B_2+\frac{2B_1}{z_0}\neq 2,{3},4\cdots.\vspace{2mm}\\
\text{Eqs. (\ref{r1a})} \text{ if } \frac{B_2}{2}+\frac{B_1}{z_0}\neq 0,\frac{1}{2}; \quad  
\text{(\ref{r2a})} \text{ if } \frac{B_2}{2}+\frac{B_1}{z_0}=\frac{1}{2}; 
\quad
\text{(\ref{r3a})} \text{ if } \frac{B_2}{2}+\frac{B_1}{z_0}= 0;
\end{array}
\end{eqnarray*}
%
%
%
\begin{eqnarray}
\label{par4}
\begin{array}{l}
\mathring{U}_{ 4}(z) =e^{i\omega z}(z-z_0)^{1-B_2-\frac{B_1}{z_0}}\displaystyle\sum_{n=0}^{\infty}
\begin{array}{l}
\frac{\mathring{b}_{n}^{4}\ [2i\omega z]^{n}}{\Gamma[2n+2-B_2-\frac{2B_1}{z_0}]}\times\end{array} \\
\hspace{3cm}
\begin{array}{l}
\Phi\left[n+i\eta+1-\frac{B_{2}}{2}-\frac{B_1}{z_0},2n+2-B_2-\frac{2B_1}{z_0};-2i\omega z\right],\end{array}\vspace{2mm}\\
\mathring{U}_{4}^{\pm}(z) =e^{\pm i\omega z}(z-z_0)^{1-B_2-\frac{B_1}{z_0}}\displaystyle\sum_{n=0}^{\infty}
\begin{array}{l}
\frac{\mathring{b}_{n}^{4}\ [-2i\omega z]^{n}}{\Gamma\left[n\mp i\eta+1-\frac{B_2}{2}-\frac{B_1}{z_0}\right]}
\times\end{array} \\
\hspace{3cm}
\begin{array}{l}
\Psi\left[n\pm i\eta+1-\frac{B_{2}}{2}-\frac{B_1}{z_0},2n+2-B_2-\frac{2B_1}{z_0};\mp 2i\omega z\right],\end{array} 
\end{array} \end{eqnarray}
\begin{eqnarray*}
\begin{array}{l}
\mathring{\alpha}_{n}^{ 4}  =  \frac{2i\omega z_{0}[n+1]
\left[n-\frac{B_{1}}{z_{0}}\right]}
{\left[2n+2-B_{2}-\frac{2B_{1}}{x_{0}}\right]
\left[2n+3-B_{2}-\frac{2B_{1}}{z_{0}}
\right]},
\vspace{.2cm} \\
%
\mathring{\beta}_{n}^{ 4}  =  -
\left[n-\frac{B_{1}}{z_{0}}\right]
\left[n+1-B_{2}-\frac{B_{1}}{z_{0}}\right]-B_{3}-\eta \omega z_{0}
-\frac{\eta \omega z_{0}\left[B_{2}-2\right]
\left[B_{2}+\frac{2B_{1}}{z_{0}}\right]}
{\left[2n-B_{2}-\frac{2B_{1}}{z_{0}}\right]
\left[2n+2-B_{2}-\frac{2B_{1}}{z_{0}}\right]},
\vspace{.2cm} \\
%
\mathring{\gamma}_{n}^{4}= -\frac{2i\omega z_{0}\left[n+1-B_{2}-\frac{B_{1}}{z_{0}}\right]
\left[n-B_{2}-\frac{2B_{1}}{z_{0}}\right]
\left[n-\frac{B_{1}}{z_{0}}-\frac{B_{2}}{2}+i\eta\right]
\left[n-\frac{B_{1}}{z_{0}}-\frac{B_{2}}{2}-i\eta \right]}
{\left[2n-1-B_{2}-\frac{2B_{1}}{z_{0}}\right]
\left[2n-B_{2}-\frac{2B_{1}}{z_{0}}\right]}.
\end{array}
\end{eqnarray*}
%
%
%
\begin{eqnarray}
\label{par8}
\begin{array}{l}
\mathring{U}_{8}(z) =e^{i\omega z}(z-z_0)^{1-B_2-\frac{B_1}{z_0}}\displaystyle\sum_{n=0}^{\infty}
\begin{array}{l}
\frac{\mathring{b}_{n}^{8}\ \left[2i\omega (z-z_0)\right]^{n}}{\Gamma\left[2n+2-B_2-\frac{2B_1}{z_0}\right]}
\times \end{array}\\
\hspace{2cm}
\begin{array}{l}
\Phi\left[n+i\eta+1-\frac{B_{2}}{2}-\frac{B_1}{z_0},2n+2-B_2-\frac{2B_1}{z_0};-2i\omega (z-z_0)\right],\end{array}
\vspace{2mm}\\
\mathring{U}_{8}^{\pm}(z) =e^{\pm i\omega z}(z-z_0)^{1-B_2-\frac{B_1}{z_0}}\displaystyle\sum_{n=0}^{\infty}
\begin{array}{l}
\frac{\mathring{b}_{n}^{6}\ \left[2i\omega (z_0-z)\right]^{n}}
{\Gamma\left[n\mp i\eta+1-\frac{B_2}{2}-\frac{B_1}{z_0}\right]}
\times\end{array} \\
\hspace{2cm}
\begin{array}{l}
\Psi\left[n\pm i\eta+1-\frac{B_2}{2}-\frac{B_1}{z_0},2n+2-B_2-\frac{2B_1}{z_0};\mp 2i\omega (z-z_0)\right],\end{array}
\end{array}
\end{eqnarray}
\begin{eqnarray*}
\begin{array}{l}
\mathring{\alpha}_{n}^{ 8}  = -\frac{2i\omega z_{0}[n+1]
\left[n+2-B_2-\frac{B_{1}}{z_{0}}\right]}
{\left[2n+2-B_{2}-\frac{2B_{1}}{x_{0}}\right]
\left[2n+3-B_{2}-\frac{2B_{1}}{z_{0}}
\right]},
\qquad
\mathring{\beta}_{n}^{8}  =  \mathring{\beta}_{n}^{4}  ,
\vspace{.2cm} \\
%
\mathring{\gamma}_{n}^{8}  = \frac{2i\omega z_{0}\left[n-1-\frac{B_{1}}{z_{0}}\right]
\left[n-B_{2}-\frac{2B_{1}}{z_{0}}\right]
\left[n-\frac{B_{1}}{z_{0}}-\frac{B_{2}}{2}+i\eta\right]
\left[n-\frac{B_{1}}{z_{0}}-\frac{B_{2}}{2}-i\eta \right]}
{\left[2n-1-B_{2}-\frac{2B_{1}}{z_{0}}\right]
\left[2n-B_{2}-\frac{2B_{1}}{z_{0}}\right]}.
\end{array}
\end{eqnarray*}
%

%
\subsection{Solutions for $\eta=0$}
The following solutions for the CHE with $\eta=0$ were used 
in Sec. 3 to discuss the one-sided solutions
for the spheroidal equation.

%
%
%
%
\noindent
{\it Second and sixth sets}: 
\begin{eqnarray*}\begin{array}{l}
\text{Conditions: }B_2+\frac{2B_1}{z_0}\neq -2,-{3},-4\cdots ;\vspace{2mm}\\
\text{Eqs. (\ref{R1a})} \text{ if } \frac{B_2}{2}+\frac{B_1}{z_0}\neq -\frac{1}{2}, \qquad  
\text{(\ref{R2a})} \text{ if } \frac{B_2}{2}+\frac{B_1}{z_0}=-\frac{1}{2}.
\end{array}
\end{eqnarray*}
\begin{eqnarray}\label{S-2}
\mathring{U}_{\ 2}^{(j)}(z)=z^{\frac{1}{2}-\frac{B_2}{2}}
\sum_{n=0}^{\infty}\mathring{a}_{ n}^{2}
Z_{n+\frac{B_2}{2}+\frac{1}{2}+\frac{B_1}{z_0}}^{(j)}(\omega z),
\quad\text{ with}
\end{eqnarray}
\begin{equation*}
\begin{array}{l}
\mathring{\alpha}_{n}^{2}=\frac{\omega z_0\left[ n+1\right]\left[n+\frac{B_1}{z_0}+2\right]}
{\left[2n+3+B_2+\frac{2B_1}{z_0}\right]},\qquad
\mathring{\beta}_{ n}^{2}=-\left[n+1+\frac{B_1}{z_0}\right]\left[n+B_2+\frac{B_1}{z_0}\right]- B_3,\vspace{2mm}\\
\mathring{\gamma}_{n}^{2}=\frac{\omega z_0\left[n+B_2+\frac{2B_1}{z_0}\right]\left[ n+B_2-1+\frac{B_1}{z_0}\right]}
{\left[2n-1+B_2+\frac{2B_1}{z_0}\right]}.
\end{array}
\end{equation*}
\begin{eqnarray}\label{S-6}
\mathring{U}_{\ 6}^{(j)}(z)=z^{1+\frac{B_1}{z_0}}(z-z_0)^{-\frac{1}{2}-\frac{B_2}{2}-\frac{B_1}{z_0}}
\sum_{n=0}^{\infty}\mathring{a}_{n}^{6}
Z_{n+\frac{B_2}{2}+\frac{1}{2}+\frac{B_1}{z_0}}^{(j)}\left[\omega (z-z_0)\right],
\quad\text{with}
\end{eqnarray}
\begin{equation*}
\begin{array}{l}
\mathring{\alpha}_{ n}^{6}=\frac{-\omega z_0\left[ n+1\right]\left[n+B_2+\frac{B_1}{z_0}\right]}
{\left[2n+3+B_2+\frac{2B_1}{z_0}\right]},\qquad
\mathring{\beta}_{ n}^{6}=\mathring{\beta}_{n}^{2},\qquad
\mathring{\gamma}_{n}^{6}=\frac{-\omega z_0\left[n+B_2+\frac{2B_1}{z_0}\right]\left[ n+1+\frac{B_1}{z_0}\right]}
{\left[2n-1+B_2+\frac{2B_1}{z_0}\right]}.
\end{array}
\end{equation*}
%
%
%
%
{\it Third and seventh sets}: 
\begin{eqnarray*}\begin{array}{l}
\text{Conditions: }B_2\neq 4,5,6,\cdots.\vspace{2mm}\\
\text{Eqs. (\ref{R1a})} \text{ if } B_2 \neq  3,\qquad
 \text{(\ref{R2a})} \text{ if } B_2=3.
\end{array}
\end{eqnarray*}
\begin{eqnarray}
\mathring{U}_{\ 3}^{(j)}(z)=(z-z_0)^{1-B_2-\frac{B_1}{z_0}}z^{\frac{B_1}{z_0}+\frac{B_2}{2}-\frac{1}{2}}
\sum_{n=0}^{\infty}\mathring{a}_{n}^{3}
Z_{n-\frac{B_2}{2}+\frac{3}{2}}^{(j)}(\omega z),
\quad\text{with}
\end{eqnarray}
\begin{equation*}
\begin{array}{l}
\mathring{\alpha}_{n}^{3}=\frac{\omega z_0\left[ n+1\right]\left[n+\frac{B_1}{z_0}+2\right]}
{\left(2n+5-B_2\right)},\qquad
\mathring{\beta}_{n}^{3}=-\left[n+2-B_2\right]\left[n+1\right]- B_3,
\vspace{2mm}\\
\mathring{\gamma}_{n}^{3}=\frac{\omega z_0\left[n+2-B_2\right]\left[ n-B_2+1-\frac{B_1}{z_0}\right]}
{\left(2n+1-B_2\right)}.
\end{array}
\end{equation*}
\begin{eqnarray}
\mathring{U}_{\ 7}^{(j)}(z)=z^{1+\frac{B_1}{z_0}}(z-z_0)^{-\frac{1}{2}-\frac{B_2}{2}-\frac{B_1}{z_0}}
\sum_{n=0}^{\infty}\mathring{a}_{n}^{7}
Z_{n-\frac{B_2}{2}+\frac{3}{2}}^{(j)}\left[\omega (z-z_0)\right],
\qquad\text{with}
\end{eqnarray}
\begin{equation*}
\begin{array}{l}
\mathring{\alpha}_{ n}^{7}=\frac{-\omega z_0\left[ n+1\right]\left[n+2-B_2-\frac{B_1}{z_0}\right]}
{\left(2n+5-B_2\right)},\qquad
\mathring{\beta}_{ n}^{7}=\mathring{\beta}_{ n}^{3},\qquad
%
\mathring{\gamma}_{n}^{7}=\frac{-\omega z_0\left[n+2-B_2\right]\left[ n+1+\frac{B_1}{z_0}\right]}
{\left(2n+1-B_2\right)}.
\end{array}
\end{equation*}

%
%
{\it Fourth and eighth sets}: 
\begin{eqnarray*}\begin{array}{l}
\text{Conditions: }B_2+\frac{2B_1}{z_0}\neq 2,{3},4\cdots.\vspace{2mm}\\
\text{Eqs. (\ref{R1a})} \text{ if } \frac{B_2}{2}+\frac{B_1}{z_0}\neq \frac{1}{2}, \qquad  
\text{(\ref{R2a})} \text{ if } \frac{B_2}{2}+\frac{B_1}{z_0}=\frac{1}{2}.
\end{array}
\end{eqnarray*}
\begin{eqnarray}
\mathring{U}_{\ 4}^{(j)}(z)=(z-z_0)^{1-B_2-\frac{B_1}{z_0}}
z^{\frac{B_1}{z_0}+\frac{B_2}{2}-\frac{1}{2}}
\sum_{n=0}^{\infty}\mathring{a}_{ n}^{4}
Z_{n-\frac{B_2}{2}+\frac{1}{2}-\frac{B_1}{z_0}}^{(j)}(\omega z),
\qquad\text{with}
\end{eqnarray}
\begin{equation*}
\begin{array}{l}
\mathring{\alpha}_{ n}^{4}=\frac{\omega z_0\left[ n+1\right]\left[n-\frac{B_1}{z_0}\right]}
{\left[2n+3-B_2-\frac{2B_1}{z_0}\right]},\qquad
\mathring{\beta}_{ n}^{4}=-\left[n+1-B_2-\frac{B_1}{z_0}\right]\left[n-\frac{B_1}{z_0}\right]- B_3,\vspace{2mm}\\
\mathring{\gamma}_{n}^{4}=\frac{\omega z_0\left[n-B_2-\frac{2B_1}{z_0}\right]\left[ n-B_2+1-\frac{B_1}{z_0}\right]}
{\left[2n-1-B_2-\frac{2B_1}{z_0}\right]}.
\end{array}
\end{equation*}
\begin{eqnarray}
\mathring{U}_{\ 8}^{(j)}(z)=(z-z_0)^{\frac{1}{2}-\frac{B_2}{2}}
\sum_{n=0}^{\infty}\mathring{a}_{ n}^{8}
Z_{n-\frac{B_2}{2}+\frac{1}{2}-\frac{B_1}{z_0}}^{(j)}\left[\omega (z-z_0)\right],
\qquad\text{with}
\end{eqnarray}
\begin{equation*}
\begin{array}{l}
\mathring{\alpha}_{n}^{8}=\frac{-\omega z_0\left[ n+1\right]\left[n+2-B_2-\frac{B_1}{z_0}\right]}
{\left[2n+3-B_2-\frac{2B_1}{z_0}\right]},\qquad
\mathring{\beta}_{n}^{8}=\mathring{\beta}_{n}^{4},
\qquad
\mathring{\gamma}_{ n}^{8}=\frac{-\omega z_0\left[n-B_2-\frac{2B_1}{z_0}\right]\left[ n-1-\frac{B_1}{z_0}\right]}
{\left[2n-1-B_2-\frac{2B_1}{z_0}\right]}.
\end{array}
\end{equation*}

%
%
%


%
%
%

%
\end{document}